\documentclass[ams, prb, twocolumn, amssymb, floatfix, longbibliography]{revtex4-2}
\usepackage{amsmath}
\usepackage{tabularx}
\usepackage{bm}
\usepackage{euscript}
\usepackage{graphicx}
\usepackage[usenames,dvipsnames]{xcolor}
\usepackage{amsfonts}
\usepackage{exscale}
\usepackage{amsbsy}
\usepackage{subfigure}
\usepackage{textcomp}
\usepackage{comment}
\usepackage{slashed}
\usepackage{hyperref}
\usepackage{braket}

\hypersetup{
	colorlinks=true,
	linkcolor=black,
	urlcolor=cyan,
	citecolor=ForestGreen
}

\newcommand{\nn}{\nonumber}

\newcommand{\beq}{\begin{equation}}
\newcommand{\eeq}{\end{equation}}
\newcommand{\be}{\begin{eqnarray}}
\newcommand{\ee}{\end{eqnarray}}

\begin{document}

\title{Neutral Excitations of Quantum Hall States: a Density Matrix Renormalization Group Study}
\author{Prashant Kumar$^{1}$ and F. D. M. Haldane$^{1}$}
\affiliation{$^1$Department of Physics, Princeton University, Princeton NJ 08544, USA
}
\date{\today}

\begin{abstract}
	We use the dynamical structure factors of the quantum Hall states at $\nu=1/3$ and $\nu=1/2$ in the lowest Landau level to study their excitation spectrum. Using the density matrix renormalization group in combination with the time-dependent variational principle on an infinite cylinder geometry, we extract the low energy properties. At $\nu=1/3$, a sharp magnetoroton mode and the two-roton continuum are present and the finite-size effects can be understood using the fractional charge of the quasi-particle. At $\nu=1/2$, we find low energy modes with linear dispersion and the static structure factor $\bar s(q) \sim (q\ell)^3$ in the limit $q\ell \rightarrow 0$. The properties of these modes agree quantitatively with the predictions of the composite-fermion theory placed on the infinite cylinder.
\end{abstract}

\maketitle

\tableofcontents
\hypersetup{linkcolor=BrickRed}


\section{Introduction}

Quantum Hall (QH) states are a beautiful set of strongly correlated phenomena in condensed matter physics exhibiting topological order, fractionally charged quasi-particles and fractional statistics. An example of such physics is obtained in the lowest Landau level around half-filling where one finds a series of fractional QH (FQH) states at Jain filling fractions $\nu_+=p/(2p+1)$ and $\nu_-=(p+1)/(2p+1)$ that accumulate at the gapless point at $\nu=1/2$ in the limit $p\rightarrow \infty$.

The theory of composite-fermions (CFs),\cite{Jain1989,Lopez91,Kalmeyer1992,Halperin1993,Simon1993,Jainbook,Fradkinbook,Son2015} traditionally formed by attaching an even number of flux quanta to fermions, has been a successful organizing principle in explaining the phenomenology of many QH states. For example, the Jain sequence at filling fractions $\nu_+$ and $\nu_-$ at can be interpreted as $p$ and $-(p+1)$ filled Landau levels (LL) of composite-fermions respectively.\footnote{In the Dirac CF description,\cite{Son2015} $\nu_{\pm}$ corresponds to $\nu_{\rm cf}=\pm(p+1/2)$ of CFs.} Their accumulation point at $\nu=1/2$ corresponds to CFs in an effective zero magnetic field. The QH state at this filling fraction corresponds to a compressible and apparently particle-hole symmetric state (within a Landau level) characterized by the presence of a CF Fermi surface.

Verifying the predictions of CF theory is of crucial importance, and various techniques such as the model wavefunctions,\cite{Jainbook} exact diagonalization\cite{Haldane1985,Rezayi1994} and density matrix renormalization group\cite{Geraedts2015} (DMRG) have been employed in previous studies. However, studying the gapless QH state at $\nu=1/2$ is challenging, since the low energy properties must be extracted. As such, the model wavefunctions may not provide a quantitative picture and the exact diagonalization is limited by the finite size. On the other hand, DMRG studies have found evidence for the presence of a circular Fermi surface and the emergent gauge field in the static properties of the ground state.\cite{Geraedts2015} Nevertheless, a quantitative comparison between the CF theory and the numerics is lacking.

In this paper, we compute the dynamical correlation functions of the QH states at $\nu=1/3$ and $\nu=1/2$ on an infinite cylinder\cite{Zaletel2015} using a combination of DMRG and the time-dependent variational principle (TDVP).\cite{Haegeman2011,Haegeman2016,Paeckel2019} Such a geometry has the advantage of having a thermodynamic limit in one of the two spatial directions. By computing the dynamical structure factor, we obtain the neutral excitation spectra at the two filling fractions and compare them with the theory. Our results for the Laughlin QH state $\nu=1/3$ can be understood in terms of fractionally charged quasi-particles that do not require the CF theory per se.\cite{Laughlin1983} On the other hand, a quasi-1D version of the CF theory\cite{Geraedts2015} makes precise quantitative predictions at low energies at $\nu=1/2$. We provide compelling evidence for CFs by showing an agreement between the theory and numerics.

At $\nu=1/3$, the primary low energy neutral excitation above the Laughlin FQH ground state is the magnetoroton mode formed by binding a quasi-hole and quasi-electron together.\cite{Girvin1986} It has been well studied\cite{Haldane1985, Laughlin1984b, Girvin1985, Girvin1986, Jainbook, Dev1992, Lopez1993, Simon1993, Kamilla1996, Jain1997, Haldane2011, Yang2012b, Golkar2016, Balram2017, Liu2018, Nguyen2021a, Nguyen2021b, Balram2022, Bartholomew2022} and shows a characteristic minimum at a finite wavevector $q_{\rm min}$. In the CF description, it can also be interpreted as a CF exciton created by exciting a CF from the zeroth LL to the first LL.\cite{Jainbook} We find that on the infinite cylinder geometry, the magnetoroton spectrum is modified due to the finite circumference equal to $L_y$. For example, the minimum lies at $q_x\ell = L_y/3\ell$ where $q_y=0$ and the $x$-direction is taken to be parallel to the axis while the $y$-direction is along the circumference of the cylinder. This effect can be understood using the fact that the quasi-particles carry fractional charge in the multiples of $e/3$. The quasi-electron and quasi-hole that constitute the roton are separated by a distance $3 q \ell^2$ transverse to the wavevector. Since the maximum separation along the circumference of the cylinder is equal to $L_y/2$, the magnetoroton minimum lies at $q_{x,\rm min.}\ell = L_y/6\ell$ when the circumference is small enough.

At $\nu=1/2$, the two-dimensional Fermi-surface theories of composite-fermions predict that the lowest energy excitations are gapless particle-hole pairs.\cite{Halperin1993,Son2015} In a field theoretical treatment of gauge-fluctuations, one obtains a marginal CF Fermi-liquid when Coulomb interactions are present and a non-Fermi liquid when the interactions are short-ranged.\cite{NayakWilczek1994short,NayakWilczeklong1994} 

When one places the CFs on the infinite cylinder, their Fermi sea splits into $N_w$ number of discrete wires since the spatial direction along the circumference is compact. We present results for the cases when the number of wires is $N_w = 2$ and $N_w = 3$ and contrast them with a quasi-1D theory of CFs that are placed on the same geometry and interact with an emergent $1+1$D gauge field.\cite{Geraedts2015} In general, the theory predicts that the emergent gauge field gaps out the total CF density mode leading to $N_w-1$ low energy modes. At $N_w=2$, we find that the dynamical structure factor contains a low energy mode with a linear dispersion that agrees quantitatively with the theory. At very low energies, it is found to have a gap that can be interpreted as a CF pairing instability. At $N_w=3$, there is only one gapless mode visible in line with the theoretical prediction that one of the two gapless modes can't appear in the density-operator at long wavelengths. Further, we show that in the limit $q_x\rightarrow 0$, the static structure factor $\bar s(q_x) \sim (q_x\ell)^3$ when the half-filled Landau level is gapless.\footnote{Notice that the usual behavior of static structure factor $\bar s(q_x) \sim (q_x\ell)^4$ is obtained under the conditions that the ground state has translational invariance, inversion symmetry, and a spectral gap.\cite{Girvin1986, Haldane2009}. The three conditions lead the static structure factor to have no momentum fluctuations, i.e. the $(q_x\ell)^2$ term vanishes, be even, and analytic in $q_x$ respectively. At $\nu=1/2$, where the excitations above the ground state are gapless, the static structure factor does not have to be an analytic function of $q_x$ and can be more singular.} In the 2D limit, it may crossover to the theoretically predicted behavior: $s(q_x) \sim (q_x\ell)^3 \log 1/q_x\ell$.\cite{Halperin1993, Read1998}

This paper is organized as follows. In section \ref{sec:model}, we explain in detail our model and methods used for performing numerical simulations. In section \ref{sec:Laughlin}, we present the dynamical structure factor for the $\nu=1/3$ Laughlin state and interpret our results based on the magnetoroton theory. In section \ref{sec:CFL}, we review the quasi-1D CF theory and compare it with our numerical results. Further numerical details are provided in Appendix \ref{sec:Laughlin_appendix} and \ref{sec:CFL_appendix}.

\section{Model and methods \label{sec:model}}
In the presence of a large perpendicular magnetic field $B$, a two-dimensional electron gas can be approximately described by a theory of electrons confined to the lowest Landau level (LLL) and interacting via density-density interactions. To this end, we define the guiding center density operator:
\begin{align}
	\bar\rho(\bm q) &= \sum_{j=1}^{N_e} e^{-i\bm q. \bm R_j}\\
	R^\alpha &\equiv r^\alpha + \ell^2\epsilon^{\alpha\beta} \Pi_\beta\\
	[R^\alpha, R^\beta] &= -i\ell^2 \epsilon^{\alpha\beta}
\end{align}
where $\ell^2 = \hbar/eB$ is the magnetic length and $B$ is the external magnetic field. $\bm R$ is the guiding center position operator, and $\bm\Pi \equiv \bm p - e\bm A$ is the kinetic momentum operator satisfying $[\Pi_\alpha, \Pi_\beta] = i\hbar^2\ell^{-2} \epsilon_{\alpha\beta},\ [\Pi_\alpha, R^\beta] =0$. We set $\hbar=e=1$ henceforth. The guiding center density operator satisfies the Girvin-MacDonald-Platzman (GMP) algebra:\cite{Girvin1986}
\begin{align}
	[\bar\rho(\bm q), \bar\rho(\bm q')] &= 2i \sin \left(\frac{\bm q\ell \times \bm q'\ell}{2}\right) \bar\rho(\bm q+ \bm q')\label{eq:GMP}
\end{align}

The Hamiltonian of the electrons projected to the LLL can be written in terms of the guiding center density operator as follows:
\begin{align}
	H = \frac{1}{2} \sum_{\bm q}\ \tilde V(\bm q) :\bar\rho(\bm q) \bar\rho(-\bm q):\label{eq:Hamiltonian}
\end{align}
where $\tilde V(\bm q) \equiv V(\bm q) e^{-q^2\ell^2/2}$ and $V(\bm q)$ is the interaction potential. In this paper, we use the Gaussian regulated Coulomb potential $V_{\rm Coul}$ and the $V_1$ Haldane pseudopotential\cite{Haldane1983} given by:
\begin{align}
	V_{\rm Coul}(r) &= \frac{e^{-r^2/2\xi^2}}{r}\label{eq:gaussian_coulomb}\\
	V_{1}(q) &= 2 L_1(q^2\ell^2)
\end{align}
where $L_1(x) = 1-x$ is the first Laguerre polynomial.

\begin{figure*}[ht!]
	\centering
	\includegraphics[width=0.9\textwidth]{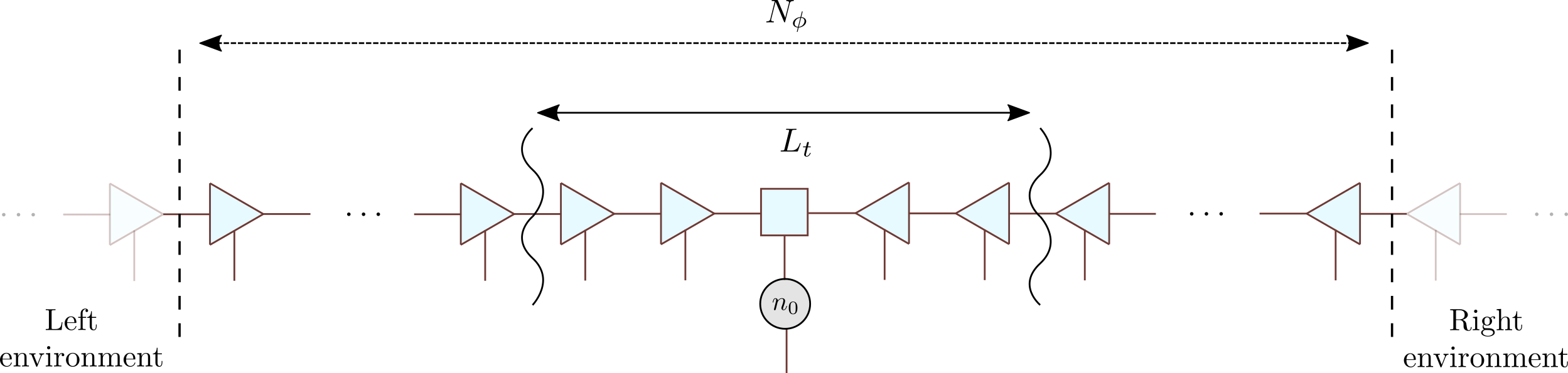}
	\caption{The initial configuration in the TDVP algorithm. A finite segment containing $N_\phi$ orbitals is embedded in the infinite quasi-1D ground state. We apply the operator $n_0$ at the mid-point to construct the excited state $\ket{\psi}$ and time-evolve it using TDVP in the region $L_t$. This region is dynamically expanded in a light cone as the disturbance spreads over time.}
	\label{fig:TDVP}
\end{figure*}

In the DMRG simulations on the infinite-cylinder geometry, we use Landau gauge with a vector potential $\bm A = (0,Bx)$. The single particle wavefunctions take the form $\psi(\bm r) \sim e^{iky} e^{-(x-k\ell^2)^2/2\ell^2}$. Assuming periodic boundary conditions along the circumference, we get $k = 2\pi n/L_y$, where $L_y$ is the circumference and $n\in \mathbb{Z}$. The guiding center density operator can be expressed in a second quantized form in this basis:
\begin{align}
	\bar\rho(\bm q) &=  e^{-iq_xq_y\ell^2/2} \sum_k e^{-i q_x k \ell^2} c^\dagger_{k} c_{k+q_y}
\end{align}
Notice that $\bar\rho(\bm q)$ is periodic under $q_x\ell \rightarrow q_x\ell + L_y/\ell$ upto a sign that depends on whether $q_yL_y/2\pi$ is an even or odd integer.

The projected static structure factor is a very useful quantity that we can extract from iDMRG simulations. It is defined by:
\begin{align}
	\bar s(\bm q) &= \frac{1}{N_\phi}\braket{0|\bar\rho(-\bm q) \bar\rho(\bm q)|0} 
\end{align}
where $N_{\phi} = N_e/\nu$ is the number of flux quanta, $N_e$ is the number of electrons, $\nu$ is the filling fraction, and $\ket{0}$ is the ground state. In this paper, we are interested in the excitation spectrum of the QH states. To this end, we define the following retarded dynamical correlation function:
\begin{align}
\bar D^R(\bm q, t) &=  \frac{i\Theta(t)}{N_\phi} \braket{0|[\bar\rho(\bm q, t),\bar\rho(-\bm q, 0)]|0}
\end{align}
where $\bar\rho(\bm q, t) = e^{iHt}\bar\rho(\bm q) e^{-iHt}$ and $\Theta(t)$ is the Heaviside step function. The imaginary part of its Fourier transform gives the projected dynamical structure factor:
\begin{align}
\bar s(\bm q, \omega) &= \lim_{\eta \rightarrow 0^+}\frac{1}{\pi} \mathrm{Im}[\bar D^R(\bm q,\omega+i\eta)]
\end{align}
The projected dynamical structure factor has the important property that it is nonzero only if there is an excited state at wavevector $\bm q$ and energy $|\omega|$. Further, the matrix element of the guiding center density operator, between the ground state and the excited state, must be nonzero. This can be seen explicitly from the following expression:
\begin{align}
\bar s(\bm q, \omega) &= \frac{1}{N_\phi}\sum_{n} |\braket{0|\bar\rho(\bm q)|n}|^2\ \delta(\omega -E_n+E_0)\nn\\
&\ \ \ \ \ \ \ \ \ \ \ \ \ \ - \{\omega \leftrightarrow -\omega\}
\end{align}
where $\ket{n}$ is an excited state with the energy $E_n-E_0$ above the ground state. It can be noticed that the integral of the projected dynamical structure factor, over positive frequencies, is equal to the projected static structure factor, i.e.,
\begin{align}
\bar s(\bm q) &= \int_0^\infty d\omega\ \bar s(\bm q, \omega)
\end{align}

In the following two subsections, we briefly review the single-mode approximation theory of GMP\cite{Girvin1986} and explain the details of our approach to the numerical computation of the dynamical structure factor. A reader interested in the results and their interpretations may skip forward to section \ref{sec:Laughlin}.

\subsection{Single-mode approximation}
The magnetoroton is a fundamental low energy neutral collective excitation of the FQH states. For example, at $\nu=1/3$, it corresponds to a bound pair of a quasi-hole and quasi-electron. GMP had proposed a single-mode approximation (SMA) to study this mode.\cite{Girvin1986} We summarize the main aspects of their theory in this subsection. 

We postulate an excited state wavefunction with momentum $\bm q$ by applying the guiding center density operator on the ground state:
\begin{align}
	\ket{\bm q} &= \frac{\bar\rho(\bm q)}{\sqrt N_\phi} \ket{0}
\end{align}
The energy of this excited state, in the presence of inversion symmetry, can be expressed as:
\begin{align}
	\Delta E(\bm q) &= \frac{\bar f(\bm q)}{\bar s(\bm q)}\label{eq:SMA_energy}\\
	\bar s(\bm q) &= \braket{\bm q|\bm q}\\
	\bar f(\bm q) &=  \braket{\bm q|(H-E_0) |\bm q} \nn\\
	&= \frac{1}{2N_\phi} \braket{0| \left[\bar\rho(\bm q)^\dagger,[H, \bar\rho(\bm q)]\right] |0}
\end{align}
where $E_0$ is the ground state energy. For the Hamiltonian in Eq. \eqref{eq:Hamiltonian}, we can use the GMP algebra \eqref{eq:GMP} and obtain:
\begin{align}
	\bar f(\bm q) &= \frac{1}{2} \sum_{\bm q'} \tilde V(\bm q)\ 4\sin^2\left(\frac{\bm q\ell \times \bm q'\ell}{2}\right) \left(\bar s(\bm q'+\bm q) - \bar s(\bm q')\right)
\end{align}
Thus, the SMA energy can be obtained purely from the ground state properties.

A general result of GMP is that $f(\bm q) \sim q^4$ in the limit $q \rightarrow 0$. Crucially, for gapped states with inversion symmetry, $\bar s(\bm q) \sim q^4$ at long wavelengths. Therefore the energy gap predicted by the SMA approaches a nonzero constant in this limit. However for gapless states, the static structure factor may vanish more slowly and the SMA may also describe certain properties of the gapless modes. 

In this paper, we compute the dynamical density-density correlation function of the quantum Hall states at $\nu=1/3$ and $\nu=1/2$. As such, we make use of the SMA in a few different ways. At $\nu=1/3$, we directly compare the SMA energy \eqref{eq:SMA_energy} with the magnetoroton spectrum. Additionally, we also use the fact that the SMA energy is equal to the average energy of excitations weighted by the dynamical structure factor $\bar s(\bm q, \omega)$, i.e.,
\begin{align}
	\Delta E(\bm q) = \int_0^\infty d\omega\ \omega \bar s(\bm q, \omega)\Big /\int_0^\infty d\omega\  \bar s(\bm q, \omega)\label{eq:SMA_avg_E}
\end{align}
Thus, it can serve as a straightforward check of the numerically computed dynamical structure factor. Notice that an implication of this relation is that the SMA gives an upper limit on the energy of excitations.

\subsection{Dynamics via TDVP}

The time-dependent variational principle\cite{Haegeman2011,Haegeman2016,Paeckel2019} can be used to time-evolve a quantum mechanical wavefunction expressed as a matrix-product state (MPS). In this paper, we use TDVP to compute the dynamical density-density correlation function. We limit ourselves to the case $q_y = 0$ where the guiding center density operator $\bar\rho(q_x) \equiv \bar\rho(q_x, q_y=0)$ takes a simple form in the Landau gauge:
\begin{align}
	\bar\rho(q_x) = \sum_k e^{-iq_x k\ell^2} n_k
\end{align}
where $n_k = c^\dagger_k c_k$ is the occupation number operator of the $k^{th}$ Landau orbital. We define the following time-dependent correlation function:
\begin{align}
	C_{pk}(t) &=\braket{0|n_p(t) n_{k}(0)|0} \nn \\
	&= \braket{0|e^{iHt}n_p e^{-iHt} n_{k}|0} \label{eq:corr_tdvp}
\end{align}

For simplicity, consider the case when the ground state is symmetric under translation by $1$-site. We have $C_{pk}(t) \equiv C_{p-k}(t)$. Therefore, using the time-evolved excited state $\ket{\psi(t)} = e^{-iHt} n_0 \ket{0}$, we can obtain $C_{k} (t)$ by measuring the following matrix element, i.e.,
\begin{align}
	C_k(t) &= e^{iE_0t}\braket{0|n_k|\psi(t)}
\end{align}

The retarded density-density correlation function can be calculated by performing a Fourier transform on the connected part of the correlation function $C^c_k(t)$:
\begin{align}
	\bar D^R(q_x, \omega+i\eta) &= \nn\\
	&\hspace{-15pt} \sum_k \int_0^\infty dt \ e^{i(\omega+i\eta) t} e^{-iq_x k\ell^2} i\left(C^c_{k}(t){-C^c_{k}}^*(t)\right)
\end{align}
where we have included a broadening factor $\eta>0$. The calculation can be readily generalized to the case when the ground state is invariant under translation by $N_u$ orbitals. In this case, we time-evolve $N_u$ excited states defined by $\ket{\psi_m(t)} = n_m \ket{0}$, where $m \in \{0, 1, \cdots, N_u-1\}$. The $N_u$ number of correlation functions $C^{(m)}_k(t) = e^{iE_0t} \braket{0|n_{k+m}|\psi_m(t)}$ are then averaged to obtain the dynamical correlation function.

Let us summarize our time-evolution approach.\cite{Milsted2013} Using iDMRG, we first converge to a ground state with a unit cell of $N_u$ orbitals. In principle, we can create an infinite 1D system by repeating this unit cell. However, since the disturbance is local and spreads in a light cone over time, we consider a geometry where a finite size segment of the 1D system containing $N_\phi$ orbitals is embedded in an otherwise uniform environment. Then the initial excited state $\ket{\psi_m(0)}$ can be constructed by applying the orbital occupation number operator $n_m$ near the mid-point of the segment. We then perform the two-site TDVP (TDVP2) with a time-step $\delta$ in a region of size $L_t$ orbitals around the midpoint and compute an MPS representation of the state $\ket{\psi_m(n\delta)}$. As the disturbance spreads, the region of time-evolution is dynamically expanded up to the maximum number of orbitals equal to $N_\phi$ and the bond-dimension is allowed to increase. This is summarized in Fig. \ref{fig:TDVP}.

The gapped ground states such as the Laughlin FQH state at $\nu=1/3$ can be well approximated in an MPS form due to a finite correlation length and the area law of entanglement. However, the excited states or gapless ground states, for example at $\nu=1/2$, are more challenging. In obtaining the dynamical properties, we necessarily deal with the latter and increasing the bond-dimension or decreasing the time-step to overcome this problem is often not practical. As such, we use a few tricks to improve our results. To this end, we define a measure of the error as follows:
\begin{align}
	\mathcal{E}(t) &= e^{iE_0t}\frac{ \braket{0|\psi(t)}}{\braket{0|\psi(0)}}\label{eq:error_tdvp}
\end{align}
For an exact time-evolution, $\mathcal{E}(t) = 1$. However, over the course of time-evolution, the TDVP algorithm adjusts the MPS basis states to the target time-evolved excited state. Therefore, the time-evolution operator $e^{-iHt}$ applied on $\ket{\psi(0)}$ is not identical to applying it on the ground state and $\mathcal{E}(t)$ becomes time-dependent. This error, for example, leads to correlations outside the light cone. Therefore, to get physically sensible results, the disconnected piece of the correlation function should be defined by inserting $\ket{0}\bra{0}$ to the left of the time evolution operator $e^{-iHt}$ in Eq. \eqref{eq:corr_tdvp}, i.e.,
\begin{align}
	C^c_k(t) &= C_k(t) - C^{dis.}_k(t)\\
	C^{dis.}_k(t) &= e^{iE_0t} \braket{0|n_k|0}\braket{0|\psi(t)}\\
	&=\mathcal{E}(t) C^{dis.}(0)
\end{align}
In addition to this, such an error induces the breaking of charge conservation when measured from the correlation function, i.e.:
\begin{align}
	N_{e,corr}(t) &= \frac{1}{\braket{0|n_0|0}} \sum_k C_k(t)\\
	&= N_e  \mathcal{E}(t)
\end{align}
This might be undesirable especially when averaging over a unit cell. We propose to fix this by dividing the measured correlation function by the error:
\begin{align}
	C^c_k(t) &= \frac{C_k(t)}{\mathcal{E}(t)} - C^{dis.}(0)\label{eq:con_corr}
\end{align}
In general, we find that this procedure makes a difference only when the bond-dimensions are not big enough to sufficiently capture the time-evolved state. Notice that the disconnected piece has now become time-independent.

\section{FQH state at $\nu=1/3$\label{sec:Laughlin}}

\begin{figure}
	\centering
	
	\includegraphics[width=0.3\textwidth]{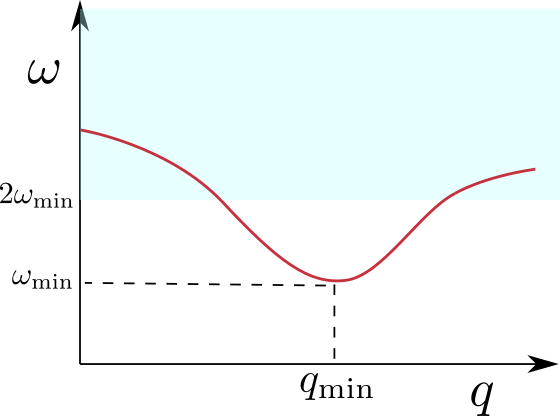}
	
	\caption{The neutral excitation spectrum of $\nu=1/3$ FQH state on an infinite plane. The red curve is the magnetoroton mode composed of a bound pair of a fractionally charged quasi-hole and quasi-electron. It has a minimum at $q=q_{\rm min}$ and $\omega = \omega_{\rm min.}$. The cyan region contains the two-roton continuum, and other states such as multiple rotons, with a minimum energy equal to twice the energy of the roton minimum.
	}
	
	\label{fig:Laughlin_2D_spectrum}
\end{figure}

\begin{figure*}
	\centering
	
	\begin{minipage}{0.7\textwidth}
		\begin{center}
			\includegraphics[width=1\textwidth]{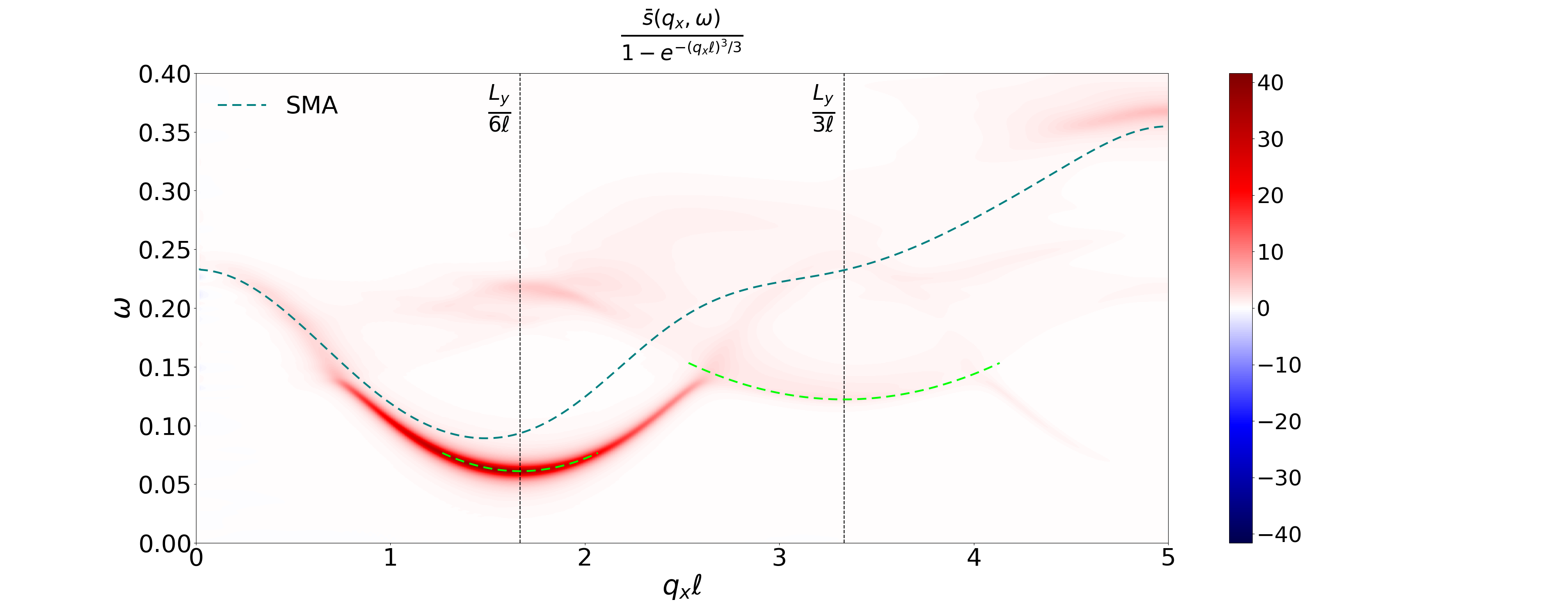}
			
			(a)
		\end{center}
	\end{minipage}%
	\begin{minipage}{0.32\textwidth}
		\begin{center}
			\includegraphics[width=1\textwidth]{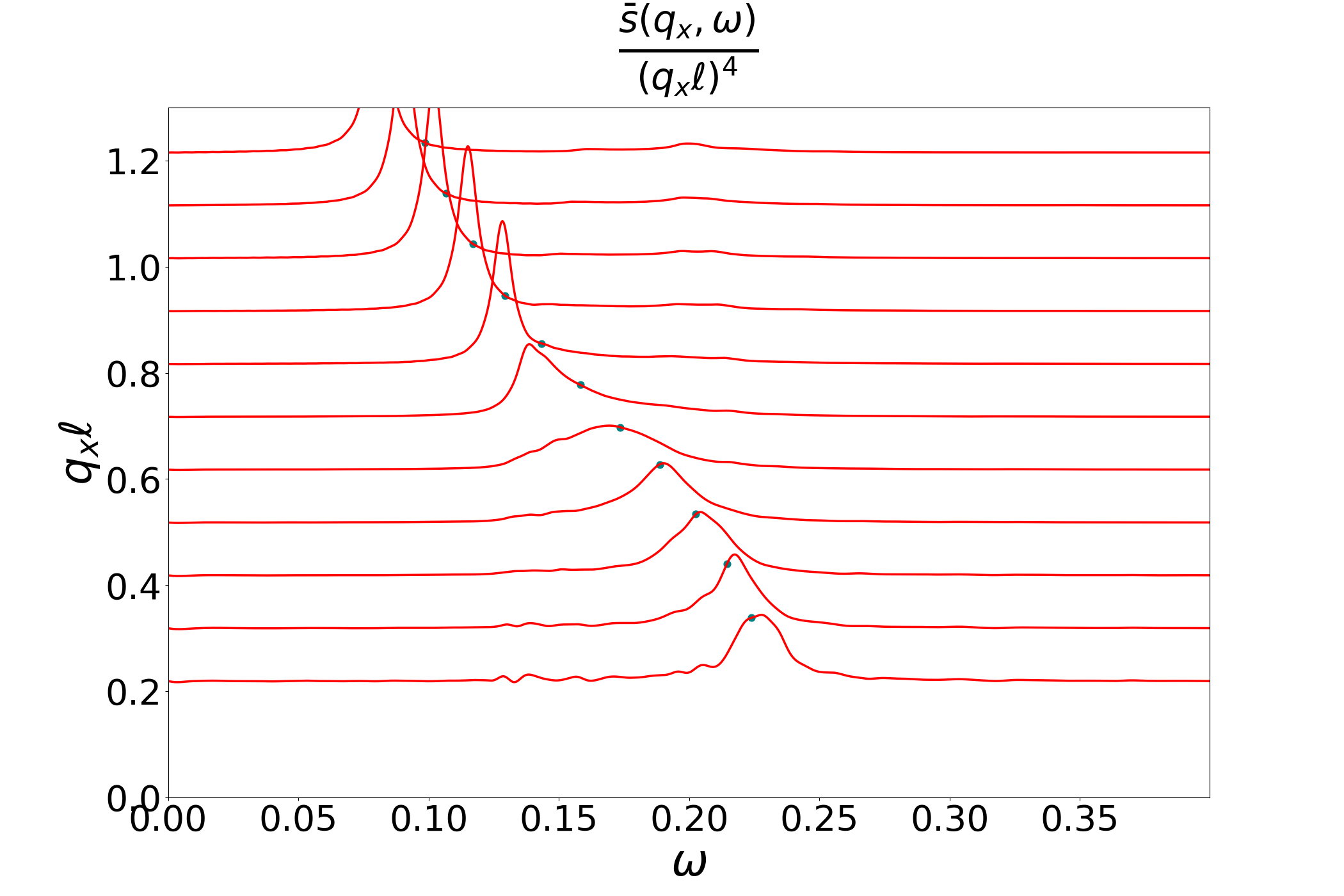}
			
			(b)
		\end{center}
	\end{minipage}
	
	\caption{(a) The dynamical structure factor of $\nu=1/3$ Laughlin FQH state at $L_y = 10\ell$ for $V_1$ Haldane pseudopotential. A sharp roton mode is visible at $0.7 \lesssim q_x\ell \lesssim 2.6$ with the roton minimum  at $q_x \ell= q_{x,\rm min}\ell = L_y/6\ell$. The energy of the roton mode is periodic under $q_x \ell \rightarrow q_x\ell + L_y/3\ell$. The two-roton continuum can be observed at around $q_x\ell = L_y/3\ell$. The dashed lime curve shows the sum of energies of two independent rotons each with a momentum $q_x\ell/2$. The dashed blue curve is the energy obtained from the single-mode approximation (SMA). (b) The dynamical structure factor at small $q_x\ell$. The sharp roton mode becomes overdamped as it enters the two-roton continuum at $q_x\ell \approx 0.7$.}
	\label{fig:Haldane_10_dynm_S_Laughlin}
\end{figure*}

\begin{figure*}
	\centering
	
	\begin{minipage}{0.7\textwidth}
		\begin{center}
			\includegraphics[width=1\textwidth]{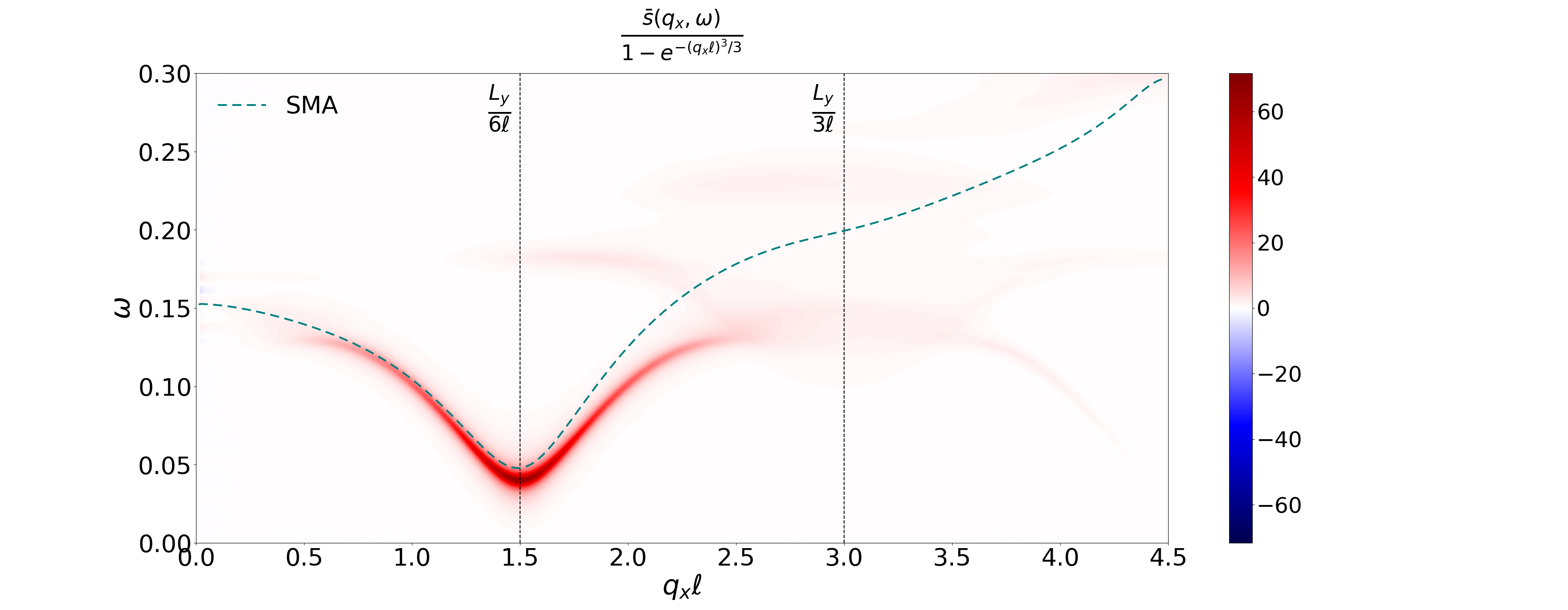}
			
			(a)
		\end{center}
	\end{minipage}%
	\begin{minipage}{0.32\textwidth}
		\begin{center}
			\includegraphics[width=1\textwidth]{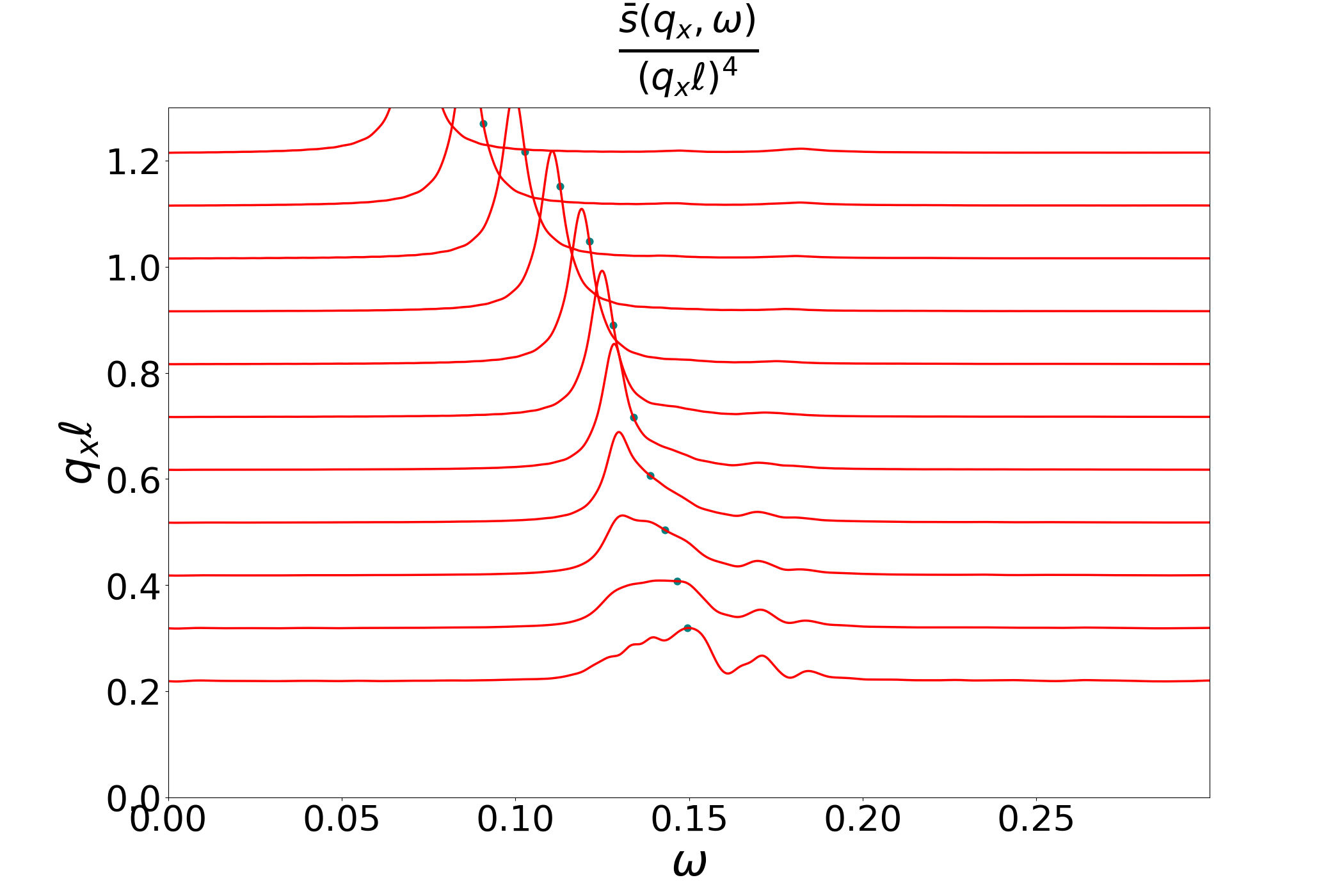}
			
			(b)
		\end{center}
	\end{minipage}%

	\caption{(a) The dynamical structure factor of $\nu=1/3$ Laughlin FQH state at $L_y = 9\ell$ for Gaussian Coulomb potential with $\xi=20\ell$. Generally, the features are similar to Fig. \ref{fig:Haldane_10_dynm_S_Laughlin}. The roton mode becomes overdamped at $q_x\ell \lesssim 0.6$ is as seen in (b).}
	\label{fig:Coulomb_9_dynm_S_Laughlin}
\end{figure*}

For the Laughlin FQH state at $\nu=1/3$, the unit cell is composed of three orbitals with the root configuration $010$.\cite{Bernevig2008,Bergholtz2008} 
In Fig. \ref{fig:Haldane_10_dynm_S_Laughlin}, we show the dynamical structure factor for the $V_1$ Haldane pseudopotential interaction at $L_y=10\ell$ and $q_y=0$. While in Fig. \ref{fig:Coulomb_9_dynm_S_Laughlin}, we plot the case of Coulomb interactions defined in Eq. \eqref{eq:gaussian_coulomb} with $\xi = 20\ell$ and $L_y=9\ell$. Additional numerical details are provided in Appendix \ref{sec:Laughlin_appendix}.

To interpret our results on the cylindrical geometry with a finite circumference, let us first describe the low energy neutral excitations of the Laughlin state in the 2D limit. As shown in Fig. \ref{fig:Laughlin_2D_spectrum}, the lowest energy mode is the magnetoroton formed by binding a quasi-hole with a quasi-electron.\cite{Laughlin1984,Haldane1985,Girvin1986,Jainbook,Simon1993} The roton dispersion has a minimum at $q = q_{\rm min}$ that represents the short distance Wigner crystal like correlations. In addition, there is a continuum of excited states formed by exciting two or more rotons above the ground state, in addition to other possible neutral modes. The continuum starts at twice the energy of the roton minimum.

We now explain the general features observed in the numerically computed dynamical structure factor at $\nu=1/3$ on the infinite cylinder geometry. In Fig. \ref{fig:Haldane_10_dynm_S_Laughlin}, we observe a sharp magnetoroton branch at $0.07 \lesssim q_x\ell \lesssim 2.6$. At other wavevectors, the roton state is overdamped or has a small weight, nevertheless, it can still be distinctly identified. The properties of this mode can be understood in terms of the fractionally charged excitations as follows. The density operator $\rho(\bm q)$ applied on the ground state kicks the guiding center of one electron by the vector $R^\alpha_{\rm eh} = -\epsilon^{\alpha\beta} q_\beta\ell^2$, creating an electron-hole pair with a dipole moment $p^\alpha = -e \epsilon^{\alpha\beta} q_\beta\ell^2$. However, the low energy excitations of the Laughlin state are the quasi-holes and quasi-electrons that carry fractional charges in multiples of $e/3$. The roton mode, being neutral, is a bound pair of a single quasi-hole and quasi-electron. Thus, the guiding centers of its constituents must be separated by the vector $R^\alpha_{\rm qeqh} = -3 \ell^2 \epsilon^{\alpha\beta}q_\beta$ to carry a dipole moment equal to $p^\alpha$. Since $q_y=0$, $R^\alpha_{\rm qeqh}$ is parallel to the circumference and therefore, we expect that the roton energy would be periodic under $R^y_{\rm qeqh} \rightarrow R^y_{\rm qeqh} + L_y$ or $q_x\ell \rightarrow q_x\ell + L_y/3\ell$. In addition, we can separate the pair of quasi-particles by a maximum distance of $L_y/2$ and thus the roton spectrum must be symmetric around $q_x\ell = L_y/6\ell$. Both of these features are observed in Figs. \ref{fig:Haldane_10_dynm_S_Laughlin} and confirm the fractional charge of the constituents of the roton. We remark that these statements pertain to the energy of the roton excitation and not the spectral weights. The dynamical structure factor involves electron operators as opposed to the quasi-particle operators and thus the spectral weight is periodic with a longer period equal to $\Delta q_x \ell = L_y/\ell$.

The finite cylinder circumference acts as a cutoff on the maximum separation of the quasi-hole and the quasi-electron when $q_y=0$. Therefore, at small circumference, the roton minimum lies at $q_{x, \rm min}\ell = L_y/\ell$. As we increase $L_y$, we expect that the position of the roton minimum would become independent of size once $q_{x, \rm min}$ reaches its true 2D-limit value. However, the energy spectrum would still remain periodic with the period $\Delta q_x\ell = L_y/3\ell$ and the period would diverge in the thermodynamic limit. 

At $q_x \lesssim q_{x, \rm min}$, the roton dispersion is close to the prediction of SMA. However, at larger wavevectors, it fails to describe the roton dispersion. This can be understood using the fact that the SMA creates an electron-hole pair. At small separations, i.e. $q_x\ell \ll 1$, the electron and the hole overlap and can describe the roton where the quasi-electron and the quasi-hole overlap as well. In this limit, the roton corresponds to a collective density mode that has a dipole moment transverse to the wavevector. However, at larger $q_x\ell$ values, the electron and the hole are composed of three quasi-electrons and three quasi-holes respectively that are well separated in space. Thus, the SMA variational state contains additional contributions from several exact excited states that contain more than one roton.

The spread out dynamical structure factor at around $q_x\ell = L_y/3\ell$ can be interpreted as the two roton continuum. As shown in the figure, the minimum energy of the continuum matches with the sum of the energies of two rotons each sharing half of the total momentum. Notice that this is different from the 2D case where one can create two rotons precisely at the minimum whose wavevectors can be added to give any $q \leq 2q_{\rm min}$ and thus the two-roton minimum is flat as a function of $q$ for $q<2q_{\rm min}$. However, $q_y$ is discrete in the infinite cylinder geometry and thus the minimum energy of the two-roton continuum acquires a curvature. Moreover, we find that as the magnetoroton enters the continuum, it becomes overdamped due to mixing with the two-roton states. We also expect the two-roton continuum to be present at around $q_x\ell = 0$ with its minimum energy equal to twice the roton minimum. However, creating two unbound rotons requires separating the quasi-particles by a large distance, while $\bar\rho(\bm q)$ operator creates an electron-hole pair separated by a small distance at small $q\ell$. Thus, the continuum at small $q_x\ell$ does not have any significant spectral weight in the dynamical structure factor. Nevertheless, we can see the effect of its presence in the overdamping of the roton mode at $q_x\ell\lesssim 0.7$.

In Fig. \ref{fig:Coulomb_9_dynm_S_Laughlin}, we plot the dynamical structure factor of the Laughlin state for Coulomb interactions and $L_y=9\ell$. In general, the features are similar to the $V_1$ Haldane pseudopotential case. There are a few minor differences such that the two-roton continuum appears with a smaller weight and the roton minimum is deeper.

\section{CFL state at $\nu=1/2$ \label{sec:CFL}}
The half-filled Landau level is a gapless QH state that exhibits properties of a Fermi surface. The theory of composite-fermions has been quite successful in understanding this phase.\cite{Kalmeyer1992,Halperin1993,Son2015} Traditionally, CFs are constructed by attaching two-flux quanta to electrons. At half-filling, these flux quanta cancel the external magnetic field and thus the CFs fill up a Fermi sea. In this section, we use an alternative Dirac CF description proposed by Son\cite{Son2015} that explicitly contains the particle-hole symmetry at $\nu=1/2$. In this language, the CFs can be interpreted as vortices and their theory is a part of the duality web in 2+1 dimensions.\cite{Seiberg:2016gmd,Karch:2016sxi,Chen2018}

\begin{figure*}
	\centering
	
	\begin{minipage}{0.5\textwidth}
		\begin{center}
			\includegraphics[width=0.7\textwidth]{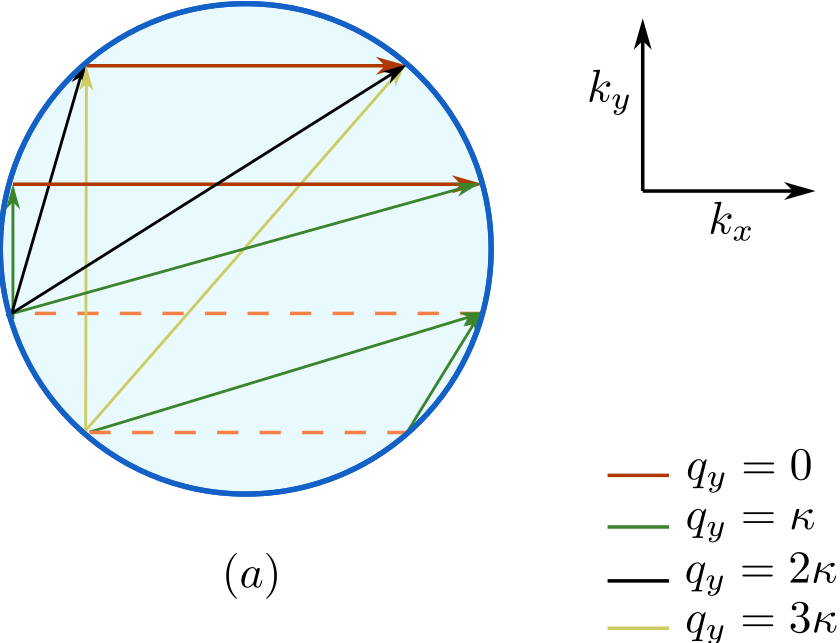}
			
		\end{center}
	\end{minipage}%
	\begin{minipage}{0.5\textwidth}
		\begin{center}
			\includegraphics[width=0.7\textwidth]{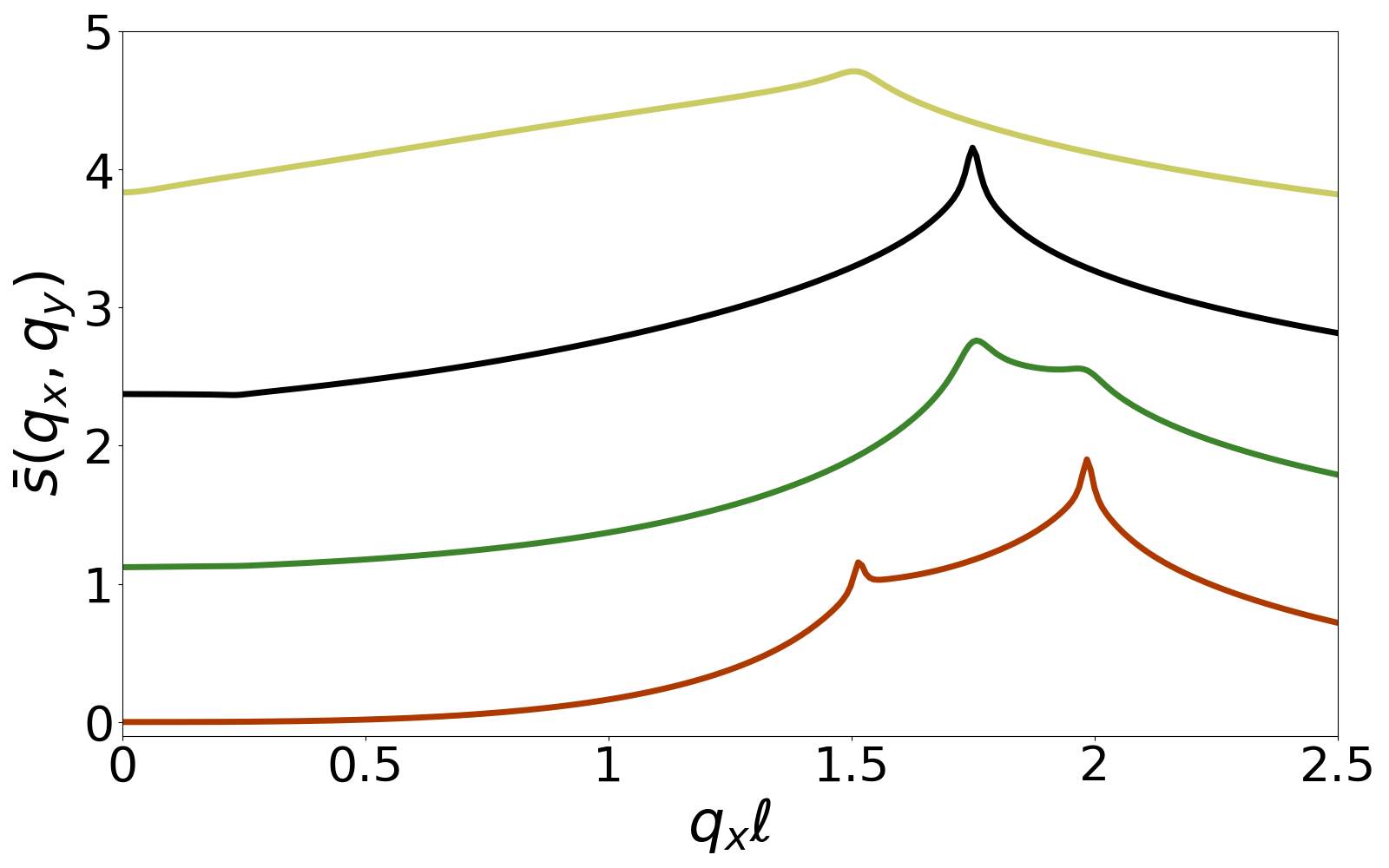}
			
			{ (b)}
		\end{center}
	\end{minipage}
	
	\caption{The Fermi sea of CFs splits into discrete wires in an infinite cylinder geometry at $k_y = n\kappa$ where $n$ is an integer if the number of wires is odd and half-integer if the number of wires is even. Also, $\kappa=2\pi/L_y$. (a) The unique scatterings between the 8 Fermi-points for the case of 4 wires (horizontal lines). (b) $\bar s(\bm q)$ calculated in iDMRG for $q_y/\kappa = 0,1,2,3$ at $L_y=14\ell$. It contains singularities at wavevectors consistent with (a).}
	\label{fig:CF_singularties}
\end{figure*}

The CF Fermi-surface description was corroborated numerically in Ref. \onlinecite{Geraedts2015} where the authors numerically computed the ground state at $\nu=1/2$ on the infinite cylinder geometry using iDMRG. Due to the finite circumference, the CF Fermi sea splits into wires at momenta $k_y =  n\kappa$, where $n$ is either an integer or a half-integer depending on whether the number of wires is odd or even and $\kappa=2\pi/L_y$. They found that the static structure factor contains singularities at wavevectors consistent with a circular Fermi surface in the 2D limit. This is summarized in Fig. \ref{fig:CF_singularties}.

In this section, we obtain the neutral excitation spectrum of the half-filled Landau level by computing the dynamical density-density correlation function. We contrast our results with the Dirac CF theory placed on the infinite cylinder and show that the theory agrees with our numerics quantitatively. Our plan for this section is as follows. We briefly describe the theory of quasi-1D CFs in subsection \ref{sec:quasi_1D_CF} and then present numerical results in subsection \ref{sec:numerics_CFL}.

\subsection{Quasi-1D theory of CFs\label{sec:quasi_1D_CF}}
In Ref. \onlinecite{Geraedts2015}, a quasi-1D theory of CFs was proposed on a torus of dimensions $L_x\times L_y$ with $L_x \rightarrow \infty$ and  $L_y \sim \ell$. The CFs interact with a gauge field and can be thought of as the quasi-1D descendants of Son's Dirac CFs.\cite{Son2015} An important advantage of the 1+1D geometry is that one can use bosonization to derive explicit expressions for the dynamical correlation functions even in the presence of interactions. In this subsection, we review its important aspects. 

We start with the 2+1D Son's theory in Euclidean time:\cite{Son2015,Son2018}
\begin{align}
	\mathcal{L} &= -\bar\psi \gamma^\tau D_\tau^a \psi - v \bar\psi \gamma^jD_j^a\psi -\mu \psi^\dagger\psi+ \frac{i}{4\pi}Ada \nn\\
	&\ \ \ \ \ \ \ - \frac{i}{8\pi}AdA + i F_{\tau j} P^j + \cdots\\
	F_{ij}P^j &= \frac{i}{2} \left(\psi^\dagger (D^a_i  \psi) - (D^a_i \psi^\dagger)\psi \right) \label{eq:dipole_term}\\
	\rho_e &=i\frac{\delta \mathcal{L}}{\delta A_\tau} = \frac{B-b}{4\pi}  - \partial_j P^j\label{eq:electric_density}
\end{align}
where $D_\mu^a \equiv \partial_\mu - i a_\mu$, the metric signature is $(+,+,+)$, $\gamma^\tau = \sigma^z, \gamma^x = \sigma^y, \gamma^y=-\sigma^x$, $\psi^\dagger = -\bar\psi \gamma^\tau$, $adc \equiv \epsilon^{\mu\nu\lambda}a_\mu \partial_\nu c_\lambda$, $F_{\mu\nu} =  \partial_\mu A_\nu-\partial_\nu A_\mu$ with $F_{ij} = \epsilon_{ij}B$. ``$a_\mu$'' is the emergent gauge field, ``$A_\mu$'' is the external gauge field and $b = \epsilon^{ij}\partial_ia_j$. Further, $\mu$ is the chemical potential of the CFs and $\rho_e$ is the electric charge density. We have introduced ``$v$'' as the velocity of the Dirac fermion. Notice that $a_\tau$ implements the constraint $\psi^\dagger\psi = 1/4\pi\ell^2$.

We have included the dipole term in the Lagrangian that was postulated to satisfy Galilean invariance.\cite{Prabhu2017,Son2018} As we'll see, the electric density-density correlation functions at $q_y =0$ will involve the electric charge density contributed by this term. We can interpret it as follows. As we can notice, the electric dipole density is proportional and perpendicular to the momentum density of the Dirac fermion. This is consistent with the lowest Landau level structure where the guiding center position in one direction is the generator of translations in the perpendicular direction. We note that this term was also proposed in Ref. \onlinecite{Geraedts2015} based on symmetries and the magnetic translation algebra.

As explained in Fig. \ref{fig:CF_singularties}, the CF Fermi sea splits into wires separated in the y-momentum direction by $\Delta k_y = m \kappa$ in the quasi-1D geometry where $m\in \mathbb{Z}$. These wires can be studied using bosonization.\cite{Giamarchibook,Delft1998BosonizationFB} Close to the Fermi-points, we denote the Dirac CF creation and annihilation operators by $f^\dagger_{nr} (k_x), f_{nr}(k_x)$ respectively. ``$n$'' indexes the $y$-momentum of the wire, i.e., $k_y = n\kappa$, while $k_x$ is the $x$-momentum. Further, $r =\pm 1$ or $r=R/L$ distinguishes the right/left mover. We can write a linearized CF Hamiltonian:
\begin{align}
	\psi(\bm r) &\sim \frac{1}{\sqrt{L_x L_y}}\sum_{ k_x}' \sum_{nr} \frac{1}{2} \begin{pmatrix}
		1 \\ \frac{k_+}{k}
	\end{pmatrix} e^{i\bm k.\bm r} f_{nr}(k_x) \\
	H_{cf}^0 &= \sum_{k_x}'\sum_{nr}v_{nr} (k_x  - {K^F_{nr}}_x) :f^\dagger_{nr}(k_x) f_{nr}(k_x):
\end{align}
where $\psi(\bm r)$ is the Dirac CF annihilation operator, the primed sum indicates that $k_x$ is near the Fermi-point labeled by ``$nr$'' with Fermi-momentum $\bm K^F_{nr}$ and $k_{\pm} = k_x \pm i k_y$. Also, $v_{nr}$ are the velocities close to the Fermi-point. 

Let us define the slow bosonic variables as follows:
\begin{align}
	f_{nr} (x) &= \frac{1}{\sqrt{L_x}}\sum_{k_x} f_{nr}(k_x)e^{i(k_x-{K^F_{nr}}_x)x}\\
	&= \frac{1}{\sqrt{2\pi\alpha}} F_{nr}e^{-ir\pi x/L_x}e^{ir\phi_{nr}(x)}
\end{align}
where $\alpha$ is a large momentum regularization with the units of distance and $F_{nr}$ are the unitary Klein factors that anti-commute between different species. 

The CFs are coupled to the emergent gauge field $a_\mu(\bm q)$. At $q_y=0$,  we assume that it corresponds to a $1+1$D gauge field. This implies that the $q_y=0$ Fourier component of the emergent magnetic field is zero, i.e. $b(q_x, q_y=0) = 0$. Thus, there is only one polarization of the gauge field important at $q_y=0$. Let us pick the gauge $a_y(\bm q) = 0$ and $a_x(q_x, q_y=0) = 0$. Our strategy for dealing with the emergent gauge field is as follows. We integrate out all of its Fourier components at finite wavevectors and assume that they generate scatterings between the CF wires. The long wavelength component, i.e., $q_y=0$ and $q_x \approx 0$, is treated separately and would be shown to result in a long range interaction that gaps out the total CF density mode.

We can write a general quadratic theory of the bosonic modes after integrating out the finite wavevector Fourier components of the emergent gauge field as follows:
\begin{align}
	\mathcal{L} &= \frac{i}{4\pi} (-1)^r\partial_x \phi_{nr} \partial_\tau \phi_{nr} + \frac{1}{4\pi} v_{nr,n'r'} \partial_x \phi_{nr} \partial_x \phi_{n'r'}\nn\\
	&\ \ \ \ \ \ + i a_\tau^{q_y=0} \frac{1}{2\pi\sqrt{L_y}} \sum_{nr} \partial_x\phi_{nr}
\end{align}
where $v_{nr,n'r'}$ is a positive-definite real matrix. Its off-diagonal components contains contributions from the scatterings between the Fermi-points due to the interactions generated by the emergent gauge field. The last term corresponds to the remaining coupling between the CF density and the emergent gauge-field at long wavelengths.

An important consequence of the presence of an emergent gauge field is that it induces a long range density-density interaction that gaps out the total charge mode of the CFs. We can see this explicitly by considering a Maxwell term at long wavelengths: $\mathcal{L} = \frac{q_x^2}{2g^2} (a_\tau^{q_y=0})^2$. Upon integrating out $a_\tau^{q_y=0}$, we generate a mass proportional to $g$ for the total density mode:
\begin{align}
	\mathcal{L}_{\rm eff}[g] &= \frac{g^2}{2(2\pi)^2L_y} \left(\sum_{nr}\phi_{nr}\right)^2
\end{align}
The massive CF plasmon is an effect of the softening of the constraint imposed by the $a_\tau$ gauge field. Its mass goes to infinity as the constraint gets hardened in the limit the Maxwell term goes to zero. Therefore, we are left with at most $N_w-1$ gapless modes. This was verified in Ref. \onlinecite{Geraedts2015} using the central charge calculated in DMRG.

We now turn to the discussion of the discrete symmetries. The theory of CFs at $\nu=1/2$ on the torus has three discrete symmetries: particle-hole (time-reversal for CFs), mirror and inversion symmetries. Let us summarize their actions on the CF operators:
\begin{enumerate}
	\item Particle-hole symmetry: 
	\begin{align}
		\mathcal{T}^{-1} f_{nr}(k_x)\mathcal{T} &= \frac{k_+}{k}f_{-n,-r}(-k_x)\\
		\mathcal{T}^{-1} i\mathcal{T} &= -i\\
		\mathcal{T}^{-1} \phi_{nr}\mathcal{T} &= \phi_{-n,-r}
	\end{align}

	\item Inversion symmetry:
	\begin{align}
		\mathcal{I}^{-1} f_{nr}(k_x)\mathcal{I} &= f_{-n,-r}(-k_x)\\
		\mathcal{I}^{-1} \phi_{nr}(x)\mathcal{I} &= -\phi_{-n,-r}(-x)
	\end{align}

	\item Mirror symmetry:
	\begin{align}
		(\mathcal{M}_x \mathcal{T})^{-1} f_{nr}(k_x)\mathcal{M}_x \mathcal{T} &= f_{n,-r}(-k_x)\\
		(\mathcal{M}_x \mathcal{T})^{-1} i\mathcal{M}_x \mathcal{T} &= -i\\
		(\mathcal{M}_x \mathcal{T})^{-1} \phi_{nr}\mathcal{M}_x \mathcal{T} &= \phi_{n,-r}
	\end{align}

\end{enumerate}
A direct consequence of these symmetries is $v_{nr,n'r'} =  v_{-n-r,-n'-r'},\ v_{nr,n'r'} =  v_{n-r,n'-r'}$. We can make the inversion and PH symmetries explicit in the bosonized theory by combining the bosonic modes in the following way:
\begin{align}
	2\Phi_{n} &= \phi_{nR} + \phi_{-nL}\\
	2\Theta_{n} &= \phi_{nR} - \phi_{-nL}
\end{align}
Notice that the pair $(\Phi_n,\Theta_n)$ is (even, odd) under PH symmetry and (odd, even) under inversion. Also, $(\mathcal{M}_x \mathcal{T})^{-1} \Phi_{n}\mathcal{M}_x \mathcal{T} = \Phi_{-n}$ and $(\mathcal{M}_x \mathcal{T})^{-1} \Theta_{n}\mathcal{M}_x \mathcal{T} = -\Theta_{-n}$.

We are interested in the electric density-density correlation function at $q_y=0$. Since there is no emergent magnetic field component at $q_y=0$, the leading contribution to the electron-density operator is the dipole term in Eq. \eqref{eq:electric_density}.
At long wavelengths, i.e. $q_y=0,\ q_x\ell \ll 1$, the density operator $\rho_e(q_x) \sim -iq_x \times\ell^2 \sum_{nr}{K^F_{nr}}_yf^\dagger_{nr}(k_x) f_{nr}(k_x+q_x)$. As such, the effect of electron-electron interactions beyond the mean-field approximation are irrelevant. To see this explicitly, let us consider the following form of the interactions: 
\begin{align}
	H_{ee} &= \frac{1}{2} \sum_{\bm q} V(\bm q)\rho_e(\bm q) \rho_e(-\bm q) 
\end{align}
As long as the interaction potential is short ranged compared to $V(q_x) \sim 1/q_x^2$, the $H_{ee}$ term has higher powers of $q_x$ than $q_x^2$. Thus, the only effects of the interactions are to give the bare velocities $v_{nr, nr}$ to the CFs and generate scatterings between the CF Fermi-points. We contrast this with the 2D limit where the gauge fluctuations induced by the Coulomb interactions, on top of the mean-field approximation, are marginal. And shorter-range interactions lead to a non-Fermi liquid behavior.\cite{NayakWilczek1994short, NayakWilczeklong1994}

Lastly, we expect the density-density correlation function to look like:
\begin{align}
	\langle \rho_e(q_x, \omega) \rho_e(-q_x, -\omega)\rangle \sim \frac{v \kappa^3(q_x\ell)^4}{\omega^2 + v^2q_x^2}+\cdots
\end{align}
where $\omega$ is the frequency corresponding to the Euclidean time. Therefore, we expect the static structure factor at long wavelengths: $\bar s(q_x) \sim (q_x\ell)^3$. This should be contrasted with the $\bar s (q) \sim (q\ell)^3 \log 1/q\ell$ behavior predicted by the 2D theory.\cite{Halperin1993,Read1998} We believe that in the limit of a large number of wires, the quasi-1D behavior would crossover to the one expected in the 2D theory. 

\subsection{Numerical results at $\nu=1/2$\label{sec:numerics_CFL}}
We present the numerical results for the case of two and three wires and contrast them with the quasi-1D theory. 

\subsubsection{Two-wires\label{sec:CFL_two_wires}}
\begin{figure*}
	\centering
	
	\begin{minipage}{0.5\textwidth}
		\begin{center}
			\includegraphics[width=1\textwidth]{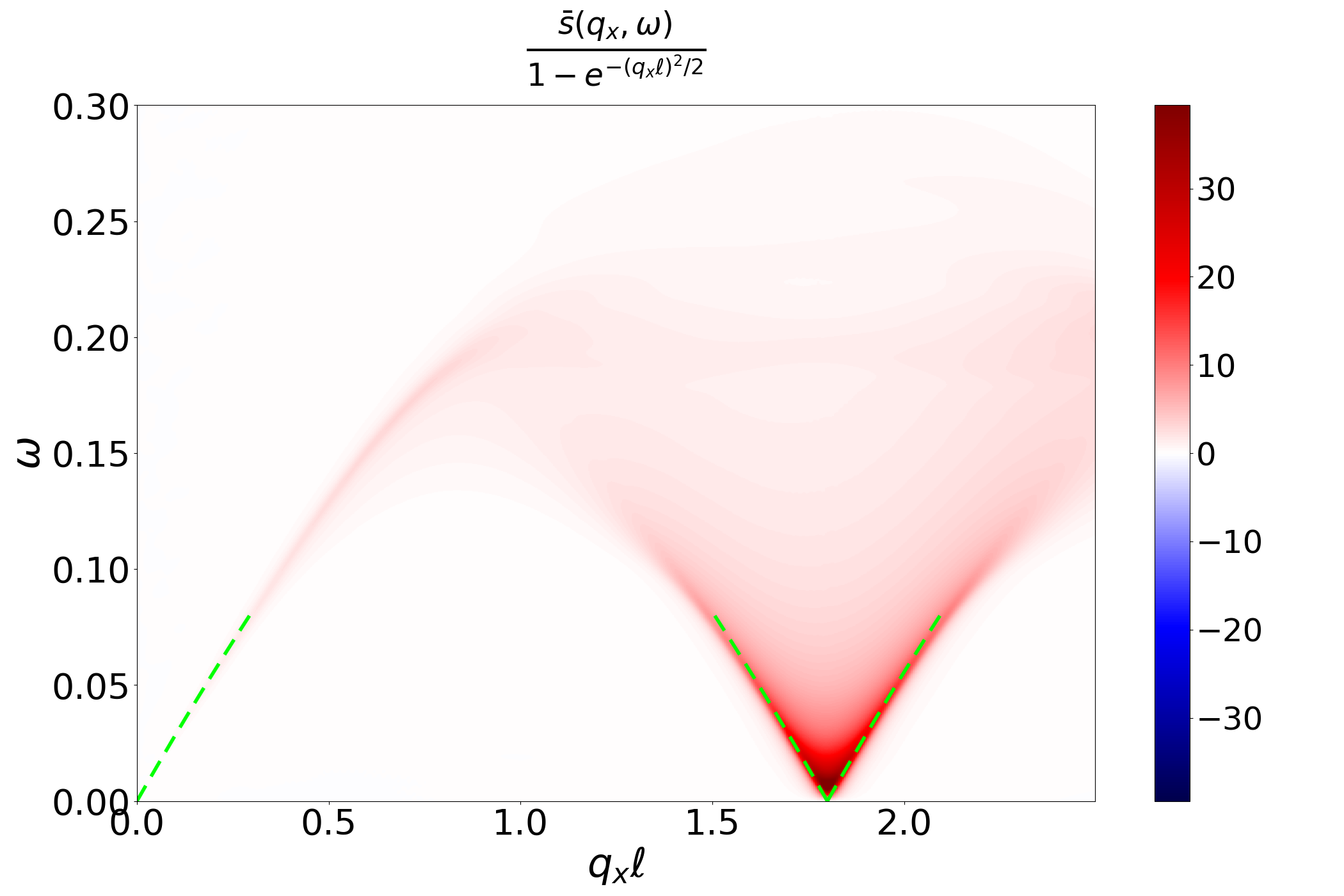}
			
			(a)
		\end{center}
	\end{minipage}%
	\begin{minipage}{0.5\textwidth}
		\begin{center}
			\includegraphics[width=0.8\textwidth]{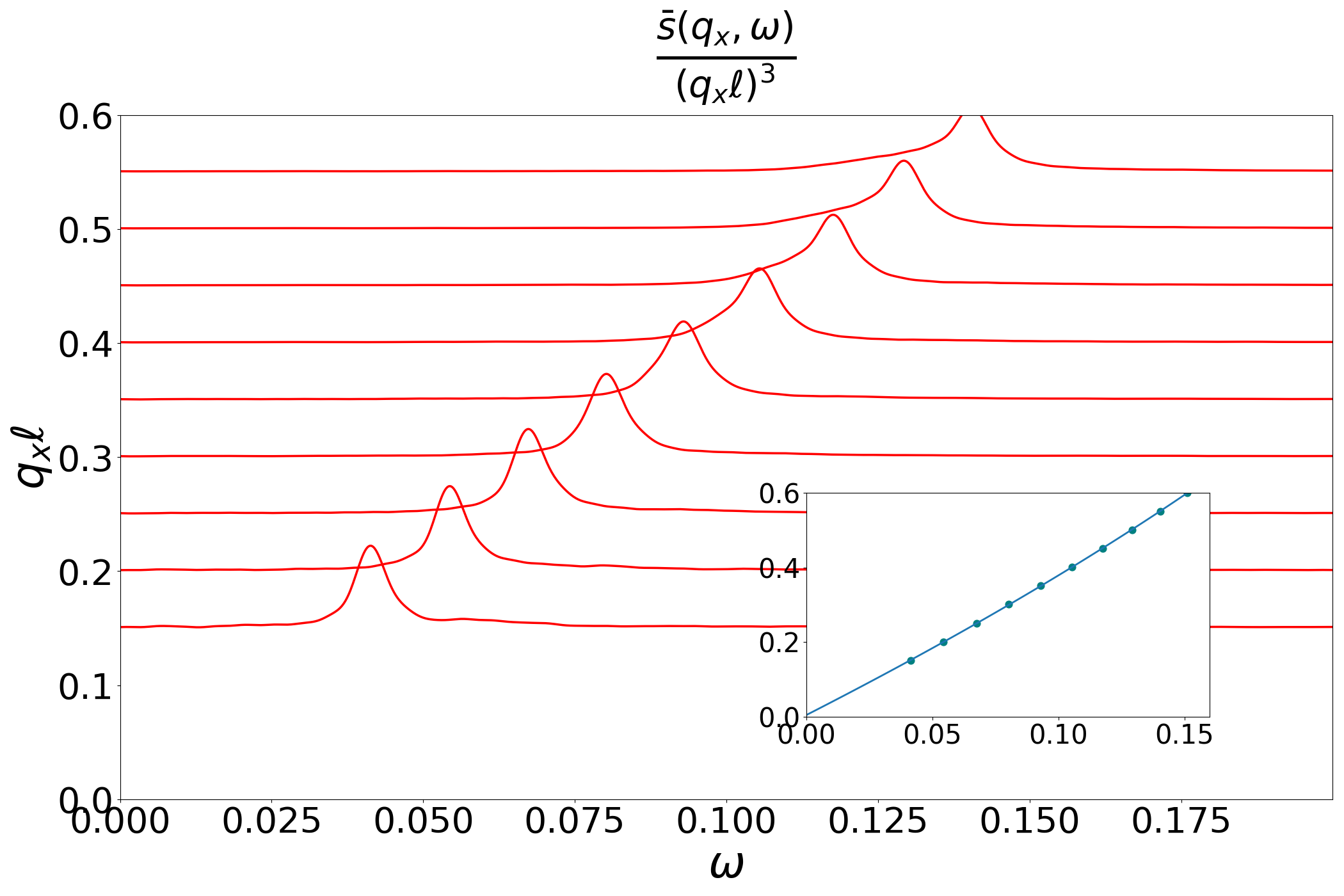}
			
			(b)
		\end{center}
	\end{minipage}
	\begin{minipage}{0.33\textwidth}
		\begin{center}
			\vspace{10pt}
			\includegraphics[width=1\textwidth]{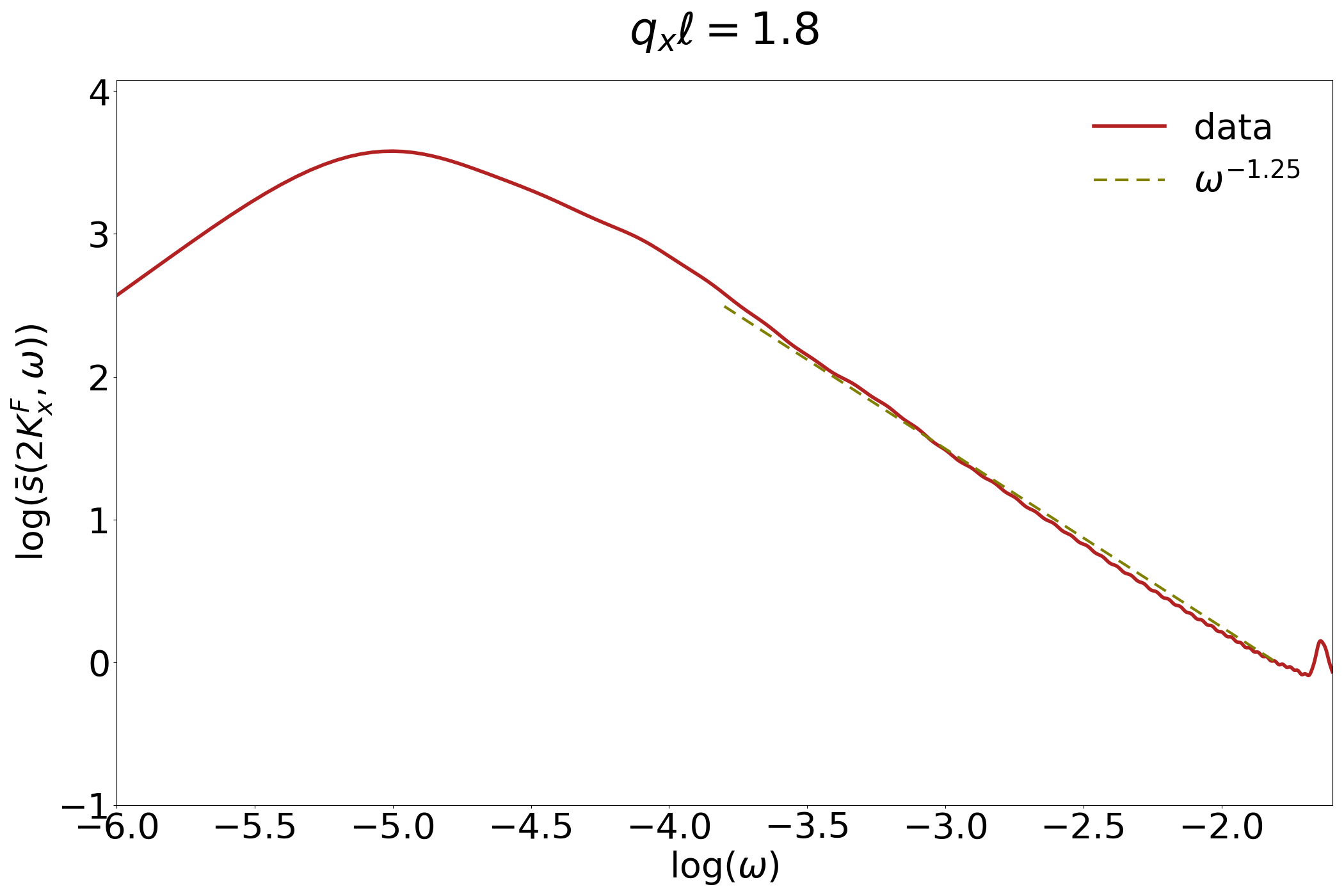}
			
			(c)
		\end{center}
	\end{minipage}%
	\begin{minipage}{0.33\textwidth}
		\begin{center}
			\includegraphics[width=1\textwidth]{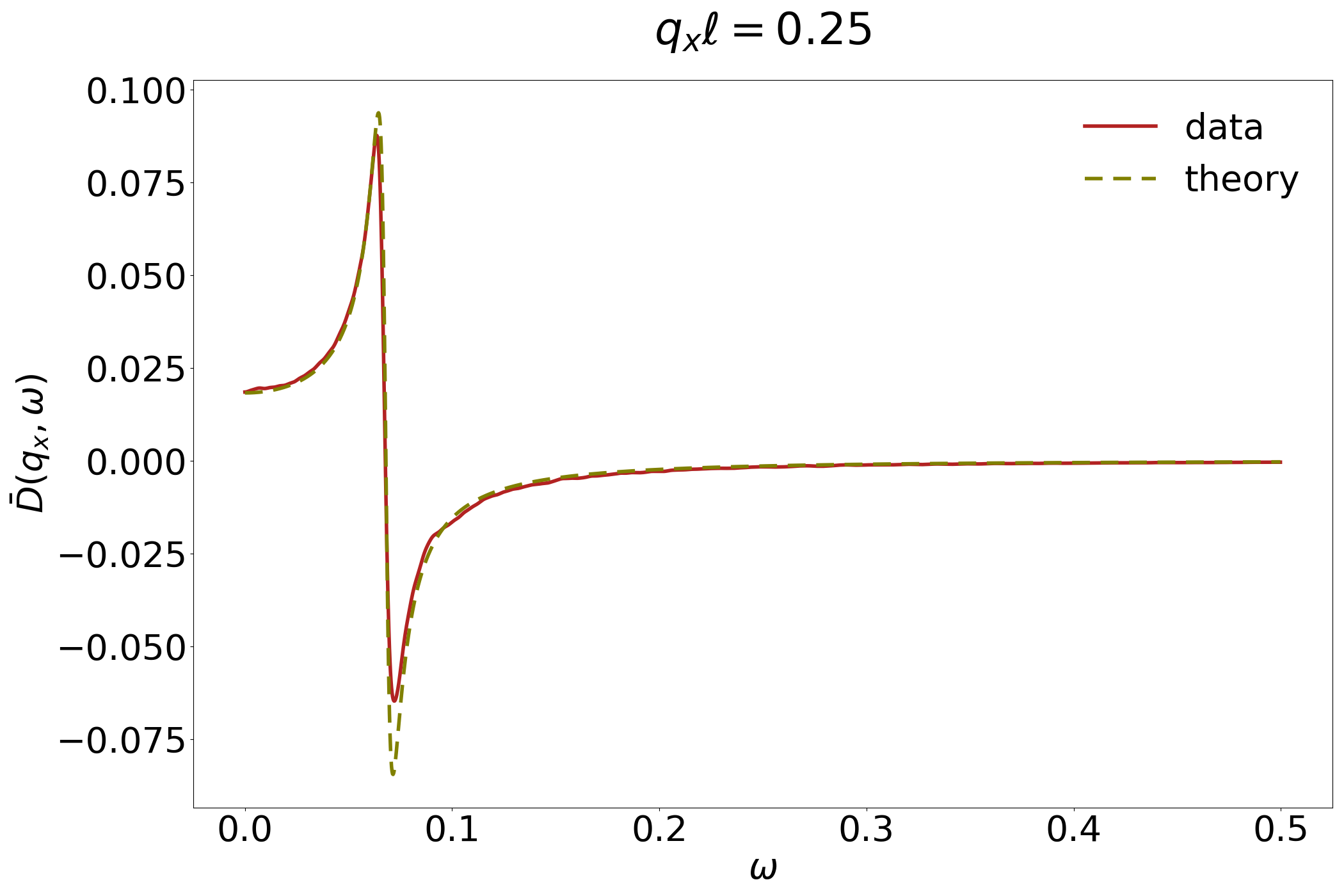}
			
			(d)
		\end{center}
	\end{minipage}%
	\begin{minipage}{0.33\textwidth}
		\begin{center}
			\includegraphics[width=1\textwidth]{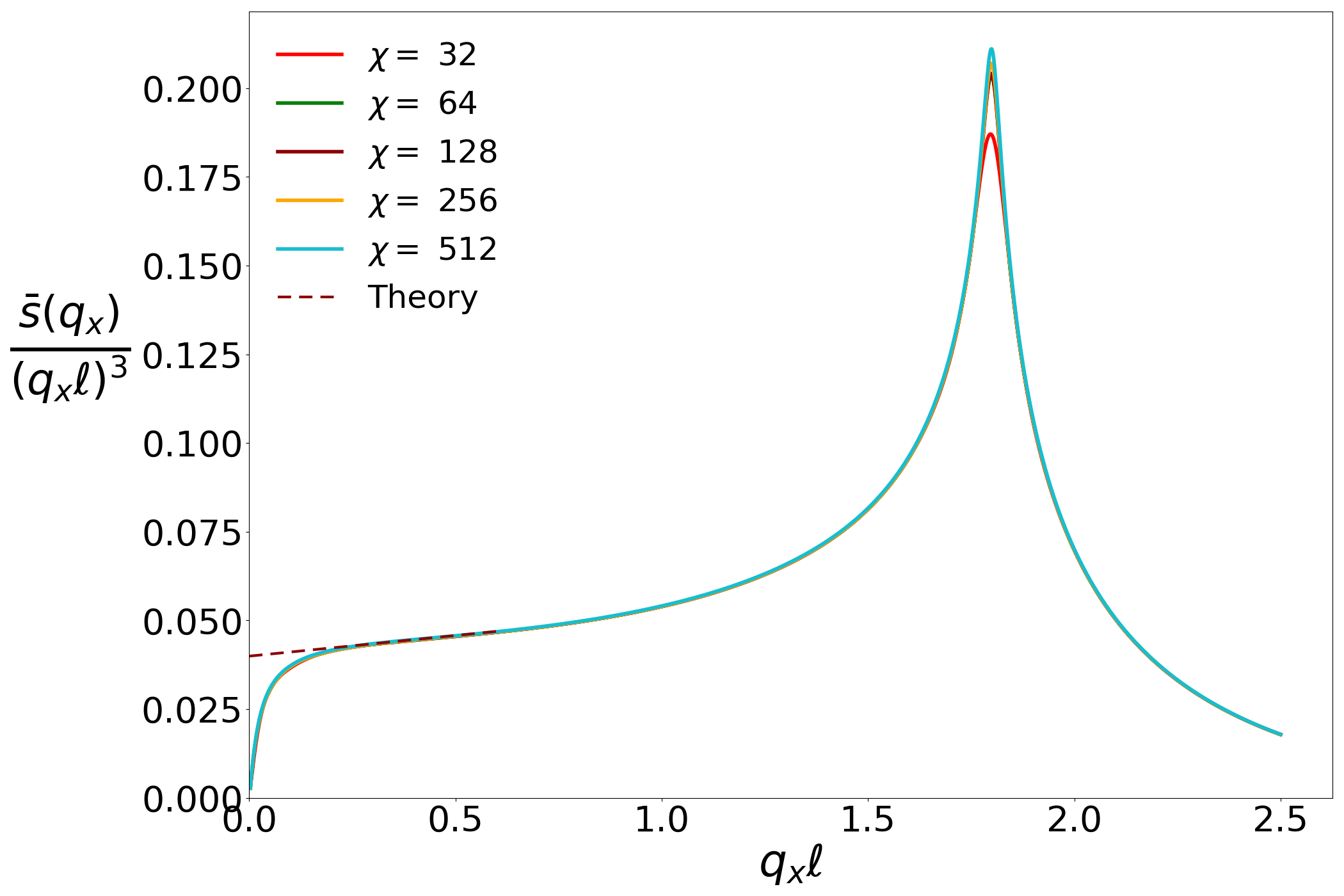}
			
			(e)
		\end{center}
	\end{minipage}
	
	\caption{The case of CF Fermi sea composed of two wires. (a) The dynamical structure factor $\bar s(q_x,\omega)$ of the $\nu=1/2$ state at $L_y=7.2\ell$ for Gaussian Coulomb interaction (Eq. \eqref{eq:gaussian_coulomb}) with $\xi=20$. There is a single low energy mode at small $q_x$ and a fan corresponding to the particle-hole excitations near $q_x\ell = 2K^F_x \ell=  L_y/4\ell = 1.8$. The dashed lime curve shows that the velocities of the two features are equal. (b) The low energy mode at small $q_x\ell$ and the linear dispersion (inset) with a velocity $u_0 = 0.29 e^2$ where $e^2$ represents the strength of the Coulomb interaction. (c) $\bar s(2K^F_x, \omega)$ vs $\omega$ on a log-log scale, (d) Real part of the density-density correlation function $\bar D(q_x,\omega)$ at $q_x\ell = 0.25$, and (e) static structure factor $\bar s(q_x)/(q_x\ell)^3$ for different bond dimensions. Dashed curves in (c), (d) and (e) show  agreement between the theory and the data with a Luttinger parameter $K_-$ equal to $1.33$. In (e), the static structure factor is converged at $q_x\ell < 1$ and it crosses over from $(q_x\ell)^3$ to $(q_x\ell)^4$ at $q_x\ell \sim 0.1$ suggesting a gap at small wavevectors. The gap is estimated to be $\omega_g = 0.01 e^2/\ell$ within SMA.}
	\label{fig:CFL_Coulomb_two_wires}
\end{figure*}

The ground state at $\nu=1/2$ is found in the root configuration 0110 when the CF Fermi sea is composed of two wires. Exploiting the mirror and PH symmetries, the unit cell averaged time-dependent quantities can be obtained by performing time-evolution of only one excited state constructed by applying the number operator on the third site of the unit cell. In Fig. \ref{fig:CFL_Coulomb_two_wires}(a)-(b), we plot the numerically computed dynamical structure factor for Coulomb interactions at $L_y=7.2\ell$. The spectrum contains one low energy mode that has a linear dispersion. Additionally, there is a continuum of excitations in a fan near $q_x\ell = 2K^F_x = L_y/4\ell=1.8$. Both features are characterized by the same velocity $u_-$ that we determine to equal $0.29e^2$, where $e^2$ is the strength of the Coulomb interaction. 

To compare the theory with our numerical results, we make a few predictions. To this end, let us consider the quasi-1D CF theory of two Fermi wires. It contains two modes which can be arranged in the linear combinations: $\sqrt 2 \Phi_{\pm} = \Phi_{1/2} \pm \Phi_{-1/2}$. The two linear combinations have different eigenvalues under the mirror symmetry and thus are decoupled. The $(\Phi_+,\Theta_+)$ mode corresponds to the total CF density mode and is gapped. The effective theory of the remaining $(\Phi_-, \Theta_-)$ mode can be written as:
\begin{align}
	\mathcal{S}_{\rm eff} &= \int dxd\tau\ \mathcal{L}_{\rm eff}\\
	\mathcal{L}_{\rm eff} &= \frac{u_-}{2\pi K_-}(\partial_x \Phi_-)^2 + \frac{u_-K_-}{2\pi} (\partial_x \Theta_-)^2 \nn\\
	&\ \ \ \ + \frac{i}{\pi} (\partial_\tau \Phi_-) (\partial_x \Theta_-) +
	\frac{g_{\rm pair}}{(2\pi\alpha)^2}\cos\left(2\sqrt 2 \Theta_-\right)\label{eq:two_wires_theory}
\end{align}
where $u_-$ and $K_-$ are the velocity and the Luttinger parameter respectively. The last term arises from scattering a pair of CFs with a net zero momentum from one pair of Fermi-points to the other. 

There are four Fermi-points at wavevectors $(\pm \kappa/2, \pm K^F_x)$, where $K^F_x>0$. The low energy electric density-density correlation functions can be calculated using the following expressions:
\begin{align}
	\rho_e(q_x,\omega) &= \rho^{(0)}_e(q_x, \omega) + \rho^{(2K^F_x)}_e(q_x,\omega)\\
	\rho^{(0)}_e(q_x, \omega) &= -\frac{q_x^2\kappa \ell^2}{\pi \sqrt{2L_y}} \Theta_-(q_x,\omega)\\
	\rho^{(2K^F_x)}_e(x, \tau) &\sim -\frac{\ell^2\kappa^2 }{\pi\alpha\sqrt{L_y}} \sin(\sqrt 2 \Theta_-)\cos(2K^F_xx + \Phi_+)\\
	\bar D^{(0)}(q_x,\omega) &=  \frac{(2\pi)^2}{L_y^3 } \frac{u_- (q_x\ell)^4/K_-}{u_-^2q_x^2-\omega^2} \label{eq:small_q_D}\\
	\bar D^{(2K^F_x)}(q_x,\omega) &\propto  \frac{1}{\left(u_-^2 (|q_x|-2K^F_x)^2-\omega^2\right)^{1-1/2K_-}}
\end{align}
where $\omega$ corresponds to the frequency in real-time, $\rho^{(0)}_e$ and $\rho^{(2K^F_x)}$  are the electric density operators near $q_x=0$ and $q_x = 2K^F_x$ respectively. Further, $\bar D(q_x, \omega)$ is the dynamical density-density correlation function.

The form of the low energy correlation functions is controlled by the two parameters $u_-$ and $K_-$. Specifically, the static structure factor $\bar s(q_x) \sim (q_x\ell)^3$ in the limit $q_x\ell\rightarrow 0$ and its coefficient is predicted to depend on the Luttinger parameter $K_-$ as follows:
\begin{align}
	\bar s^{(0)}(q_x) &= \frac{2\pi^2\ell^3}{L_y^3 K_-} (q_x\ell)^3
\end{align}

In Fig. \ref{fig:CFL_Coulomb_two_wires}(e), we determine $K_-=1.33$ from the intercept of  a linear fit of $\bar s(q_x)/(q_x\ell)^3$ to $q_x\ell$ in the limit $q_x\ell \rightarrow 0$. The Luttinger parameter obtained from the long wavelength limit also controls the behavior at finite $q_x \ell$. To see this, note that the dynamical structure factor at $q_x\ell = 2K^F_x$ at low energies is given by:
\begin{align}
\bar s(2K^F_x, \omega) \propto \frac{\mathrm{sgn}[\omega]}{|\omega|^{1-1/2K_-}}
\end{align}
In Fig. \ref{fig:CFL_Coulomb_two_wires}(c), we show that the behavior predicted from $K_- = 1.33$ agrees with the numerically computed $ \bar s(2K^F_x, \omega)$. In addition, we show that the precise form of real part of the dynamical correlation function $\bar D(q_x\ell, \omega)$ predicted by the theory at $q_x\ell = 0.25 \lesssim 1$ (Eq. \eqref{eq:small_q_D}) is consistent with our numerics.

At $q_x\ell \approx 0.1$, we observe that the $\bar s(q_x) \sim (q_x\ell)^3$ behavior breaks down and the static structure factor crosses over to $\bar s(q_x) \sim (q_x\ell)^4$. This indicates that the low energy mode is, in fact, gapped. We estimate the gap to be around $\omega_g \approx 0.01e^2/\ell$ using SMA which is too small to be resolved in our numerics. Therefore, we only observe the physics at energies above the gap where, as we have shown, the quasi-1D CF theory is quantitatively consistent with our numerics.

The nature of the ground state at energies below the gap can be understood using the $g_{\rm pair}$ term in Eq. \eqref{eq:two_wires_theory}. When $K_- >1$, this term is relevant and leads to a gapped phase as long as $g_{\rm pair} \neq 0$. Therefore, in fact, the gap is predicted by the CF theory since we obtained $K_-=1.33>1$ and the numerical observation confirms it. Since the pairing term scatters a pair of fermions with a net zero momentum, the resulting phase is a quasi-1D version of a CF superconductor. We determine that the state breaks particle-hole symmetry since the unit cell has a $0110$ CDW pattern. Thus, the ground state is a quasi-1D version of the Moore-Read state.\cite{Moore1991}

\subsubsection{Three-wires}
\begin{figure*}
	\centering
	
	\begin{minipage}{0.5\textwidth}
		\begin{center}
			\includegraphics[width=1\textwidth]{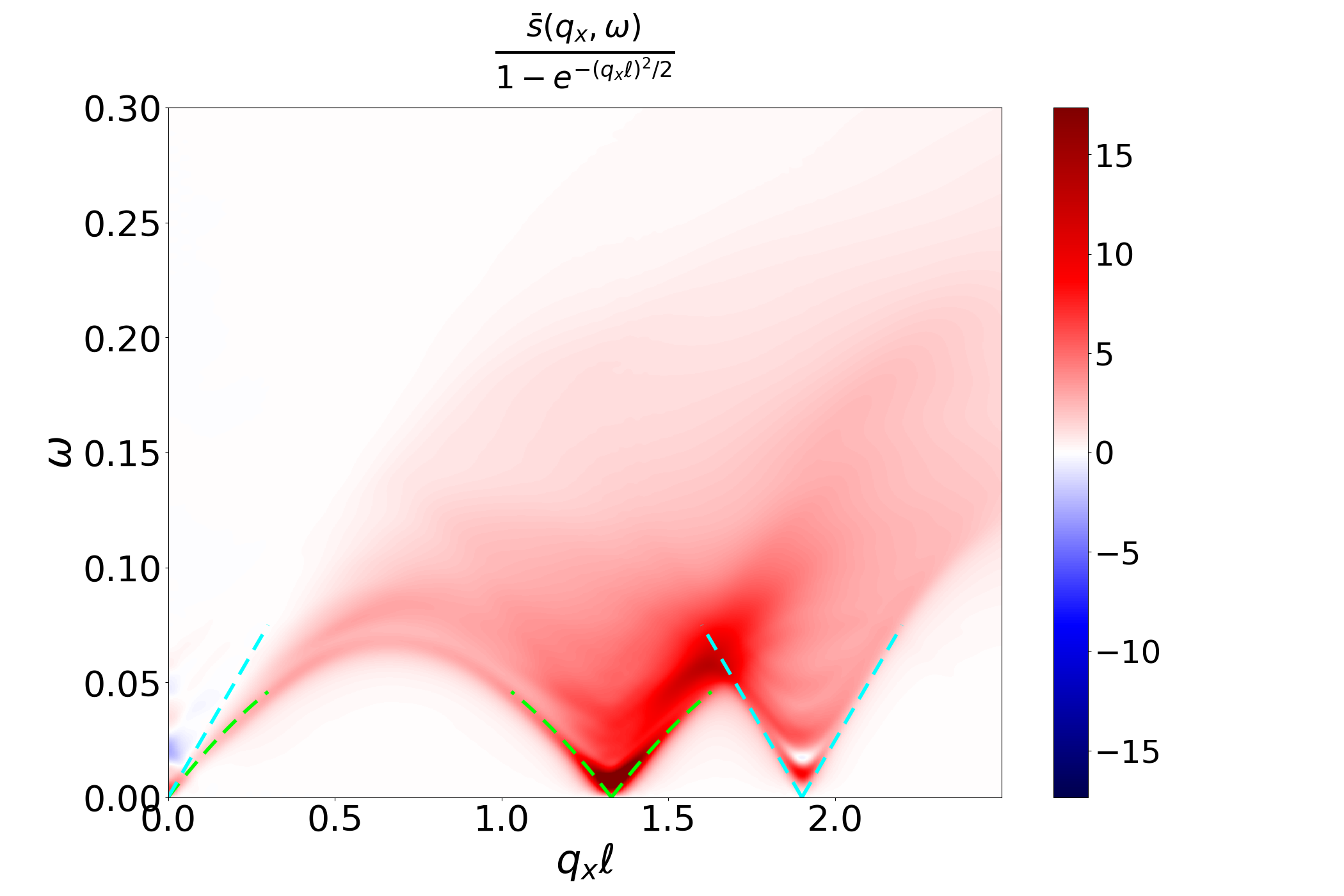}
			
			(a)
		\end{center}
	\end{minipage}%
	\begin{minipage}{0.5\textwidth}
		\begin{center}
			\includegraphics[width=0.9\textwidth]{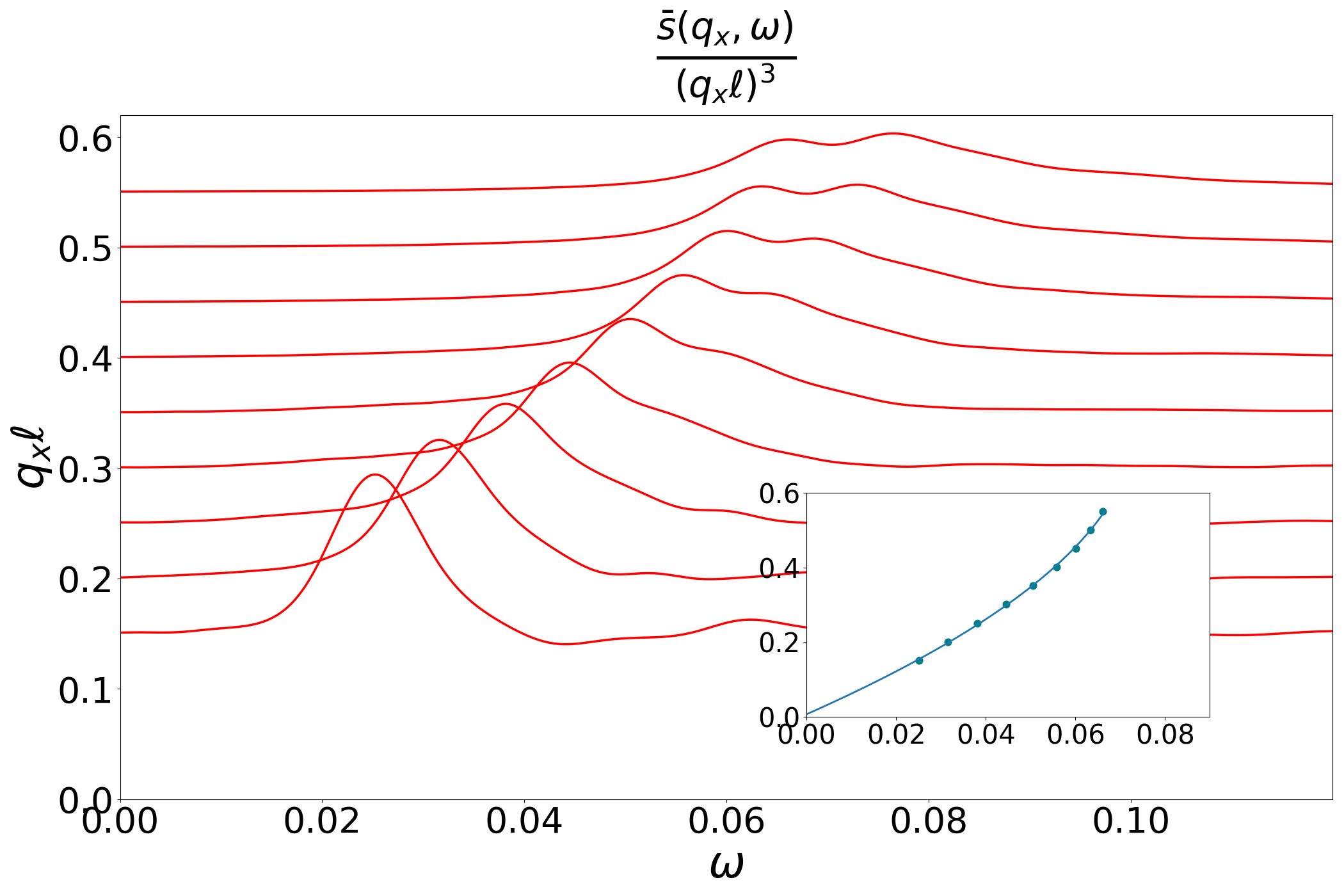}\vspace{10pt}
			
			(b)
		\end{center}
	\end{minipage}
	\begin{minipage}{0.5\textwidth}
		\begin{center}
			\vspace{10pt}
			\includegraphics[width=0.75\textwidth]{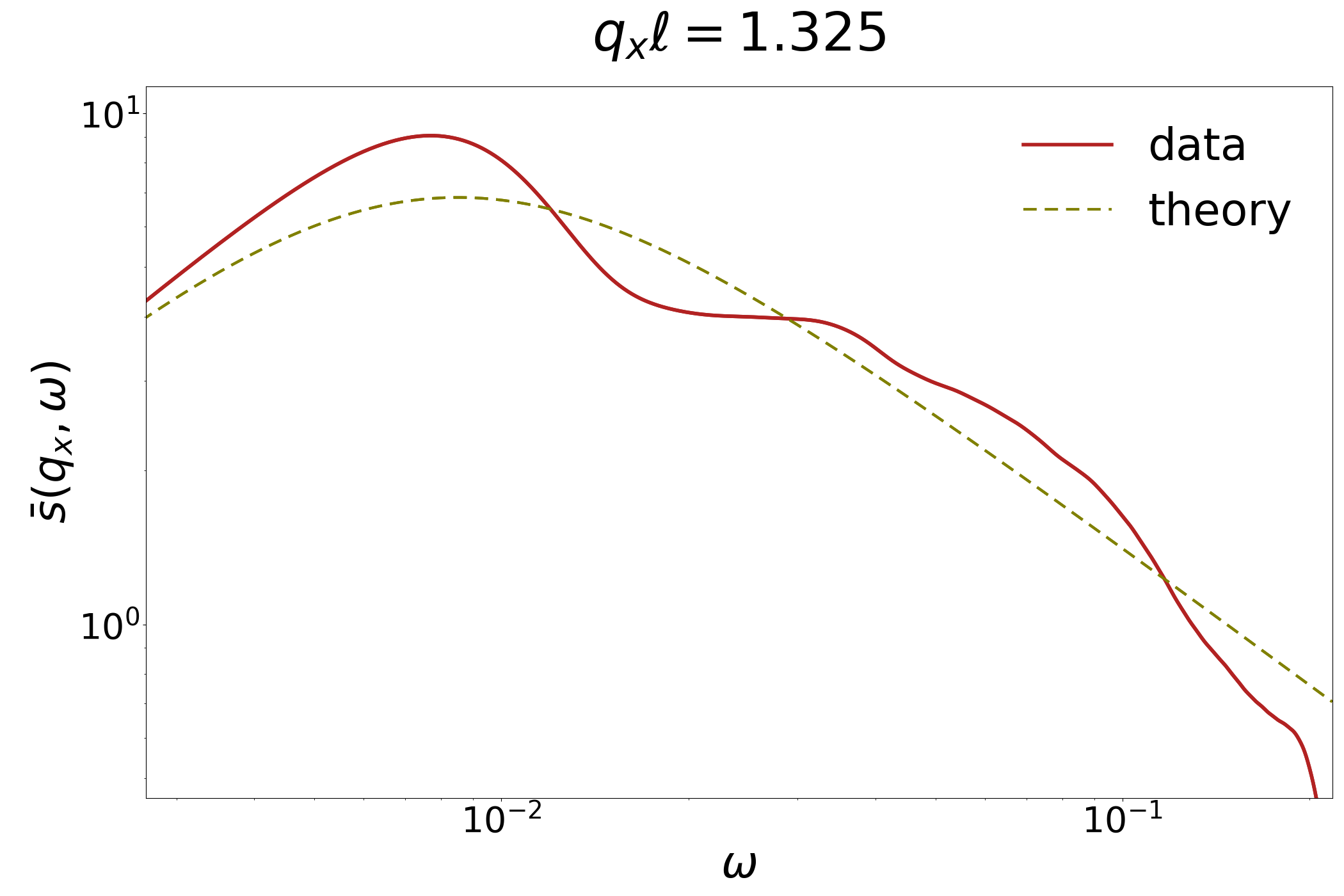}
			
			(c)
		\end{center}
	\end{minipage}%
	\begin{minipage}{0.5\textwidth}
		\begin{center}
			\vspace{10pt}
			\includegraphics[width=0.75\textwidth]{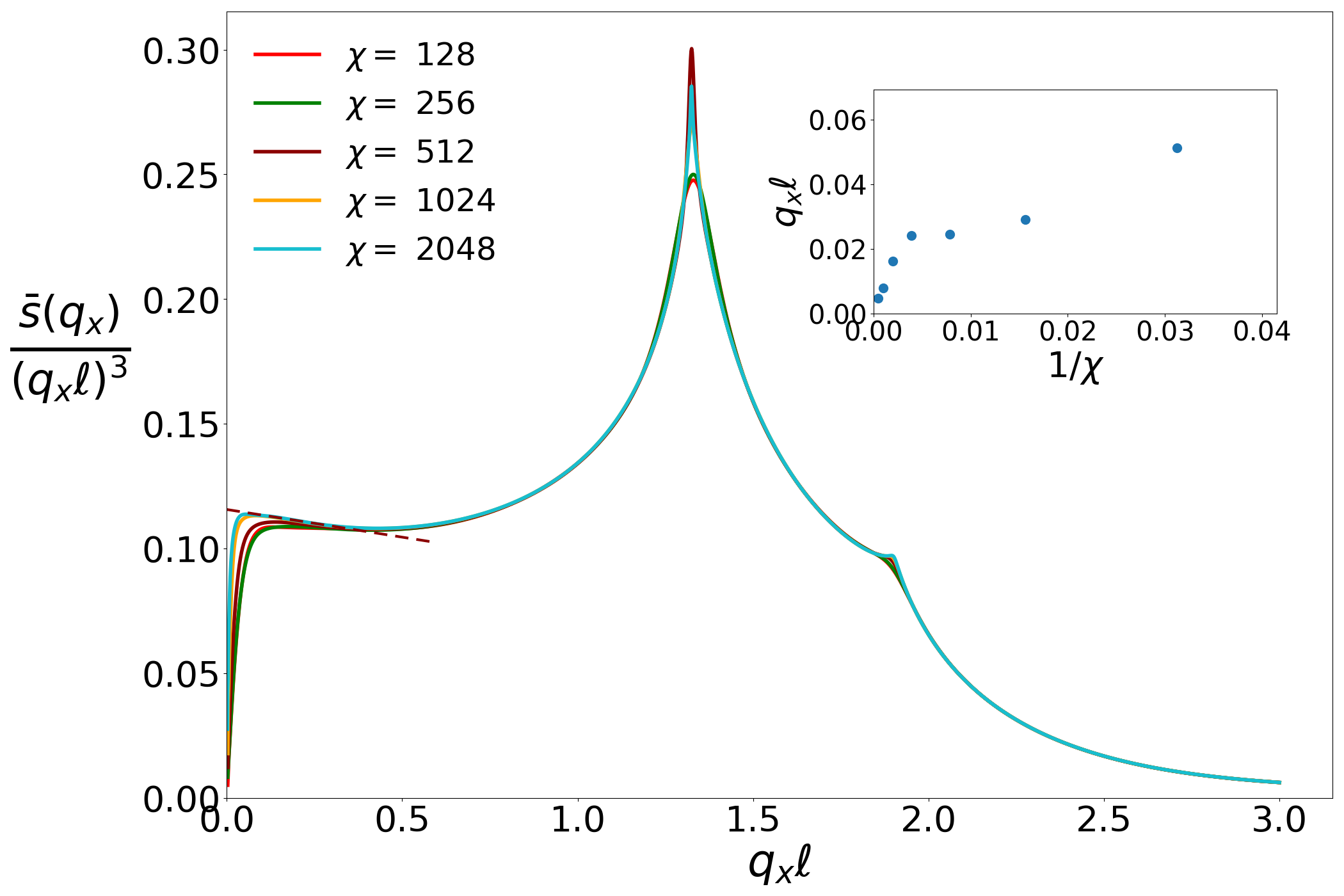}
			
			(d)
		\end{center}
	\end{minipage}
	
	\caption{The case of CF Fermi sea composed of three wires. (a) The dynamical structure factor $\bar s(q_x,\omega)$ of the $\nu=1/2$ state at $L_y=9.1\ell$ for Gaussian Coulomb interaction (Eq. \eqref{eq:gaussian_coulomb}) with $\xi=20$. There are two gapless degrees of freedom that produce the fans near $q_x\ell= 1.325$ and $q_x\ell=1.9$. At small $q_x\ell$, we observe one gapless mode that has a velocity equal to the fan at $q_x=1.325$ (dashed lime curve). There is no spectral weight at the velocity of the second fan at $q_x \ell=1.9$ (dashed cyan curve) in agreement with the quasi-1D CF theory. (b) The single gapless mode at small $q_x\ell$. The inset shows a fit to a linear dispersion with a quadratic correction with velocity $u_- = 0.19e^2$. Another mode is visible at $q_x\ell\approx 0.5$, however, it disappears at long wavelengths. (c) A comparison of the numerically computed dynamical structure factor and the theoretical prediction with $K_- = 0.91$ at the first fan, i.e. $q_x\ell = 2K^F_{x,-}=1.325$. (d) The static structure factor $\bar s(q_x)/(q_x\ell)^3$ vs. $q_x \ell$ for various bond dimensions $\chi$. As shown in the inset, the point of the crossover between $\bar s(q_x) \sim (q_x\ell)^3$ and $\bar s(q_x) \sim (q_x\ell)^4$ goes to $q\ell = 0$ as $\chi \rightarrow \infty$.
	}
	\label{fig:CFL_Coulomb_three_wires}
\end{figure*}

Similar to the two-wires case, we perform numerical simulation of the three-wires CF Fermi sea. The root configuration is $01$ and $L_y=9.1\ell$ in the presence of Coulomb interaction. To minimize errors, we take an explicit average of the correlations over the unit cell. The dynamical structure factor is shown in Fig. \ref{fig:CFL_Coulomb_three_wires}. We observe two fans at $q_x\ell=1.325$ and $q_x\ell=1.9$ that correspond to the two unique $2K^F_x$ values of the three CF wires. However, there is only one gapless mode visible in the limit $q_x\ell\rightarrow 0$. Its dispersion agrees with the fan near the smaller $2K^F_x$ value, i.e. $q_x\ell = 1.325$. As we explain, the predictions of the quasi-1D theory are consistent with these features observed in the numerical results.

Let us analyze the three-wire quasi-1D CF theory that is composed of 3 bosonic modes. We can form the following linear combinations, chosen to be the eigenstates of the mirror symmetry:
\begin{align}
	\sqrt{3}\Phi_{\rm T} &= \Phi_{1} + \Phi_{0} + \Phi_{-1}\\
	\sqrt{6} \Phi_{\rm c} &= 2\Phi_{0} - \Phi_{1} - \Phi_{-1}\\
	\sqrt{2} \Phi_{-} &= \Phi_{1} - \Phi_{-1}
\end{align}
$\Phi_T$ is gapped and $\Phi_{\rm c}, \Phi_{-}$ are the two linearly dispersing low energy modes. Notice that the pairs $(\Phi_c,\Theta_c)$ and $(\Phi_T,\Theta_T)$ have the same symmetries and thus can mix. On the other hand, the pair $(\Phi_-,\Theta_-)$ is decoupled from them.

At long wavelengths, the electric density operator is given by $\rho_e(x) = -\partial_x^2\sum {K^F_{nr}}_y \phi_{nr}$. Since ${K^F_{nr}}_y = \mathrm{sgn}[n]|{K^F_{nr}}_y|$, $\rho_e(x)$ involves linear combinations of $\Theta_n$ that change sign under $n\rightarrow -n$. As such, only $\Theta_-$ can contribute to the electron density operator at long wavelengths and the theory predicts that only one gapless mode should be visible. Further, since the $(\Phi_-, \Theta_-)$ mode involves the $n=\pm 1$ wires, it produces a fan at $q_x = {2K^F_{1R}}_x$ that is at a smaller wavevector than the fan produced by the $n=0$ central wire at $q_x=2{K^F_{0R}}_x$. This is consistent with the observations in Fig. \ref{fig:CFL_Coulomb_three_wires} where the gapless mode at $q_x\ell \ll 1$ has the same velocity as the fan at the smaller wavevector. 

The static structure factor at long wavelengths is predicted to be:
\begin{align}
	\bar s^{(0)}(q_x) &= \frac{8\pi^2\ell^3}{L_y^3 K_-} (q_x\ell)^3
\end{align}
where $K_-$ represents the Luttinger parameter of the $(\Phi_-,\Theta_-)$ pair. Using the $q_x\ell \rightarrow 0$ limit of $\bar s(q_x)/(q_x\ell)^3$ in Fig. \ref{fig:CFL_Coulomb_three_wires}(d), we obtain $K_- = 0.91$. As shown in Fig. \ref{fig:CFL_Coulomb_three_wires} (c), the dynamical structure factor $\bar s(q_x, \omega)$ at the first fan, i.e. $q_x\ell = 2K^F_{x,-} = 1.325$ agrees with the theoretical prediction based on this value of $K_-$. This quantitatively confirms the quasi-1D CF theory in the three-wires case.

A plot of the $\bar s (q_x)/(q_x\ell)^3$ vs $q_x\ell$ shows that the crossover from $s (q_x) \sim (q_x\ell)^3$ to $s (q_x) \sim (q_x\ell)^4$ behavior occurs at smaller and smaller $q_x\ell$ values as the bond dimension goes to infinity. Therefore, the $(\Phi_-,\Theta_-)$ mode is gapless at long wavelengths. Since $K_- < 1$, the theory predicts that this mode is stable to a weak $g_{\rm pair,1}$ with the interaction term:
\begin{align}
	\mathcal{L}_{\rm eff}^{\rm int,1} = \frac{g_{\rm pair, 1}}{(2\pi\alpha)^2} \cos\left(2\sqrt{ 2} \Theta_-\right)\label{eq:three_wires_pair1}
\end{align}
Thus the theory is consistent with the numerical observation that the linearly dispersing mode at $q_x\ell \ll 1$ is gapless. Analogous to the two-wires case, this interaction term corresponds to the scattering of a pair of CFs with net zero momentum within the $n=\pm 1$ wires. 

In addition, the theory contains an interaction that scatters a pair of electrons with a net zero momentum from the central wire at $n=0$ to one of the the two other wires at $n = \pm 1$. It can be expressed as follows in terms of the bosonic modes:
\begin{align}
	\mathcal{L}_{\rm eff}^{\rm int,2} = \frac{g_{\rm pair, 2}}{(2\pi\alpha)^2} \cos\left(\sqrt{ 2} \Theta_-\right)\cos\left(\sqrt{6}\Theta_c\right)
\end{align}
This term is irrelevant if $K_c < 3K_-/(4K_--1)\approx 1.03$, where $K_c$ is the Luttinger parameter of the canonically conjugate pair $(\Phi_c, \Theta_c)$ obtained after integrating out $(\Phi_T,\Theta_T)$. However, at present, we are unable to reliably calculate $K_c$ since the $(\Phi_c, \Theta_c)$ mode is not visible at long wavelengths. We believe that the dynamical correlation function of an operator, that is even under PH symmetry, may facilitate a more quantitative understanding of the $(\Phi_c, \Theta_c)$ mode.

\section{Conclusion \label{sec:conclusion}}
In summary, we have computed at the neutral excitation spectra of the $\nu=1/3$ and $\nu=1/2$ QH states in the lowest Landau level using their dynamical structure factors in the infinite-cylinder geometry. At $\nu=1/3$, we found a magnetoroton mode and the two roton continuum that are modified from the 2D limit due to the finite circumference of the cylinder. We showed that these finite size effects can be understood in terms of the quasi-particles that carry fractional charges in multiples of $e/3$.

At $\nu=1/2$, we found composite-fermion (CF) modes at low energies and confirmed the predictions of the quasi-1D CF theory. When the CF Fermi sea is composed of two wires, we observe only one low energy mode. Its properties were shown to be quantitatively consistent with the quasi-1D CF theory. Further, it was found to be unstable to an interaction that corresponds to the scattering of a pair of CFs with a net zero momentum. Therefore, the ground state can be understood as a CF superconductor, i.e., the quasi-1D analogue of Moore-Read state at energies below the gap. 

In the case when the Fermi-sea is composed of three wires, we observed one gapless linearly dispersing mode at low energies, but two fans at finite wavevectors. This is consistent with the theoretical prediction that, even though there are two low energy modes, only one contributes to the dynamical density-density correlation function at long wavelengths. Nevertheless, they produce two fans at the two distinct $2K^F$ values of the three wires. Our results provide a compelling evidence for the CF theory at the half-filled Landau level.

At present, it is not clear to us how the quasi-1D CF theory crosses over to the 2D limit. The static structure factor behaves as $\bar s(q_x) \sim (q_x\ell)^3$ in the long wavelength limit in the former. However, the 2D theories predict $\bar s (q) \sim (q\ell)^3 \log 1/q\ell$.\cite{Halperin1993,Read1998} In the limit of a large number of wires, we expect that the gauge fluctuations that scatter CFs between different wires in the quasi-1D theory would become relevant. In addition, wires with vanishingly small lengths would appear near the top and bottom of the Fermi sea. Understanding how these effects modify the long wavelength behavior is an interesting problem for future research.

An obvious generalization of our work would be to study the Jain states at filling fraction $\nu=p/(2mp+1)$ with $p,m\geq 2$.\cite{Nguyen2021a,Nguyen2021b,Balram2022} Especially, determining how the series of fractional quantum Hall states leads to the $\nu=1/2$ gapless state is an interesting problem. We leave these questions for future work.

\section{Acknowledgment}
We thank Shinsei Ryu, Olexei Motrunich, Ajit Balram, Srinivas Raghu and Hart Goldman for discussions.
The TDVP codes used in the paper were built on top of the iDMRG libraries developed by Roger Mong, Michael Zaletel and the TenPy collaboration. This work was supported by DOE BES Grant No. DE-SC0002140.

\bibliographystyle{utphys}
\bibliography{bigbib}

\providecommand{\href}[2]{#2}\begingroup\raggedright\begin{thebibliography}{10}

\bibitem{Jain1989}
J.~K. Jain, ``{Composite-fermion approach for the fractional quantum Hall
  effect},'' \href{http://dx.doi.org/10.1103/PhysRevLett.63.199}{{\em Phys.
  Rev. Lett.} {\bfseries 63} (1989) 199}.

\bibitem{Lopez91}
A.~Lopez and E.~Fradkin, ``Fractional quantum hall effect and chern-simons
  gauge theories,'' \href{http://dx.doi.org/10.1103/PhysRevB.44.5246}{{\em
  Phys. Rev. B} {\bfseries 44} (Sep, 1991) 5246--5262}.
  \url{https://link.aps.org/doi/10.1103/PhysRevB.44.5246}.

\bibitem{Kalmeyer1992}
V.~Kalmeyer and S.-C. Zhang, ``{Metallic phase of the quantum Hall system at
  even-denominator filling fractions},''
  \href{http://dx.doi.org/10.1103/PhysRevB.46.9889}{{\em Phys. Rev. B}
  {\bfseries 46} (Oct, 1992) 9889--9892}.
  \url{http://link.aps.org/doi/10.1103/PhysRevB.46.9889}.

\bibitem{Halperin1993}
B.~I. Halperin, P.~A. Lee, and N.~Read, ``{Theory of the half-filled Landau
  level},'' \href{http://dx.doi.org/10.1103/PhysRevB.47.7312}{{\em Phys. Rev.
  B} {\bfseries 47} (Mar, 1993) 7312--7343}.
  \url{http://link.aps.org/doi/10.1103/PhysRevB.47.7312}.

\bibitem{Simon1993}
S.~H. Simon and B.~I. Halperin, ``Finite-wave-vector electromagnetic response
  of fractional quantized hall states,''
  \href{http://dx.doi.org/10.1103/PhysRevB.48.17368}{{\em Phys. Rev. B}
  {\bfseries 48} (Dec, 1993) 17368--17387}.
  \url{https://link.aps.org/doi/10.1103/PhysRevB.48.17368}.

\bibitem{Jainbook}
J.~K. Jain, {\em {Composite Fermions}}.
\newblock Cambridge University Press, 2007.

\bibitem{Fradkinbook}
E.~Fradkin, {\em {Field Theories of Condensed Matter Physics}}.
\newblock Cambridge University Press, 2013.

\bibitem{Son2015}
D.~T. Son, ``{Is the Composite Fermion a Dirac Particle?},''
  \href{http://dx.doi.org/10.1103/PhysRevX.5.031027}{{\em Phys. Rev. X}
  {\bfseries 5} (2015) 031027}.

\bibitem{Note1}
In the Dirac CF description,\cite {Son2015} $\nu _{\pm }$ corresponds to $\nu
  _{\protect \rm cf}=\pm (p+1/2)$ of CFs.

\bibitem{Haldane1985}
F.~D.~M. Haldane and E.~H. Rezayi, ``Finite-size studies of the incompressible
  state of the fractionally quantized hall effect and its excitations,''
  \href{http://dx.doi.org/10.1103/PhysRevLett.54.237}{{\em Phys. Rev. Lett.}
  {\bfseries 54} (Jan, 1985) 237--240}.
  \url{https://link.aps.org/doi/10.1103/PhysRevLett.54.237}.

\bibitem{Rezayi1994}
E.~Rezayi and N.~Read, ``{Fermi-liquid-like state in a half-filled Landau
  level},'' {\em Phys. Rev. Lett.} {\bfseries 72} (1994) 900.

\bibitem{Geraedts2015}
S.~D. {Geraedts}, M.~P. {Zaletel}, R.~S.~K. {Mong}, M.~A. {Metlitski},
  A.~{Vishwanath}, and O.~I. {Motrunich}, ``{The half-filled Landau level: The
  case for Dirac composite fermions},''
  \href{http://dx.doi.org/10.1126/science.aad4302}{{\em Science} {\bfseries
  352} (2016) 197}, \href{http://arxiv.org/abs/1508.04140}{{\ttfamily
  arXiv:1508.04140 [cond-mat.str-el]}}.

\bibitem{Zaletel2015}
M.~P. Zaletel, R.~S.~K. Mong, F.~Pollmann, and E.~H. Rezayi, ``Infinite density
  matrix renormalization group for multicomponent quantum hall systems,''
  \href{http://dx.doi.org/10.1103/PhysRevB.91.045115}{{\em Phys. Rev. B}
  {\bfseries 91} (Jan, 2015) 045115}.
  \url{https://link.aps.org/doi/10.1103/PhysRevB.91.045115}.

\bibitem{Haegeman2011}
J.~Haegeman, J.~I. Cirac, T.~J. Osborne, I.~Pi\ifmmode~\check{z}\else
  \v{z}\fi{}orn, H.~Verschelde, and F.~Verstraete, ``Time-dependent variational
  principle for quantum lattices,''
  \href{http://dx.doi.org/10.1103/PhysRevLett.107.070601}{{\em Phys. Rev.
  Lett.} {\bfseries 107} (Aug, 2011) 070601}.
  \url{https://link.aps.org/doi/10.1103/PhysRevLett.107.070601}.

\bibitem{Haegeman2016}
J.~Haegeman, C.~Lubich, I.~Oseledets, B.~Vandereycken, and F.~Verstraete,
  ``Unifying time evolution and optimization with matrix product states,''
  \href{http://dx.doi.org/10.1103/PhysRevB.94.165116}{{\em Phys. Rev. B}
  {\bfseries 94} (Oct, 2016) 165116}.
  \url{https://link.aps.org/doi/10.1103/PhysRevB.94.165116}.

\bibitem{Paeckel2019}
S.~Paeckel, T.~Köhler, A.~Swoboda, S.~R. Manmana, U.~Schollwöck, and
  C.~Hubig, ``Time-evolution methods for matrix-product states,''
  \href{http://dx.doi.org/https://doi.org/10.1016/j.aop.2019.167998}{{\em
  Annals of Physics} {\bfseries 411} (2019) 167998}.
  \url{https://www.sciencedirect.com/science/article/pii/S0003491619302532}.

\bibitem{Laughlin1983}
R.~B. Laughlin, ``Anomalous quantum hall effect: An incompressible quantum
  fluid with fractionally charged excitations,''
  \href{http://dx.doi.org/10.1103/PhysRevLett.50.1395}{{\em Phys. Rev. Lett.}
  {\bfseries 50} (May, 1983) 1395--1398}.
  \url{https://link.aps.org/doi/10.1103/PhysRevLett.50.1395}.

\bibitem{Girvin1986}
S.~M. Girvin, A.~H. MacDonald, and P.~M. Platzman, ``Magneto-roton theory of
  collective excitations in the fractional quantum hall effect,''
  \href{http://dx.doi.org/10.1103/PhysRevB.33.2481}{{\em Phys. Rev. B}
  {\bfseries 33} (Feb, 1986) 2481--2494}.
  \url{https://link.aps.org/doi/10.1103/PhysRevB.33.2481}.

\bibitem{Laughlin1984b}
R.~B. {Laughlin}, ``{Excitons in the fractional quantum hall effect},''
  \href{http://dx.doi.org/10.1016/0378-4363(84)90172-4}{{\em Physica B+C}
  {\bfseries 126} no.~1, (Nov., 1984) 254--259}.

\bibitem{Girvin1985}
S.~M. Girvin, A.~H. MacDonald, and P.~M. Platzman, ``Collective-excitation gap
  in the fractional quantum hall effect,''
  \href{http://dx.doi.org/10.1103/PhysRevLett.54.581}{{\em Phys. Rev. Lett.}
  {\bfseries 54} (Feb, 1985) 581--583}.
  \url{https://link.aps.org/doi/10.1103/PhysRevLett.54.581}.

\bibitem{Dev1992}
G.~Dev and J.~K. Jain, ``Band structure of the fractional quantum hall
  effect,'' \href{http://dx.doi.org/10.1103/PhysRevLett.69.2843}{{\em Phys.
  Rev. Lett.} {\bfseries 69} (Nov, 1992) 2843--2846}.
  \url{https://link.aps.org/doi/10.1103/PhysRevLett.69.2843}.

\bibitem{Lopez1993}
A.~Lopez and E.~Fradkin, ``Response functions and spectrum of collective
  excitations of fractional-quantum-hall-effect systems,''
  \href{http://dx.doi.org/10.1103/PhysRevB.47.7080}{{\em Phys. Rev. B}
  {\bfseries 47} (Mar, 1993) 7080--7094}.
  \url{https://link.aps.org/doi/10.1103/PhysRevB.47.7080}.

\bibitem{Kamilla1996}
R.~K. Kamilla, X.~G. Wu, and J.~K. Jain, ``Excitons of composite fermions,''
  \href{http://dx.doi.org/10.1103/PhysRevB.54.4873}{{\em Phys. Rev. B}
  {\bfseries 54} (Aug, 1996) 4873--4884}.
  \url{https://link.aps.org/doi/10.1103/PhysRevB.54.4873}.

\bibitem{Jain1997}
J.~K. {Jain} and R.~K. {Kamilla}, ``{Composite Fermions in the Hilbert Space of
  the Lowest Electronic Landau Level},''
  \href{http://dx.doi.org/10.1142/S0217979297001301}{{\em International Journal
  of Modern Physics B} {\bfseries 11} no.~22, (Jan., 1997) 2621--2660},
  \href{http://arxiv.org/abs/cond-mat/9704031}{{\ttfamily
  arXiv:cond-mat/9704031 [cond-mat.mes-hall]}}.

\bibitem{Haldane2011}
F.~D.~M. Haldane, ``Geometrical description of the fractional quantum hall
  effect,'' \href{http://dx.doi.org/10.1103/PhysRevLett.107.116801}{{\em Phys.
  Rev. Lett.} {\bfseries 107} (Sep, 2011) 116801}.
  \url{https://link.aps.org/doi/10.1103/PhysRevLett.107.116801}.

\bibitem{Yang2012b}
B.~Yang, Z.-X. Hu, Z.~Papi\ifmmode~\acute{c}\else \'{c}\fi{}, and F.~D.~M.
  Haldane, ``Model wave functions for the collective modes and the magnetoroton
  theory of the fractional quantum hall effect,''
  \href{http://dx.doi.org/10.1103/PhysRevLett.108.256807}{{\em Phys. Rev.
  Lett.} {\bfseries 108} (Jun, 2012) 256807}.
  \url{https://link.aps.org/doi/10.1103/PhysRevLett.108.256807}.

\bibitem{Golkar2016}
S.~Golkar, D.~X. Nguyen, M.~M. Roberts, and D.~T. Son, ``Higher-spin theory of
  the magnetorotons,''
  \href{http://dx.doi.org/10.1103/PhysRevLett.117.216403}{{\em Phys. Rev.
  Lett.} {\bfseries 117} (Nov, 2016) 216403}.
  \url{https://link.aps.org/doi/10.1103/PhysRevLett.117.216403}.

\bibitem{Balram2017}
A.~C. {Balram} and S.~{Pu}, ``{Positions of the magnetoroton minima in the
  fractional quantum Hall effect},''
  \href{http://dx.doi.org/10.1140/epjb/e2017-80177-5}{{\em European Physical
  Journal B} {\bfseries 90} no.~6, (June, 2017) 124},
  \href{http://arxiv.org/abs/1609.01703}{{\ttfamily arXiv:1609.01703
  [cond-mat.str-el]}}.

\bibitem{Liu2018}
Z.~Liu, A.~Gromov, and Z.~Papi\ifmmode~\acute{c}\else \'{c}\fi{}, ``Geometric
  quench and nonequilibrium dynamics of fractional quantum hall states,''
  \href{http://dx.doi.org/10.1103/PhysRevB.98.155140}{{\em Phys. Rev. B}
  {\bfseries 98} (Oct, 2018) 155140}.
  \url{https://link.aps.org/doi/10.1103/PhysRevB.98.155140}.

\bibitem{Nguyen2021a}
D.~X. Nguyen and D.~T. Son, ``Dirac composite fermion theory of general jain
  sequences,'' \href{http://dx.doi.org/10.1103/PhysRevResearch.3.033217}{{\em
  Phys. Rev. Research} {\bfseries 3} (Sep, 2021) 033217}.
  \url{https://link.aps.org/doi/10.1103/PhysRevResearch.3.033217}.

\bibitem{Nguyen2021b}
D.~X. Nguyen, F.~D.~M. Haldane, E.~H. Rezayi, D.~T. Son, and K.~Yang,
  ``Multiple magnetorotons and spectral sum rules in fractional quantum hall
  systems,'' 2021.
\newblock \url{https://arxiv.org/abs/2111.10593}.

\bibitem{Balram2022}
A.~C. Balram, Z.~Liu, A.~Gromov, and Z.~Papi\ifmmode~\acute{c}\else \'{c}\fi{},
  ``Very-high-energy collective states of partons in fractional quantum hall
  liquids,'' \href{http://dx.doi.org/10.1103/PhysRevX.12.021008}{{\em Phys.
  Rev. X} {\bfseries 12} (Apr, 2022) 021008}.
  \url{https://link.aps.org/doi/10.1103/PhysRevX.12.021008}.

\bibitem{Bartholomew2022}
B.~Andrews and G.~Möller, ``Self-similarity of spectral response functions for
  fractional quantum hall states with long-range interactions,'' 2022.
\newblock \url{https://arxiv.org/abs/2201.04704}.

\bibitem{NayakWilczek1994short}
C.~Nayak and F.~Wilczek, ``{Non-Fermi liquid fixed point in 2 + 1
  dimensions},'' {\em Nucl. Phys. B} {\bfseries 417} (1994) 359.

\bibitem{NayakWilczeklong1994}
C.~Nayak and F.~Wilczek, ``{Renormalization group approach to low temperature
  properties of a non-Fermi liquid metal},'' {\em Nucl. Phys. B} {\bfseries
  430} (1994) 534.

\bibitem{Note2}
Notice that the usual behavior of static structure factor $\protect \bar s(q_x)
  \sim (q_x\ell )^4$ is obtained under the conditions that the ground state has
  translational invariance, inversion symmetry, and a spectral gap.\cite
  {Girvin1986, Haldane2009}. The three conditions lead the static structure
  factor to have no momentum fluctuations, i.e. the $(q_x\ell )^2$ term
  vanishes, be even, and analytic in $q_x$ respectively. At $\nu =1/2$, where
  the excitations above the ground state are gapless, the static structure
  factor does not have to be an analytic function of $q_x$ and can be more
  singular.

\bibitem{Read1998}
N.~Read, ``{Lowest-Landau-level theory of the quantum Hall effect: The
  Fermi-liquid-like state of bosons at filling factor one},'' {\em Phys. Rev.
  B} {\bfseries 58} (1998) 16262.

\bibitem{Haldane1983}
F.~D.~M. Haldane, ``Fractional quantization of the hall effect: A hierarchy of
  incompressible quantum fluid states,''
  \href{http://dx.doi.org/10.1103/PhysRevLett.51.605}{{\em Phys. Rev. Lett.}
  {\bfseries 51} (Aug, 1983) 605--608}.
  \url{https://link.aps.org/doi/10.1103/PhysRevLett.51.605}.

\bibitem{Milsted2013}
A.~Milsted, J.~Haegeman, T.~J. Osborne, and F.~Verstraete, ``Variational matrix
  product ansatz for nonuniform dynamics in the thermodynamic limit,''
  \href{http://dx.doi.org/10.1103/PhysRevB.88.155116}{{\em Phys. Rev. B}
  {\bfseries 88} (Oct, 2013) 155116}.
  \url{https://link.aps.org/doi/10.1103/PhysRevB.88.155116}.

\bibitem{Bernevig2008}
B.~A. Bernevig and F.~D.~M. Haldane, ``Model fractional quantum hall states and
  jack polynomials,''
  \href{http://dx.doi.org/10.1103/PhysRevLett.100.246802}{{\em Phys. Rev.
  Lett.} {\bfseries 100} (Jun, 2008) 246802}.
  \url{https://link.aps.org/doi/10.1103/PhysRevLett.100.246802}.

\bibitem{Bergholtz2008}
E.~J. Bergholtz and A.~Karlhede, ``Quantum hall system in tao-thouless limit,''
  \href{http://dx.doi.org/10.1103/PhysRevB.77.155308}{{\em Phys. Rev. B}
  {\bfseries 77} (Apr, 2008) 155308}.
  \url{https://link.aps.org/doi/10.1103/PhysRevB.77.155308}.

\bibitem{Laughlin1984}
R.~B. Laughlin, ``Levitation of extended-state bands in a strong magnetic
  field,'' \href{http://dx.doi.org/10.1103/PhysRevLett.52.2304}{{\em Phys. Rev.
  Lett.} {\bfseries 52} (Jun, 1984) 2304--2304}.
  \url{https://link.aps.org/doi/10.1103/PhysRevLett.52.2304}.

\bibitem{Seiberg:2016gmd}
N.~{Seiberg}, T.~{Senthil}, C.~{Wang}, and E.~{Witten}, ``{A duality web in 2 +
  1 dimensions and condensed matter physics},''
  \href{http://dx.doi.org/10.1016/j.aop.2016.08.007}{{\em Annals of Physics}
  {\bfseries 374} (Nov., 2016) 395--433},
  \href{http://arxiv.org/abs/1606.01989}{{\ttfamily arXiv:1606.01989
  [hep-th]}}.

\bibitem{Karch:2016sxi}
A.~Karch and D.~Tong, ``{Particle-Vortex Duality from 3d Bosonization},''
\href{http://arxiv.org/abs/1606.01893}{{\ttfamily arXiv:1606.01893 [hep-th]}}.

\bibitem{Chen2018}
J.-Y. Chen, J.~H. Son, C.~Wang, and S.~Raghu, ``Exact boson-fermion duality on
  a 3d euclidean lattice,''
  \href{http://dx.doi.org/10.1103/PhysRevLett.120.016602}{{\em Phys. Rev.
  Lett.} {\bfseries 120} (Jan, 2018) 016602}.
  \url{https://link.aps.org/doi/10.1103/PhysRevLett.120.016602}.

\bibitem{Son2018}
D.~T. Son, ``The dirac composite fermion of the fractional quantum hall
  effect,''
  \href{http://dx.doi.org/10.1146/annurev-conmatphys-033117-054227}{{\em Annual
  Review of Condensed Matter Physics} {\bfseries 9} no.~1, (2018) 397--411}.
  \url{https://doi.org/10.1146/annurev-conmatphys-033117-054227}.

\bibitem{Prabhu2017}
K.~Prabhu and M.~M. Roberts, ``Electrons and composite dirac fermions in the
  lowest landau level,'' \href{http://arxiv.org/abs/1709.02814}{{\ttfamily
  arXiv:1709.02814 [cond-mat.mes-hall]}}.

\bibitem{Giamarchibook}
T.~Giamarchi, {\em Quantum Physics in One Dimension}.
\newblock International Series of Monographs on Physics (Book 121). Oxford
  University Press, 2004.

\bibitem{Delft1998BosonizationFB}
J.~von Delft and H.~Schoeller, ``Bosonization for beginners —
  refermionization for experts,'' {\em Annalen der Physik} {\bfseries 7} (1998)
  225--305.

\bibitem{Moore1991}
G.~Moore and N.~Read, ``{Nonabelions in the fractional quantum hall effect},''
  \href{http://dx.doi.org/http://dx.doi.org/10.1016/0550-3213(91)90407-O}{{\em
  Nuclear Physics B} {\bfseries 360} no.~2--3, (1991) 362}.
  \url{http://www.sciencedirect.com/science/article/pii/055032139190407O}.

\bibitem{Haldane2009}
F.~D.~M. Haldane, ``"hall viscosity" and intrinsic metric of incompressible
  fractional hall fluids,''. \url{https://arxiv.org/abs/0906.1854}.

\end{thebibliography}\endgroup


\providecommand{\href}[2]{#2}\begingroup\raggedright\endgroup

\appendix

\section{Additional numerical details at $\nu=1/3$\label{sec:Laughlin_appendix}}

\begin{figure*}
	\centering
	
	\begin{minipage}{0.33\textwidth}
		\begin{center}
			\includegraphics[width=0.9\textwidth]{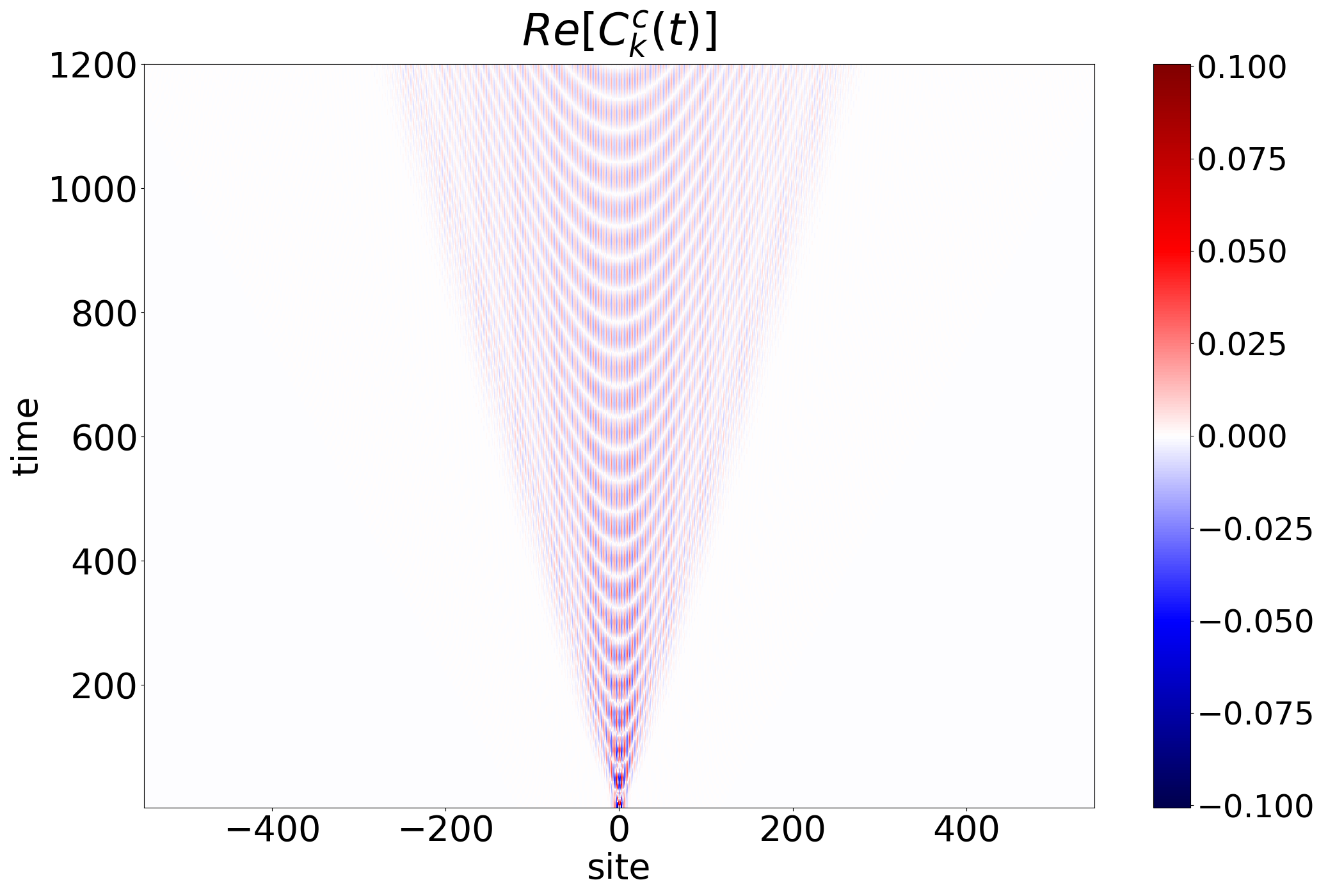}
			
			(a)
		\end{center}
	\end{minipage}%
	\begin{minipage}{0.33\textwidth}
		\begin{center}
			\includegraphics[width=0.9\textwidth]{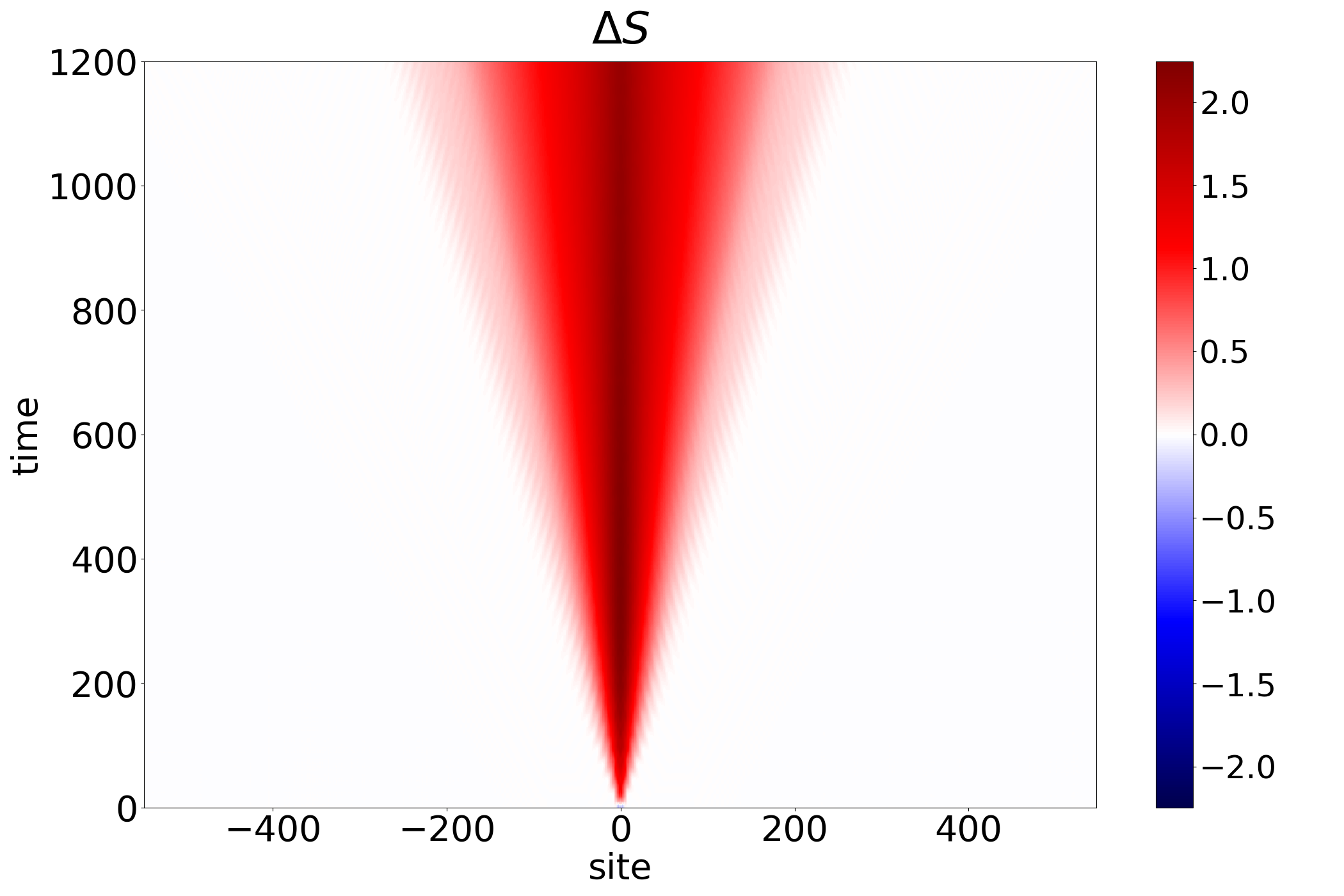}
			
			(b)
		\end{center}
	\end{minipage}%
	\begin{minipage}{0.33\textwidth}
		\begin{center}
			\includegraphics[width=0.9\textwidth]{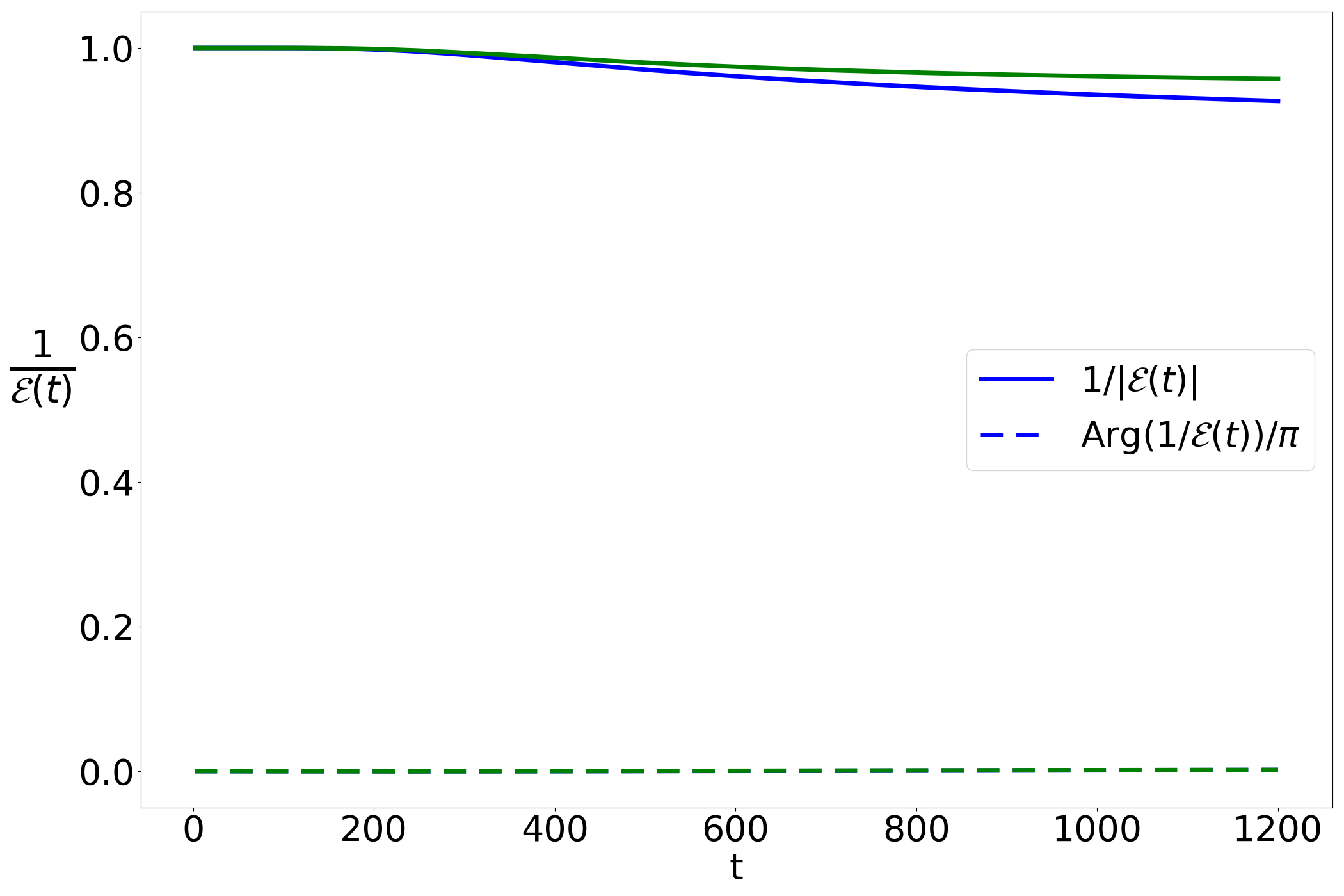}
			
			(c)
		\end{center}
	\end{minipage}
	\begin{minipage}{0.33\textwidth}
		\begin{center}
			\vspace{10pt}
			\includegraphics[width=0.9\textwidth]{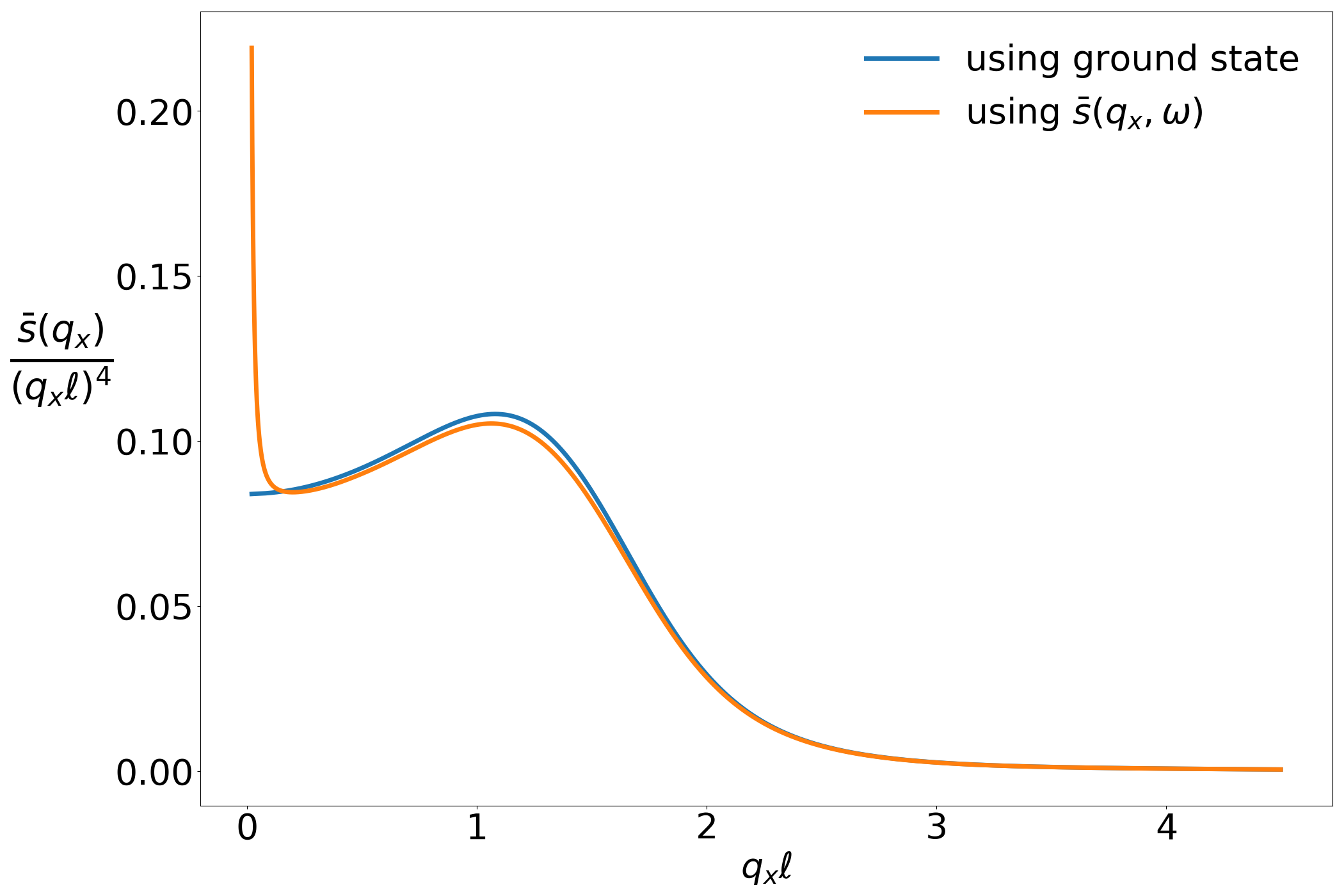}
			
			(d)
		\end{center}
	\end{minipage}%
	\begin{minipage}{0.33\textwidth}
		\begin{center}
			\vspace{10pt}
			\includegraphics[width=0.9\textwidth]{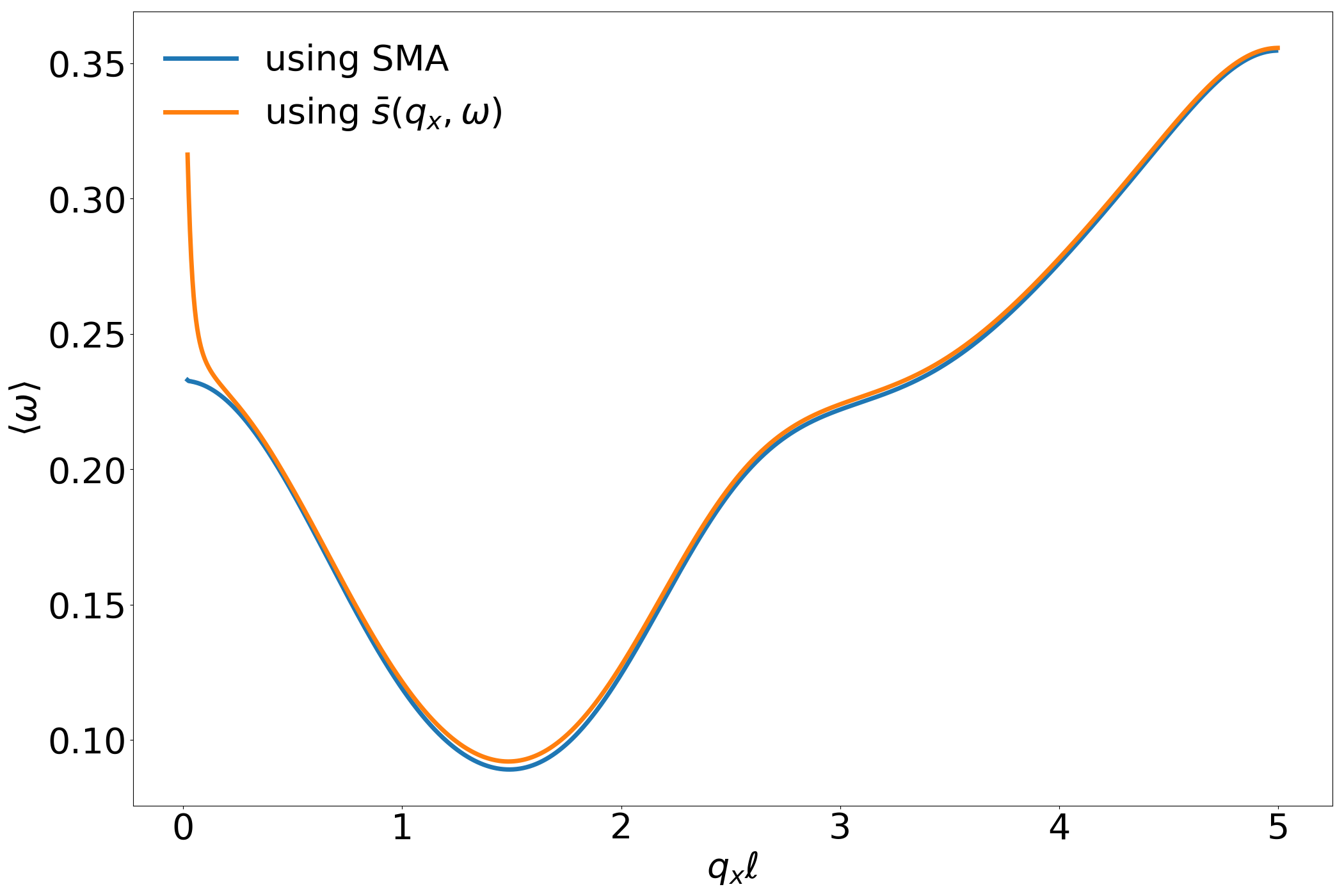}
			
			(e)
		\end{center}
	\end{minipage}
	
	\caption{Additional figures for Laughlin state at $\nu=1/3$, $L_y=10\ell$ in the presence of $V_1$ Haldane pseudopotential interaction. (a) The connected correlation function defined in Eq. \eqref{eq:con_corr} (b) The excess bipartite von Neumann entanglement entropy $\Delta S$ generated during time-evolution. It is obtained by computing the von Neumann entanglement entropy of the time-evolved excited state when the cylinder is cut into two halves to the left of the given site and then subtracting the ground state value.  (c) The error $\mathcal{E}(t)$ vs time. The blue and the green curve correspond to the excited states obtained by applying the occupation number operator $n_k$ on the $2^{\rm nd}$ and $3^{\rm rd}$ orbital in the unit cell respectively. (d) The comparison of the static structure factor obtained from the ground state vs. the integral of the dynamical structure factor over $\omega>0$. (e) First moment of the dynamical structure factor $\langle \omega\rangle$ vs. SMA.}
	\label{fig:Laughlin_additional_figures}
\end{figure*}

In Fig. \ref{fig:Laughlin_additional_figures}, we plot additional figures showing the details of the time evolution for the $V_1$ Haldane pseudopotential interaction at $L_y=10\ell$. The bond dimension of the ground state is $\chi=80$ and the maximum bond dimension in the time-evolved region is set to be $\chi = 512$.

The correlation and entanglement entropy spread out in a light cone with nearly equal speeds. The error $\mathcal{E}(t)$ defined in Eq. \eqref{eq:error_tdvp} remains small until long times. Further, the static structure factor calculated from the ground state agrees with the one calculated by integrating the dynamical structure factor. Moreover, as explained in Eq. \eqref{eq:SMA_avg_E}, the SMA energy is close to the first moment of the dynamical structure factor $\bar s(q_x,\omega)$. 

\section{Additional numerical details at $\nu=1/2$\label{sec:CFL_appendix}}

\subsection{Two-wires}
\begin{figure*}
	\centering
	
	\begin{minipage}{0.33\textwidth}
		\begin{center}
			\includegraphics[width=0.9\textwidth]{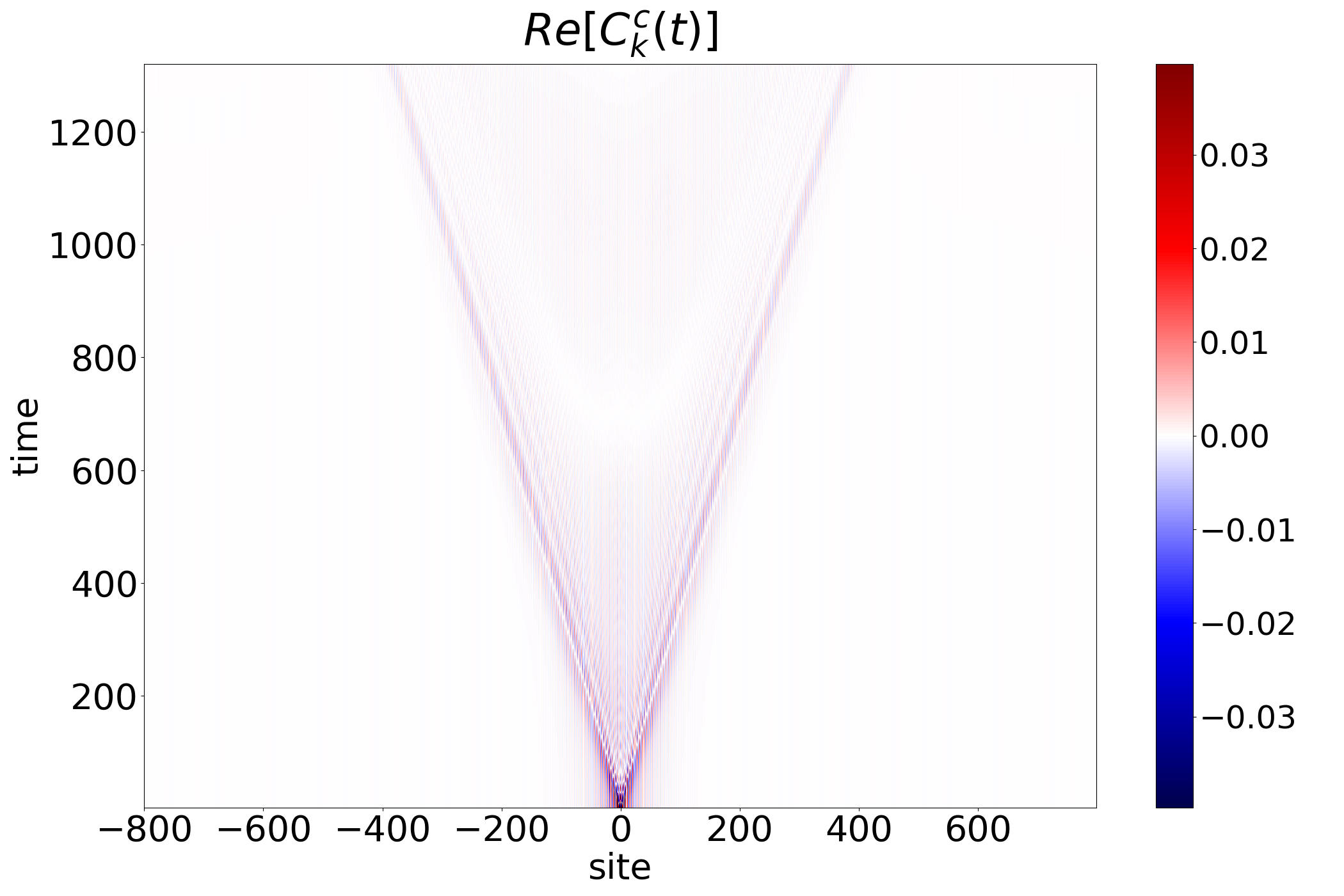}
			
			(a)
		\end{center}
	\end{minipage}%
	\begin{minipage}{0.33\textwidth}
		\begin{center}
			\includegraphics[width=0.9\textwidth]{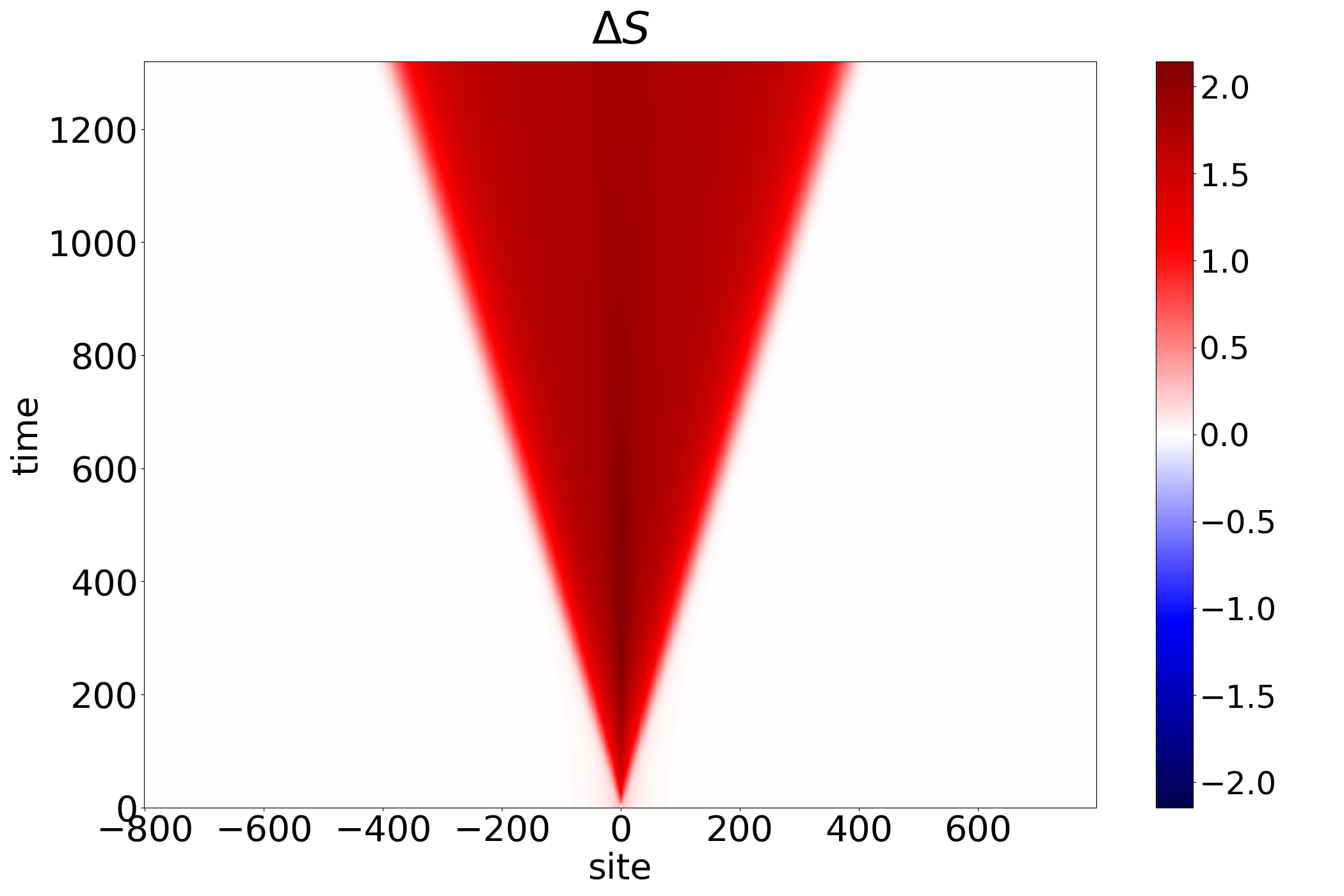}
			
			(b)
		\end{center}
	\end{minipage}%
	\begin{minipage}{0.33\textwidth}
		\begin{center}
			\includegraphics[width=0.9\textwidth]{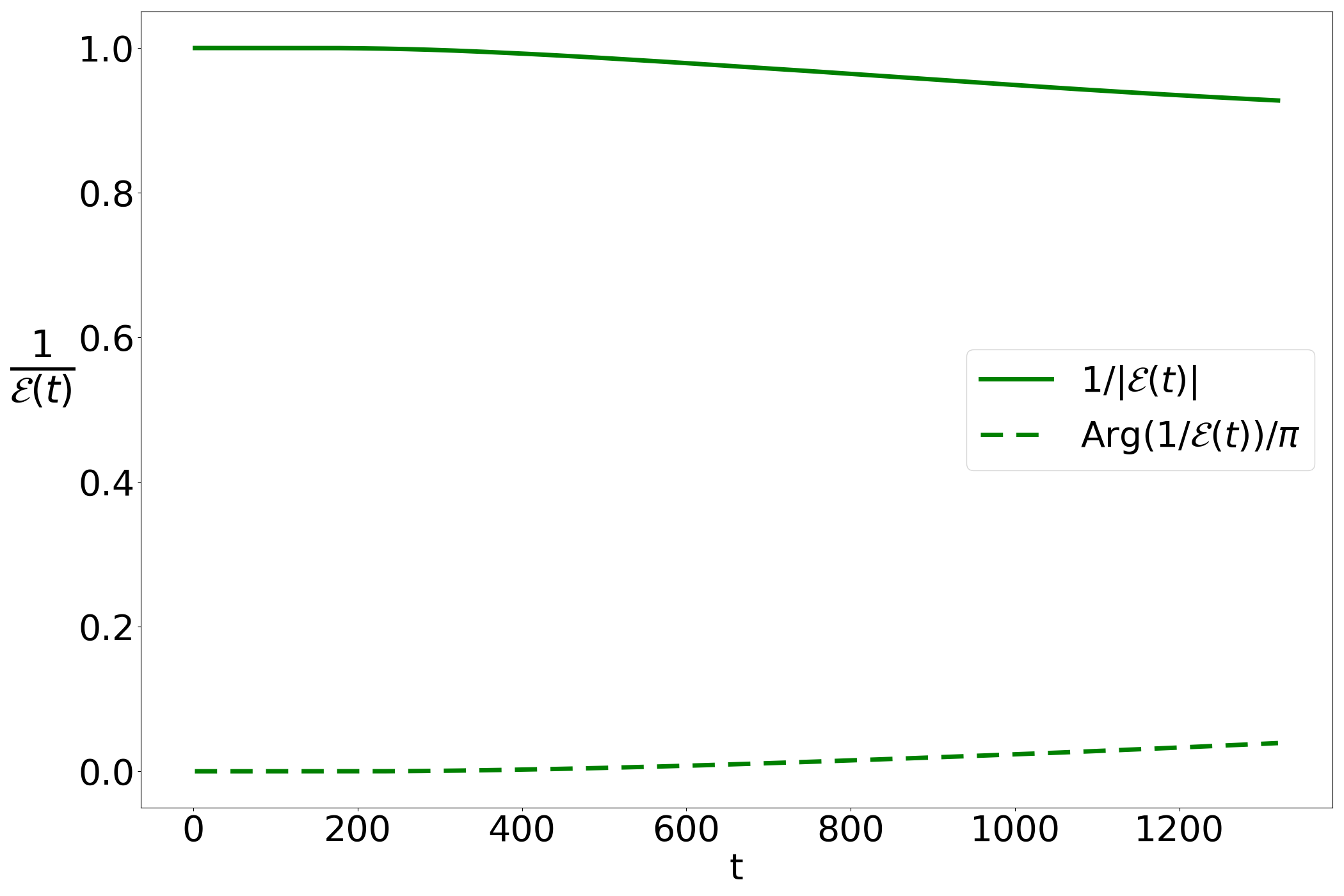}
			
			(c)
		\end{center}
	\end{minipage}
	\begin{minipage}{0.33\textwidth}
		\begin{center}
			\vspace{10pt}
			\includegraphics[width=0.9\textwidth]{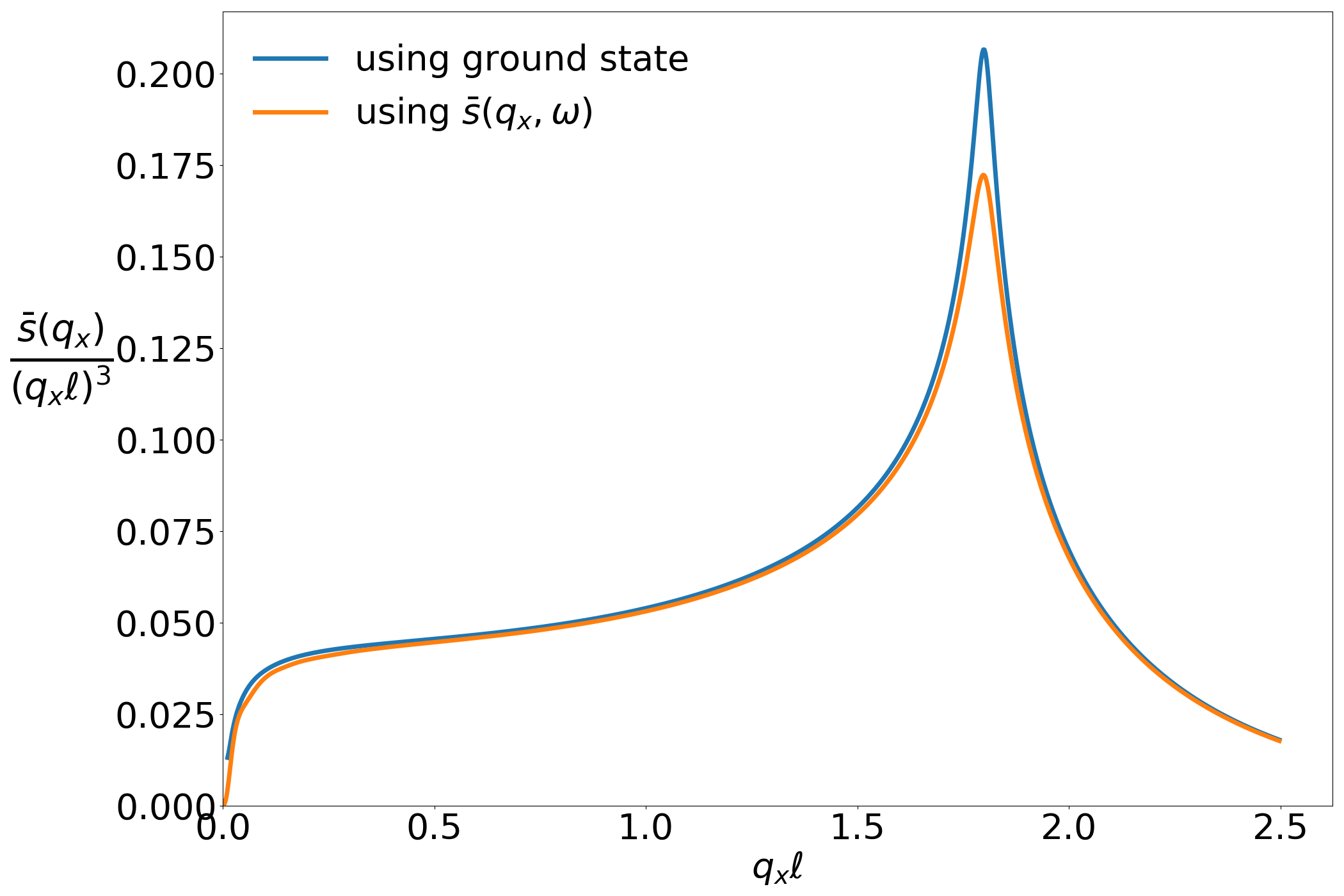}
			
			(d)
		\end{center}
	\end{minipage}%
	\begin{minipage}{0.33\textwidth}
		\begin{center}
			\vspace{10pt}
			\includegraphics[width=0.9\textwidth]{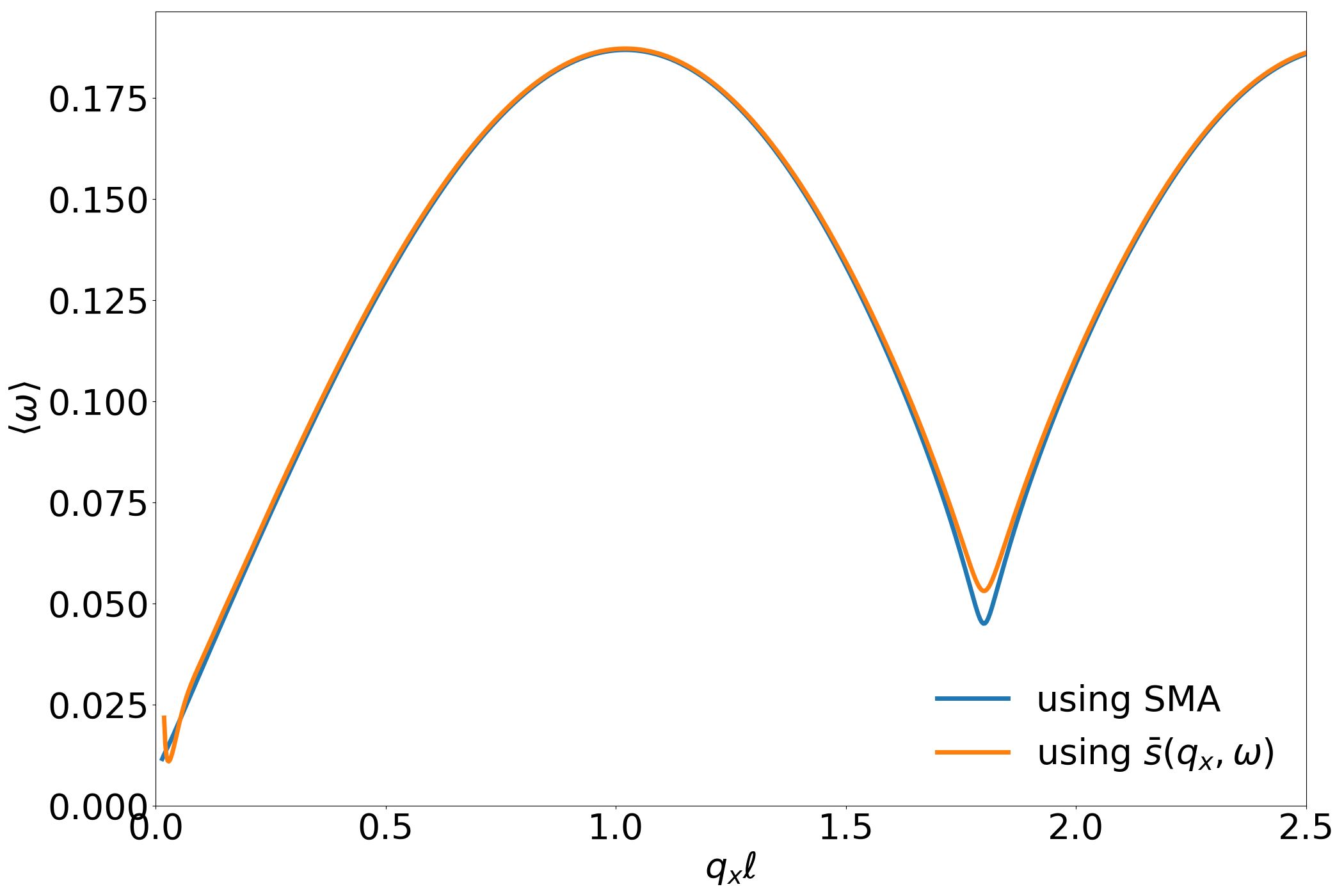}
			
			(e)
		\end{center}
	\end{minipage}
	
	\caption{Additional figures for CFL state with a Fermi sea composed of two-wires, $L_y=7.2\ell$ in the presence of Coulomb interaction. (a) The connected correlation function, (b) the excess bipartite von Neumann entanglement entropy, (c) The error $\mathcal{E}(t)$ vs. time. (d) The static structure factor obtained from the ground state vs. the integral of the dynamical structure factor over $\omega>0$. (e) First moment of the dynamical structure factor $\langle \omega\rangle$ vs. SMA.}
	\label{fig:CFL_2_additional_figures}
\end{figure*}
\begin{figure*}
	\centering
	
	\begin{minipage}{0.5\textwidth}
		\begin{center}
			\includegraphics[width=0.8\textwidth]{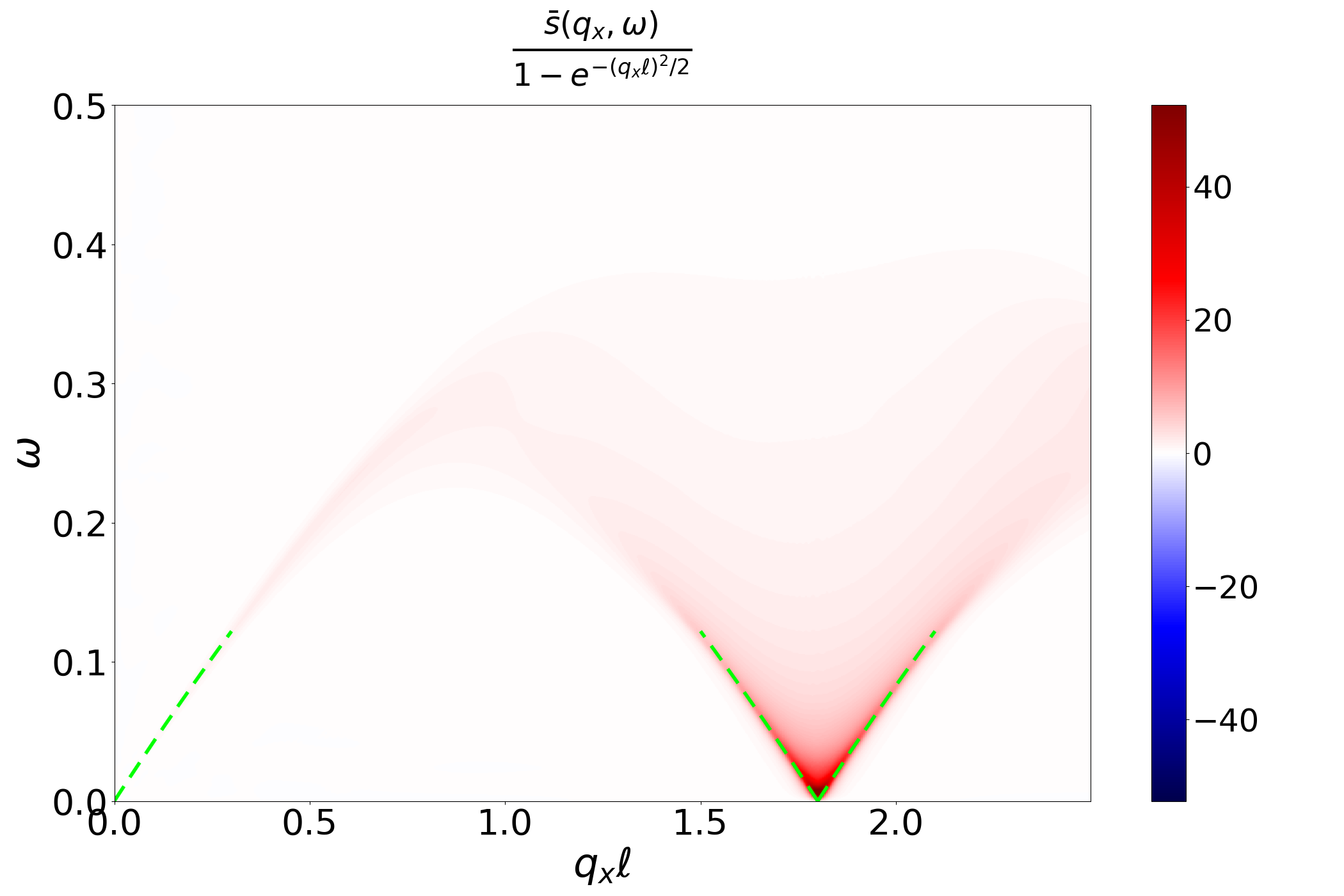}
			
			(a)
		\end{center}
	\end{minipage}%
	\begin{minipage}{0.5\textwidth}
		\begin{center}
			\includegraphics[width=0.64\textwidth]{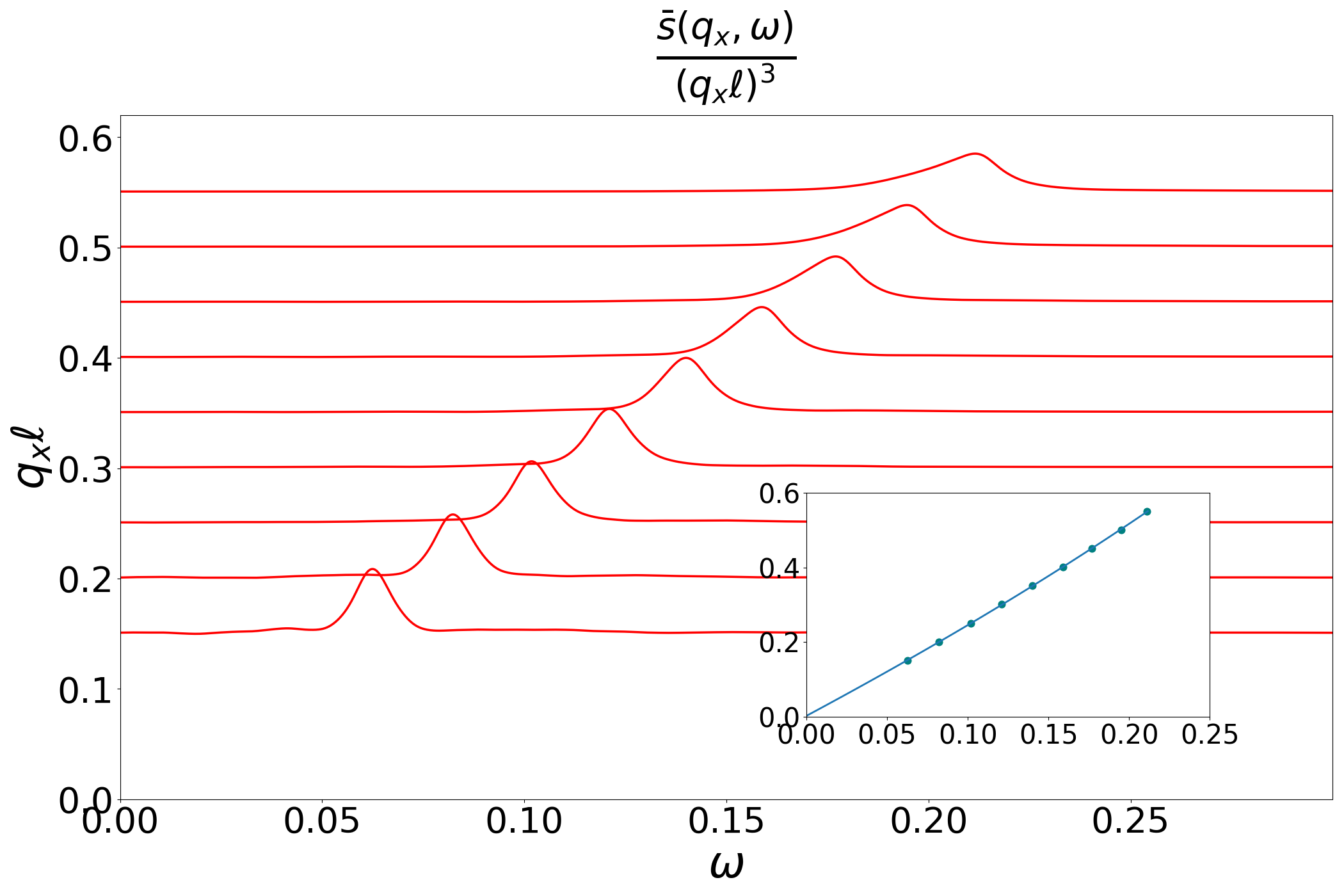}
			
			(b)
		\end{center}
	\end{minipage}
	\begin{minipage}{0.33\textwidth}
		\begin{center}
			\vspace{10pt}
			\includegraphics[width=0.9\textwidth]{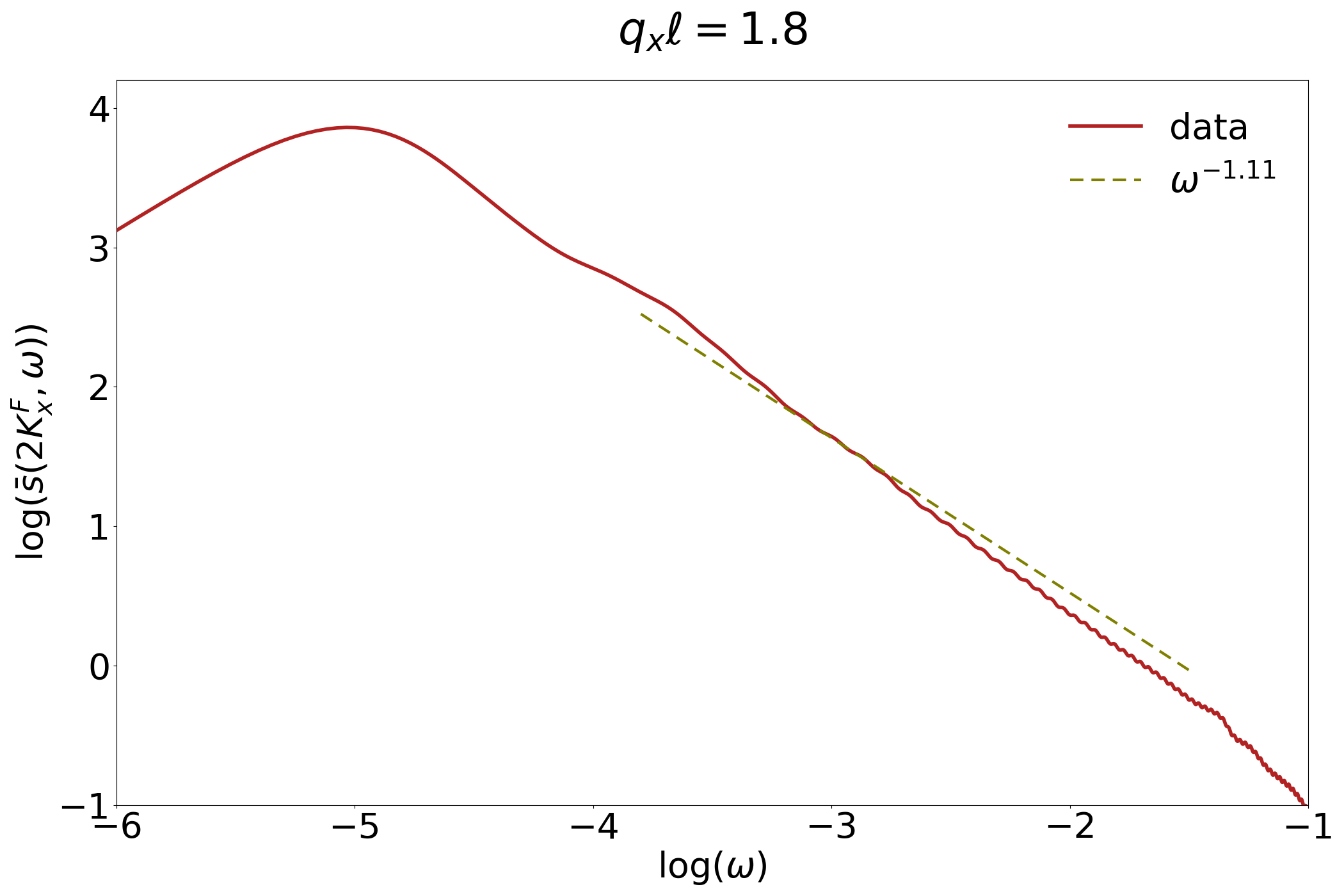}
			
			(c)
		\end{center}
	\end{minipage}%
	\begin{minipage}{0.33\textwidth}
		\begin{center}
			\includegraphics[width=0.9\textwidth]{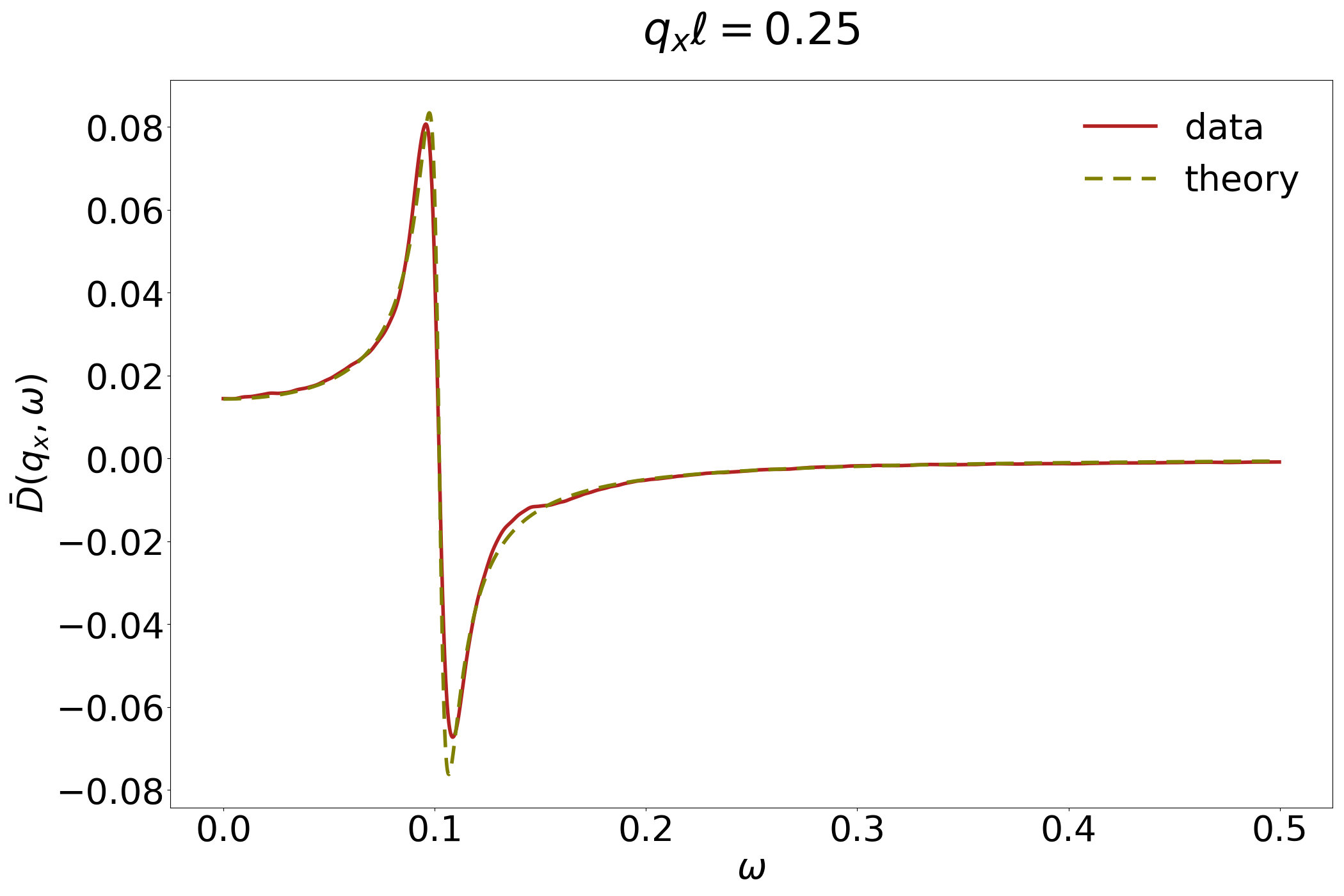}
			
			(d)
		\end{center}
	\end{minipage}%
	\begin{minipage}{0.33\textwidth}
		\begin{center}
			\includegraphics[width=0.9\textwidth]{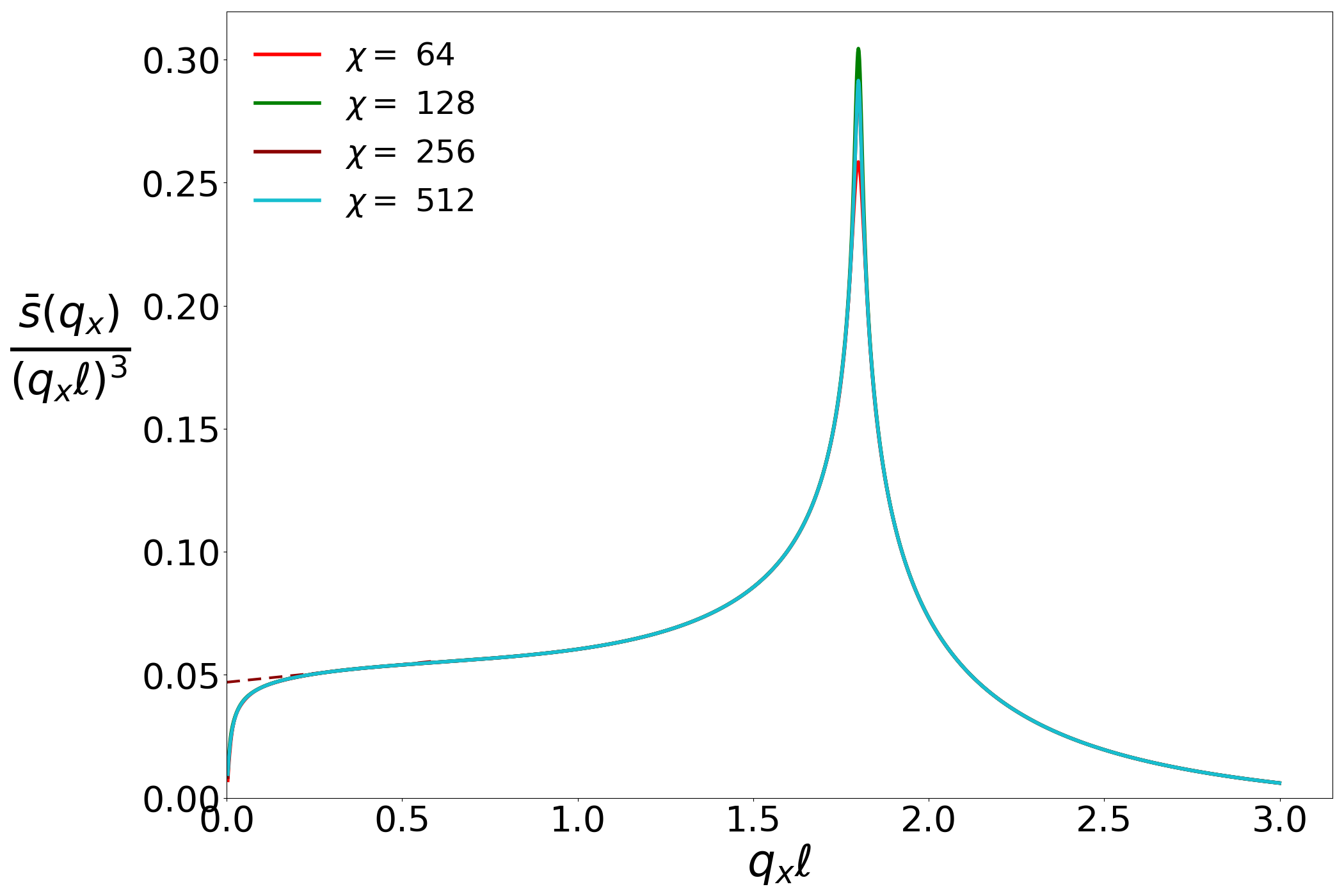}
			
			(e)
		\end{center}
	\end{minipage}
	
	\caption{(a) The dynamical structure factor $\bar s(q_x,\omega)$ of the $\nu=1/2$ state at $L_y=7.2\ell$ for Gaussian Coulomb interaction \eqref{eq:gaussian_coulomb} with $\xi=20$. The results are similar to Fig. \ref{fig:CFL_Coulomb_two_wires}. (b) The low energy mode at small $q_x\ell$ and the inset shows the linear dispersion with a velocity $u_- = 0.43 V_1 \ell$ where $V_1$ represents the strength of the interaction. (c) $\bar s(2K^F_x, \omega)$ vs $\omega$ on a log-log scale, (d) Real part of the density-density correlation function $\bar D(q_x,\omega)$ at $q_x\ell = 0.25$, and (e) static structure factor $\bar s(q_x)/(q_x\ell)^3$ for different bond dimensions. Dashed curves in (c), (d) and (e) show theoretical prediction with Luttinger parameter $K_- = 1.125$. In (e), We estimate a gap $\omega_g$ equal to $0.01 e^2/\ell$ using SMA.}
	\label{fig:CFL_Haldane_two_wires}
\end{figure*}
In Fig. \ref{fig:CFL_2_additional_figures}, we plot additional figures similar to the case of $\nu=1/3$ for $L_y=7.2\ell$ in the presence of Coulomb interaction. The bond dimension of the ground state is $\chi=128$, while the maximum bond dimension in the region of time-evolution is $\chi=512$. The error $\mathcal{E}(t)$ is more significant than the $\nu=1/3$ case at late times, however, other quantities such as the first moment and the frequency integral of the dynamical structure factor are accurate except at small $q_x\ell$.

As a complement to the Coulomb interaction, we show the results for $L_y=7.2\ell$ in the presence of $V_1$ Haldane pseudopotential interaction in Fig. \ref{fig:CFL_Haldane_two_wires}. The results are similar to the former case. The Luttinger parameter is found to be $K_- = 1.125$ and there is a small gap at small $q_x\ell$ due to the CF pair instability. The agreement between the theory and data for the power law $\bar s(2K_x^F, \omega) \sim 1/\omega^{2-1/K_-}$ is slightly worse when compared with the Coulomb interaction case presented in the main text. It may be a result of the inaccuracies caused by larger entanglement entropies observed for $V_1$ Haldane pseudopotential interaction.

\subsection{Three-wires}

\begin{figure*}
	\centering
	
	\begin{minipage}{0.33\textwidth}
		\begin{center}
			\includegraphics[width=0.9\textwidth]{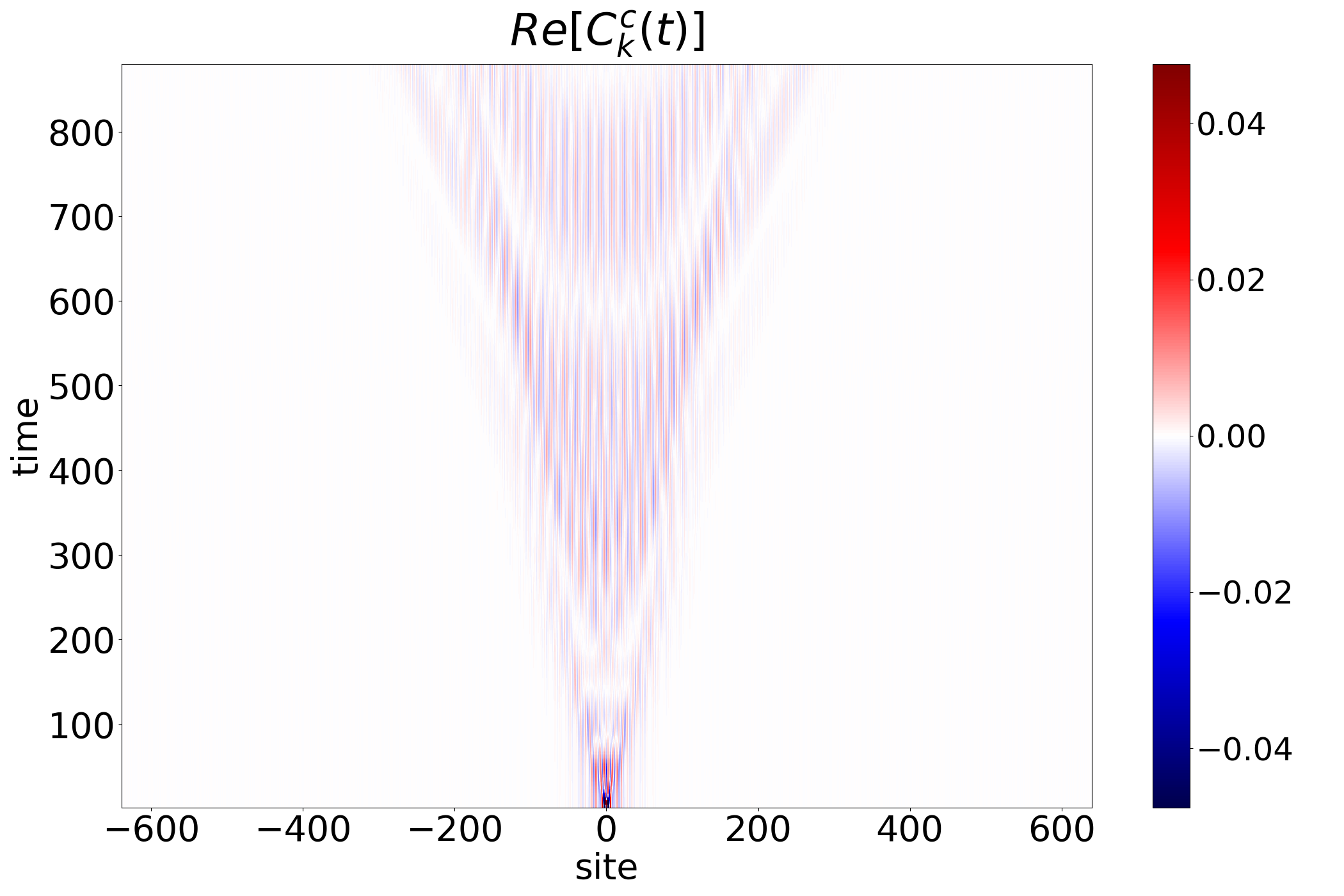}
			
			(a)
		\end{center}
	\end{minipage}%
	\begin{minipage}{0.33\textwidth}
		\begin{center}
			\includegraphics[width=0.9\textwidth]{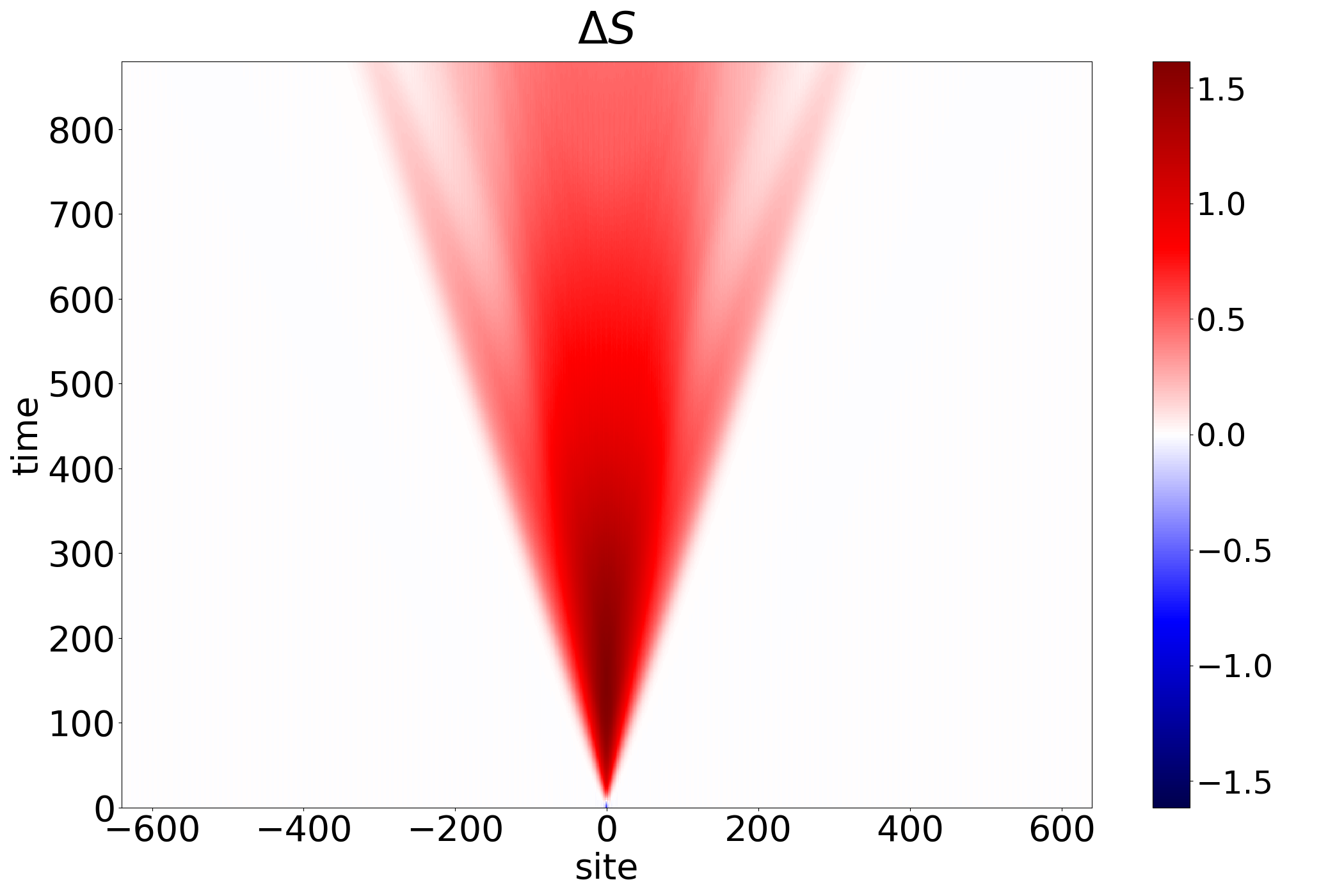}
			
			(b)
		\end{center}
	\end{minipage}%
	\begin{minipage}{0.33\textwidth}
		\begin{center}
			\includegraphics[width=0.9\textwidth]{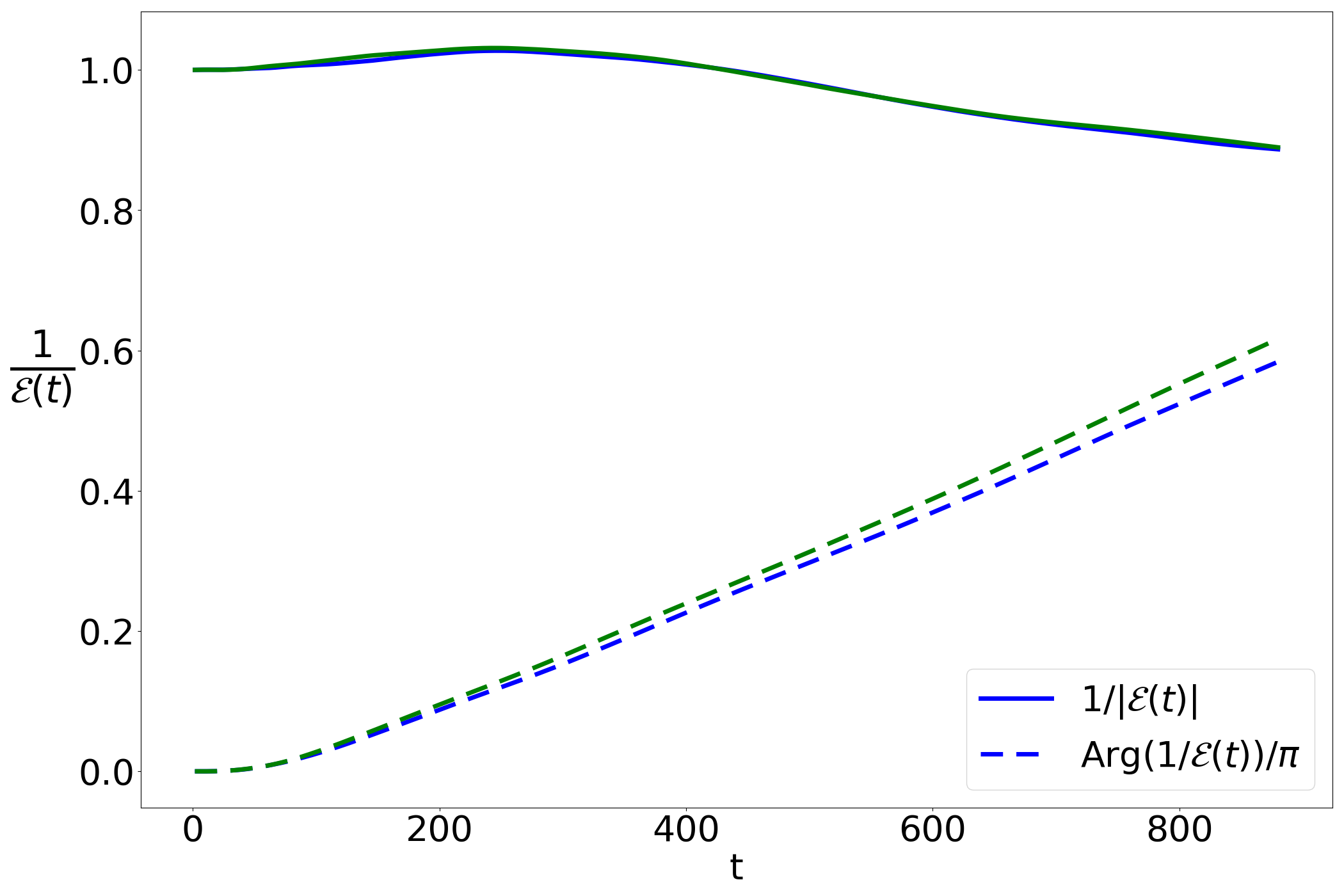}
			
			(c)
		\end{center}
	\end{minipage}
	\begin{minipage}{0.33\textwidth}
		\begin{center}
			\vspace{10pt}
			\includegraphics[width=0.9\textwidth]{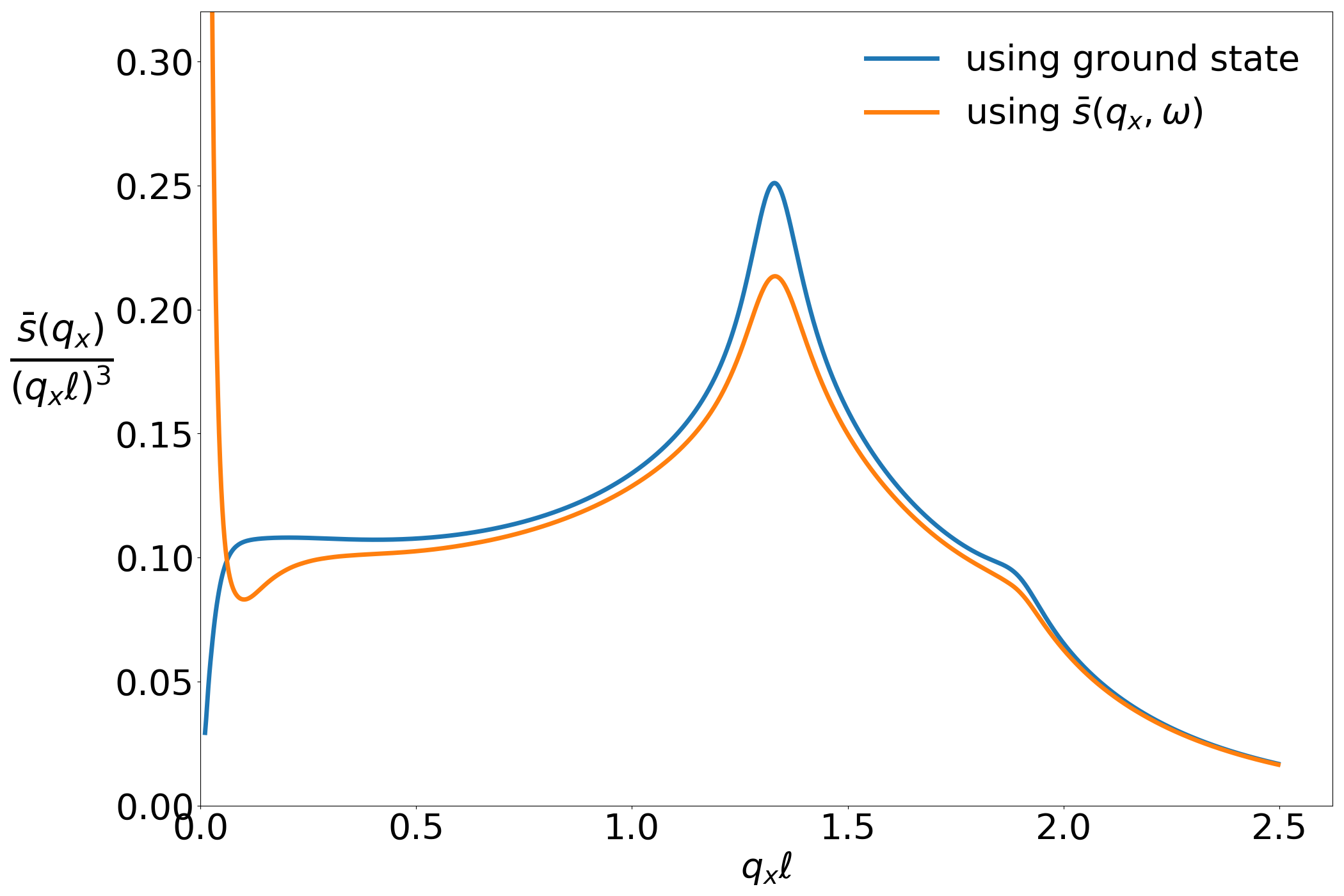}
			
			(d)
		\end{center}
	\end{minipage}%
	\begin{minipage}{0.33\textwidth}
		\begin{center}
			\vspace{10pt}
			\includegraphics[width=0.9\textwidth]{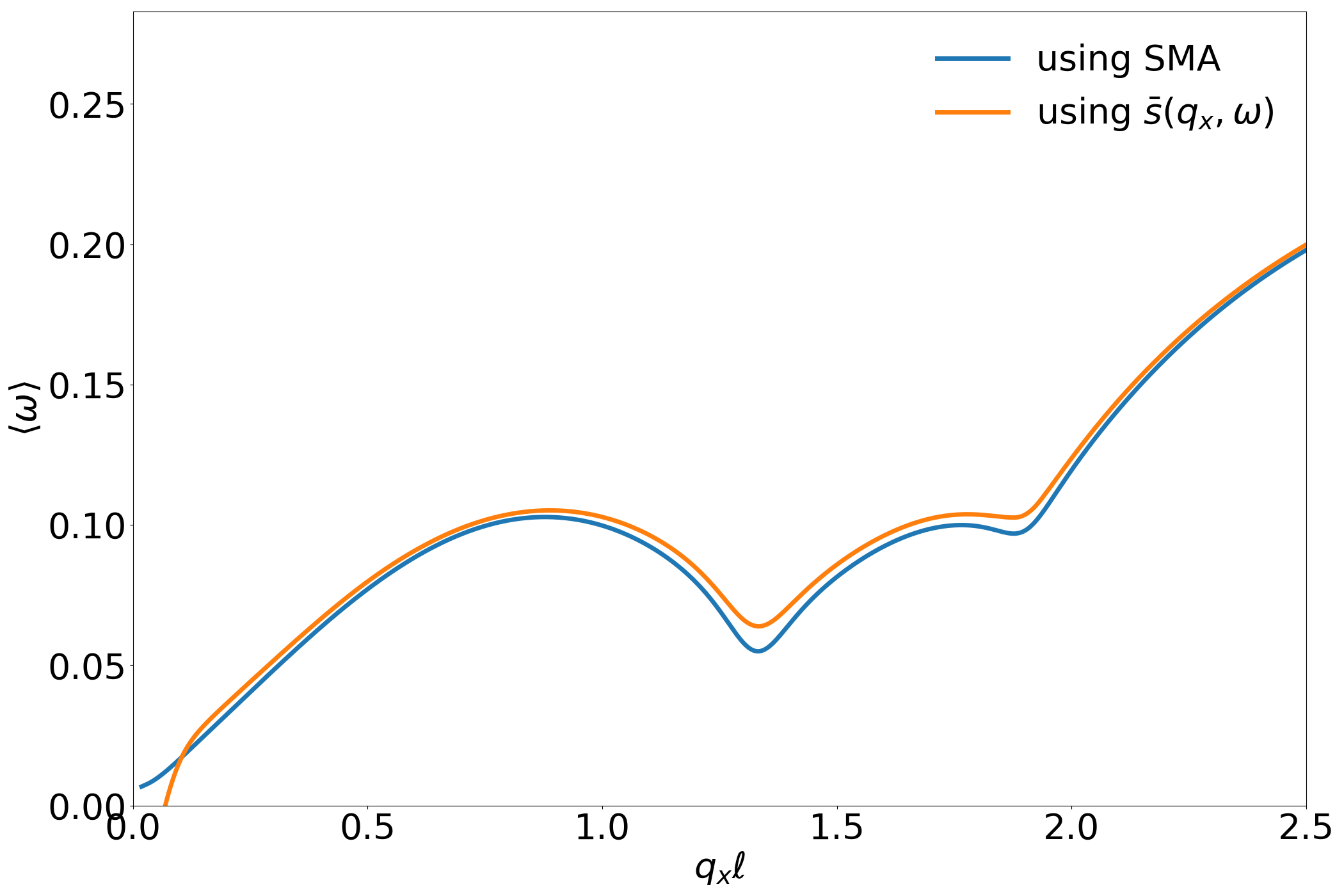}
			
			(e)
		\end{center}
	\end{minipage}
	
	\caption{Additional figures for CFL state with a Fermi sea composed of three-wires, $L_y=9.1\ell$ in the presence of Coulomb interaction. (a) The connected correlation function, (b) the excess bipartite von Neumann entanglement entropy, (c) The error $\mathcal{E}(t)$ vs time. The blue and green curves correspond to the excited states obtained by applying the occupation number operator on the two sites inside the unit cell. (d) The static structure factor obtained from the ground state vs. the integral of the dynamical structure factor over $\omega>0$. (e) First moment of the dynamical structure factor $\langle \omega\rangle$ vs. SMA.}
	\label{fig:CFL_3_additional_figures}
\end{figure*}

The additional figures for the three-wires case at $L_y=9.1\ell$ in the presence of Coulomb interaction are shown in Fig. \ref{fig:CFL_3_additional_figures}. The bond-dimension of the ground state is $\chi=200$ while the maximum bond-dimension of the time-evolved state is chosen to be $\chi=700$. The errors are significantly larger when compared with the CFL composed of two-wires.

\end{document}